\pdfoutput=1
\documentclass[11pt,twoside,a4paper,cmspaper,final,collab]{cms-tdr}
\def\svnVersion{aeb128c}\def\svnDate{2024/06/12}\def\cmsCernNoTag{CERN-EP-2024-035}\def\cmsCernDate{\today}\def\cmsMessage{Published in the European Physical Journal C as \href{http://dx.doi.org/10.1140/epjc/s10052-024-12925-0}{\doi{10.1140/epjc/s10052-024-12925-0}.}}
\begin{document}\cmsNoteHeader{HIG-22-008}

\newcommand{\WW}{\ensuremath{\PW\PW}\xspace}  
\newcommand{\ZZ}{\ensuremath{\PZ\PZ}\xspace}
\newcommand{\HWW}{\ensuremath{\PH\PW\PW}\xspace} 
\newcommand{\HtoWW}{\ensuremath{\PH\to\PW\PW}\xspace} 
\newcommand{\HtoZZ}{\ensuremath{\PH\to\PZ\PZ}\xspace}
\newcommand{\HtoTT}{\ensuremath{\PH\to\PGt\PGt}\xspace}
\newcommand{\HVV}{\ensuremath{\PH\PV\PV}\xspace}
\newcommand{\Hgg}{\ensuremath{\PH\Pg\Pg}\xspace}
\newcommand{\glgl}{\ensuremath{\Pg\Pg}\xspace}
\newcommand{\ggH}{\ensuremath{\Pg\Pg\PH}\xspace}
\newcommand{\ttH}{\ensuremath{\ttbar{}\PH}\xspace}
\newcommand{\bbH}{{\PQb{}\PAQb{}\PH}\xspace} 
\newcommand{\VH}{\ensuremath{\PV\PH}\xspace}
\newcommand{\cw}{\ensuremath{c_\text{w}}\xspace}
\newcommand{\sw}{\ensuremath{s_\text{w}}\xspace}
\newcommand{\sss}{\vspace{2 mm}\noindent}
\newcommand{\njet}{\ensuremath{N_\text{jet}}\xspace} 
\newcommand{\wg}{\ensuremath{\PW{}\PGg}\xspace} 
\newcommand{\wgs}{\ensuremath{\PW{}{\PGg}^{*}}\xspace} 
\newcommand{\mll}{\ensuremath{m_{\Pell\Pell}}\xspace} 
\newcommand{\mTH}{\ensuremath{\mT^{\PH}}\xspace} 
\newcommand{\mTl}{\ensuremath{\mT^{\Pell 2}}\xspace} 
\newcommand{\mjj}{\ensuremath{m_{\text{jj}}}\xspace} 
\newcommand{\mj}{\ensuremath{m_{\text{J}}}\xspace} 
\newcommand{\ptll}{\ensuremath{\pt^{\Pell\Pell}}\xspace} 
\newcommand{\tautau}{\ensuremath{\PGt{}\PGt}\xspace} 
\newcommand{\fai}{\ensuremath{f_{ai}}\xspace}
\newcommand{\fathree}{\ensuremath{f_{a3}}\xspace}
\newcommand{\fatwo}{\ensuremath{f_{a2}}\xspace}
\newcommand{\fL}{\ensuremath{f_{\Lambda 1}}\xspace}
\newcommand{\fLzg}{\ensuremath{f_{\Lambda 1}^{\PZ\PGg}}\xspace}
\newcommand{\fagg}{\ensuremath{f_{a3}^{\ggH}}\xspace}
\newcommand{\ptlone}{\ensuremath{\pt^{\Pell 1}}\xspace} 
\newcommand{\ptltwo}{\ensuremath{\pt^{\Pell 2}}\xspace} 
\newcommand{\tausj}{\ensuremath{\tau_{2}/\tau_{1}}\xspace} 
\newlength\cmsTabSkip\setlength{\cmsTabSkip}{1ex}
\ifthenelse{\boolean{cms@external}}{\providecommand{\cmsLeft}{upper\xspace}}{\providecommand{\cmsLeft}{left\xspace}}
\ifthenelse{\boolean{cms@external}}{\providecommand{\cmsRight}{lower\xspace}}{\providecommand{\cmsRight}{right\xspace}}

\cmsNoteHeader{HIG-22-008}

\title{Constraints on anomalous Higgs boson couplings from its production and decay using the \texorpdfstring{\WW}{WW} channel in proton-proton collisions at \texorpdfstring{$\sqrt{s} = 13\TeV$}{sqrt(s) = 13 TeV}}

\date{\today}

\abstract{
A study of the anomalous couplings of the Higgs boson to vector bosons, including  $CP$-violation effects, has been conducted using its production and decay in the \WW channel. This analysis is performed on proton-proton collision data collected with the CMS detector at the CERN LHC during 2016--2018 at a center-of-mass energy of 13\TeV, and corresponds to an integrated luminosity of 138\fbinv.  The different-flavor dilepton ($\Pe\PGm$) final state is analyzed, with dedicated categories targeting gluon fusion, electroweak vector boson fusion, and associated production with a \PW or \PZ boson. Kinematic information from associated jets is combined using matrix element techniques to increase the sensitivity to anomalous effects at the production vertex. A simultaneous measurement of four Higgs boson couplings to electroweak vector bosons is performed in the framework of a standard model effective field theory. All measurements are consistent with the expectations for the standard model Higgs boson and constraints are set on the fractional contribution of the anomalous couplings to the Higgs boson production cross section.
}

\hypersetup{
pdfauthor={CMS Collaboration},
pdftitle={Constraints on anomalous Higgs boson couplings from its production and decay using the WW channel in proton-proton collisions at sqrt(s) = 13 TeV},
pdfsubject={CMS},
pdfkeywords={CMS, Higgs bosons, WW, anomalous couplings, EFT}} 

\titlerunning{Constraints on anomalous Higgs boson couplings from its production and decay using the \texorpdfstring{\WW}{WW} channel}
\maketitle 

\section{Introduction}
{\tolerance=800 After the discovery of the Higgs boson (\PH) by the ATLAS and CMS Collaborations in 2012~\cite{Aad:2012tfa,Chatrchyan:2012xdj,Chatrchyan:2013lba}, the CMS~\cite{Chatrchyan:2012jja,Chatrchyan:2013mxa,Khachatryan:2014kca,Khachatryan:2015mma,Khachatryan:2016tnr, Sirunyan:2017tqd,Sirunyan:2019twz,Sirunyan:2019nbs} and ATLAS~\cite{Aad:2013xqa,Aad:2015mxa,Aad:2016nal,Aaboud:2017oem,Aaboud:2017vzb,Aaboud:2018xdt,Aad:2020mnm} experiments set constraints on the spin-parity properties of the Higgs boson and its couplings with gluons and electroweak (EW) gauge bosons, denoted here as \Hgg and \HVV, respectively. The Higgs boson quantum numbers are consistent with the standard model (SM) expectation $J^{PC} = 0^{++}$, but the possibility of small, anomalous couplings is not yet ruled out.
In beyond-the-SM (BSM) theories, interactions with the Higgs boson may occur through several anomalous couplings, which lead to new tensor structures in the interaction terms that can be both $CP$-even or $CP$-odd. The $CP$-odd anomalous couplings between the Higgs boson and BSM particles may generate $CP$ violation in the interactions of the Higgs boson.\par}
In this paper, we study the tensor structure of the \Hgg and \HVV couplings, and we search for several anomalous effects, including $CP$ violation, using the different-flavor dilepton ($\Pe\PGm$) final state from \HtoWW decays. The Higgs boson production processes include gluon fusion (\ggH), EW vector boson fusion (VBF), and associated production with a \PW or \PZ boson (\VH).
Higgs boson production and decay processes are sensitive to certain anomalous contributions, which can be described by higher-dimensional operators in an effective field theory (EFT)~\cite{deFlorian:2227475} that can modify the kinematic distributions of the Higgs boson decay products and the particles from associated production.

Each production process of the Higgs boson is identified using its kinematic features, and events are assigned to corresponding production categories. The matrix element likelihood approach (MELA)~\cite{Gao:2010qx,Bolognesi:2012mm,Anderson:2013afp,Gritsan:2016hjl,Gritsan:2020pib} is employed to construct observables that are optimal for the measurement of anomalous couplings, or EFT operators, at the production vertex.
These and other decay-based variables are used to explore all kinematic features of the events, giving the analysis sensitivity to simultaneous anomalous effects at the Higgs boson production and decay vertices.
Fully simulated signal samples that include anomalous couplings incorporate the detector response into the analysis.

The analysis is based on the proton-proton ($\Pp\Pp$) collision data collected at the CERN LHC from 2016 to 2018, at a center-of-mass energy of 13\TeV, corresponding to an integrated luminosity of 138\fbinv. This paper builds on a previous analysis conducted by the CMS Collaboration in the \HtoWW channel~\cite{CMS:2022uhn}, which focused on measuring the Higgs boson production cross sections and coupling parameters in the so-called $\kappa$ framework~\cite{Heinemeyer:1559921}. We follow a formalism used in previous CMS analyses of anomalous couplings in Run 1 and Run 2~\cite{Chatrchyan:2012jja,Chatrchyan:2013mxa,Khachatryan:2014kca,Khachatryan:2015mma,Khachatryan:2016tnr, Sirunyan:2017tqd,Sirunyan:2019twz,Sirunyan:2019nbs,Sirunyan:2020sum,Sirunyan:2017tqd,hzz_ac_2021}, focusing on the case where the Higgs boson is produced on-shell. The coupling parameters are extracted using the signal strength and the fractional contributions of the couplings to the cross section. A general study of the \HVV interaction is performed with four anomalous couplings analyzed individually. Through SU(2) x U(1) symmetry considerations, the anomalous \HVV couplings are reduced in number to three and analyzed simultaneously.
The primary \HVV coupling measurements are performed in terms of cross section fractions with additional interpretations in terms of EFT couplings included.
A study of the \Hgg interaction is also performed in terms of a $CP$-odd anomalous coupling cross section fraction.

This paper is organized as follows. The phenomenology of anomalous couplings is discussed in Section~\ref{sec:phenomenology}.
Section~\ref{sec:cmsdet} gives a brief overview of the CMS apparatus.
Data sets and Monte Carlo (MC) simulation samples are discussed in Section~\ref{sec:datasets}. The event reconstruction and selection are outlined in Sections~\ref{sec:evreco} and~\ref{sec:evsel}, respectively. Methods to estimate backgrounds are given in Section~\ref{sec:background}.
In Section~\ref{sec:KD}, we discuss the kinematic variables associated with Higgs boson production and decay.
Sources of systematic uncertainties are presented in Section~\ref{sec:systematics}. The results are presented and discussed in Section~\ref{sec:results}.
Finally, a summary is given in Section~\ref{sec:summary}.
Tabulated results are provided in the {HEPD}ata record for this analysis~\cite{hepdata}.

\section{Phenomenology}
\label{sec:phenomenology}

In this analysis, we investigate anomalous coupling effects in gluon fusion or electroweak Higgs boson production, as well as in its decay to \WW pairs. A detailed discussion of the theoretical considerations can be found in Refs.~\cite{hzz_ac_2021, Anderson:2013afp, Gritsan:2020pib}. The interaction of the spin-zero Higgs boson with two spin-one gauge bosons $\PV_{1}\PV_{2}$, such as \WW, \ZZ, $\PZ\PGg$, $\PGg\PGg$, or \glgl, can be parametrized by the scattering amplitude
\begin{equation} 
\label{eq:formfactfullampl} 
\begin{aligned}
A(\PH\PV_{1}\PV_{2}) 
\sim 
\left[ a_{1}^{\PV\PV} 
+ \frac{\kappa_1^{\PV\PV}q_{\PV 1}^2 + \kappa_2^{\PV\PV} q_{\PV 2}^{2}}{\left(\Lambda_{1}^{\PV\PV} \right)^{2}} \right]
m_{\sss\PV 1}^2 \epsilon_{\sss\PV 1}^* \epsilon_{\sss\PV 2}^*  \\
+ \frac{1}{v} a_{2}^{\PV\PV}  f_{\mu \nu}^{*(1)}f^{*(2),\mu\nu} 
+ \frac{1}{v} a_{3}^{\PV\PV}  f^{*(1)}_{\mu \nu} {\tilde f}^{*(2),\mu\nu}, 
\end{aligned}
\end{equation} 
where $q_{\PV i}$ and $\epsilon_{\sss\PV i}$ are the spin-one gauge boson four-momentum and polarization vectors, $m_{\sss\PV 1}$ is the pole mass of the boson,  $f^{(i),{\mu \nu}} = \epsilon_{{\sss\PV}i}^{\mu}q_{{\sss\PV}i}^{\nu} - \epsilon_{{\sss\PV}i}^\nu q_{{\sss\PV}i}^{\mu}$ and ${\tilde f}^{(i)}_{\mu \nu} = \frac{1}{2} \epsilon_{\mu\nu\rho\sigma} f^{(i),\rho\sigma}$ (with $\epsilon_{\mu\nu\rho\sigma}$ the Levi-Civita symbol), $\Lambda_{1}^{\PV\PV}$ is the scale of BSM physics, and $v$ is the Higgs field vacuum expectation value. 

{\tolerance=800 The only leading tree-level contributions in the scattering amplitude are $a_{1}^{\PZ\PZ}\ne 0$ and $a_{1}^{\PW\PW} \ne 0$; other $a_{1}$ coupling parameters ($\PZ\PGg$, $\PGg\PGg$, \glgl) do not contribute because the pole mass vanishes. Additional \ZZ and \WW couplings are considered anomalous contributions. Anomalous terms arising in the SM via loop effects are typically small and are not yet accessible experimentally. The BSM contributions, however, could yield larger coupling parameters. Among the anomalous contributions, considerations of symmetry and gauge invariance require 
$\kappa_1^{\PZ\PZ} = \kappa_2^{\PZ\PZ}$, 
$\kappa_1^{\PGg\PGg} = \kappa_2^{\PGg\PGg} = 0$, 
$\kappa_1^{\Pg\Pg} = \kappa_2^{\Pg\Pg} = 0$, 
and $\kappa_1^{\PZ\PGg} = 0$~\cite{Gritsan:2020pib}. 
The presence of $CP$-odd $a^{\PV\PV}_{3}$ couplings together with any of the other couplings (all of them $CP$-even), will result in $CP$ violation. We reduce the number of independent parameters by assuming that $a_2^{\PGg\PGg}$, $a_3^{\PGg\PGg}$, $a_2^{\PZ\PGg}$ and $a_3^{\PZ\PGg}$ are constrained in direct decays of $\PH\to\PGg\PGg$ and $\PZ\PGg$, therefore fixing them to be zero. The $a^{\glgl}_{2}$ term results from loop effects in the SM.\par}
The relationship between the \ZZ and \WW couplings is mostly relevant for VBF production. There are no kinematic differences between the \ZZ and \WW fusion processes; therefore, it is not possible to disentangle the couplings. One possibility is to set the \ZZ and \WW couplings to be equal, $a_{i} = a_{i}^{\PZ\PZ} = a_{i}^{\PW\PW}$, leaving four \HVV anomalous couplings to be measured: $a_2$, $a_3$, $\kappa_1/(\Lambda_1)^2$, and $\kappa_2^{\PZ\PGg}/(\Lambda_1^{\PZ\PGg})^2$. The $a_{1}^{\PZ\PZ} = a_{1}^{\PW\PW}$ relationship also appears under custodial symmetry. This approach provides a general test of the Higgs boson Lagrangian tensor structure and a search for $CP$ violation in \HVV interactions. In an alternative approach, the SU(2) $\times$ U(1) symmetry reduces the number of independent \HVV anomalous couplings to three ($a_2$, $a_3$, and $\kappa_1/(\Lambda_1)^2$) through the introduction of the following coupling parameter relationships~\cite{deFlorian:2227475} :
\begin{align}
a_1^{\WW} &= a_1^{\ZZ},  \label{eq:EFT1}  \\
a_2^{\WW} &= \cw^2 a_2^{\ZZ}, \label{eq:EFT2} \\
a_3^{\WW} &= \cw^2 a_3^{\ZZ}, \label{eq:EFT3} \\
\frac{\kappa_1^{\WW}}{(\Lambda_1^{\WW})^2} &= \frac{1}{\cw^2-\sw^2} \left(\frac{\kappa_1^{\ZZ}}{(\Lambda_1^{\ZZ})^2} -2 \sw^2 \frac{a_2^{\ZZ}}{m_\PZ^2}\right), \label{eq:EFT4} \\
\frac{\kappa_2^{\PZ\PGg}}{(\Lambda_1^{\PZ\PGg})^2} &= \frac{ 2 \sw \cw}{\cw^2-\sw^2} \left(\frac{\kappa_1^{\ZZ}}{(\Lambda_1^{\ZZ})^2} - \frac{a_2^{\ZZ}}{m_\PZ^2}\right), \label{eq:EFT5}
\end{align}
where $\cw$ and $\sw$ are the cosine and sine of the weak mixing angle, respectively, and $m_\PZ$ is the $\PZ$ boson mass. With this approach, there is a linear relationship between the scattering amplitude couplings and the SM EFT (SMEFT) couplings in the Higgs basis~\cite{deFlorian:2227475} :  
\begin{align}
\delta c_\text{z} & = \frac12 a_1^{\ZZ} - 1,   \label{eq:EFTpar1}  \\
c_\text{zz} &= -\frac{2 \sw^2 \cw^2}{e^2} a_2^{\ZZ},  \label{eq:EFTpar2}  \\
\tilde{c}_\text{zz} &= -\frac{2 \sw^2 \cw^2}{e^2} a_3^{\ZZ},  \label{eq:EFTpar3} \\
c_{\text{z} \Box} &= \frac{m_\PZ^2 \sw^2}{e^2} \, \frac{\kappa_1^{\ZZ}}{(\Lambda_1^{\ZZ})^2},   \label{eq:EFTpar4}
\end{align}
where $e$ is the electron charge. The amplitude couplings may also be related to the SMEFT Warsaw basis~\cite{Grzadkowski_2010,deFlorian:2227475} couplings through the following translation~\cite{hzz_ac_2021,Davis:2021tiv} :
\begin{align}
\delta a_1^{\ZZ} &= \frac{v^2}{\Lambda^2} \left(2c_{\text{H}\Box} + \frac{6e^2}{\sw^2}c_\text{HWB} + (\frac{3\cw^2}{2\sw^2} -\frac{1}{2})c_\text{HD} \right), \label{eq:EFTWpar1}  \\
\kappa_1^{\ZZ} &= \frac{v^2}{\Lambda^2} \left(-\frac{2e^2}{\sw^{2}}c_\text{HWB} + (1-\frac{1}{2\sw^2})c_\text{HD}  \right), \label{eq:EFTWpar2}  \\
a_2^{\ZZ} &= -2\frac{v^2}{\Lambda^2} \left( \sw^2 c_\text{HB} + \cw^2 c_\text{HW} + \sw \cw c_\text{HWB} \right), \label{eq:EFTWpar3}  \\
a_3^{\ZZ} &= -2\frac{v^2}{\Lambda^2} \left( \sw^2 c_{\text{H}\tilde{\text{B}}} + \cw^2 c_{\text{H}\tilde{\text{W}}} + \sw \cw c_{\text{H}\tilde{\text{W}}\text{B}} \right), \label{eq:EFTWpar4}
\end{align}
where $\Lambda$ is the UV cutoff of the theory (set to 1\TeV), and $\delta a_1^{\ZZ}$ is a correction to the SM value of $a_1^{\ZZ}$.
Further discussion on the EFT operators corresponding to the couplings considered here may be found in Chapter 2.2 of Ref.~\cite{deFlorian:2227475}.
The assumed constraints on $a_2^{\PGg\PGg}$, $a_3^{\PGg\PGg}$, $a_2^{\PZ\PGg}$ and $a_3^{\PZ\PGg}$ imply that only one of the three coupling parameters $c_\text{HW}$ , $c_\text{HWB}$, and $c_\text{HB}$ is independent; the same is also true for their $CP$-odd counterparts $c_{\text{H}\tilde{\text{W}}}$, $c_{\text{H}\tilde{\text{W}}\text{B}}$, and $c_{\text{H}\tilde{\text{B}}}$. Therefore, we have four independent \HVV couplings in both the Higgs and Warsaw basis. All the EFT couplings are expected to be zero in the SM.

We thus adopt two approaches to the \HVV coupling study. 
In Approach 1, we use the $a_{i}^{\PZ\PZ} = a_{i}^{\PW\PW}$ relationship and individually analyze each of the four anomalous couplings. 
In Approach 2, we enforce the SU(2) x U(1) relationships from Eqs.~(\ref{eq:EFT1}--\ref{eq:EFT5}) and analyze the three independent anomalous couplings both individually and simultaneously. Approach 1 may be considered to follow the relationships from Eqs.~(\ref{eq:EFT1}--\ref{eq:EFT4}) in the limiting case $\cw = 1$.

It is convenient to measure the fractional contribution of the anomalous couplings to the Higgs boson cross section rather than the anomalous couplings themselves.
For the anomalous \HVV couplings, the effective fractional cross sections \fai are defined as 
\begin{equation} 
\fai = \frac{\abs{a_i}^2 \sigma_i}{\sum_{j}{\abs{a_{j}}^2 \sigma_{j}}} \, \text{sign}\left(\frac{a_i}{a_{1}}\right),  
\label{eq:fa_definitions_hvv} 
\end{equation} 
where $\sum_{j}$ sums over all the coupling parameters considered, including $a_1$, and $\sigma_{i}$ is the cross section for the process corresponding to $a_{i} = 1$ and $a_{j \neq i} = 0$. Many systematic uncertainties cancel out in the ratio, and the physical range is conveniently bounded between $-1$ and $+1$. Our primary measurements are performed in terms of cross section fractions, with additional interpretations in terms of the SMEFT Higgs and Warsaw basis couplings also included.
For consistency with previous CMS measurements, the $\sigma_i$ coefficients used to define the fractional cross sections correspond to the $\Pg\Pg\to\PH\to\PV\PV\to2\Pe2\PGm$ process~\cite{hzz_ac_2021}. The numerical values are given in Table~\ref{tab:ac_crossections} as calculated using the \textsc{JHUGen} simulation~\cite{Gao:2010qx,Bolognesi:2012mm,Anderson:2013afp,Gritsan:2016hjl}. Two sets of values are shown corresponding to the different coupling relationships adopted in Approach 1 and 2.   

It has been shown that the angular correlations of the associated jets in the \ggH + 2 jets process are sensitive to anomalous \Hgg coupling effects at the production vertex~\cite{jhu2020}. The quark-quark initiated process, $\PQq\PQq\to \PQq\PQq \PH$, corresponds to the gluon scattering topology sensitive to anomalous effects.
For the anomalous \Hgg coupling, the effective fractional cross section can be defined as 
\begin{equation}
\fagg = \frac{\abs{a_3^{\glgl}}^2 \sigma_{3}^{\glgl}}{\abs{a_2^{\glgl}}^2 \sigma_{2}^{\glgl} + \abs{a_3^{\glgl}}^2 \sigma_{3}^{\glgl}} \, \text{sign}\left(\frac{a_{3}^{\glgl}}{a_{2}^{\glgl}}\right). \\ 
\label{eq:fa_definitions} 
\end{equation}
The $\sigma_{3}^{\glgl}$ and $\sigma_{2}^{\glgl}$ cross sections correspond to $a_3^{\glgl} = 1, a_2^{\glgl} = 0$ and $a_2^{\glgl} = 1, a_3^{\glgl} = 0$, respectively, and are equal. With this analysis it is not possible to distinguish the top quark, bottom quark, and heavy BSM particle contributions in the gluon fusion loop. As such, the \Hgg coupling is treated as an effective coupling with heavy degrees of freedom integrated out. 

\begin{table}
\centering 
\topcaption{ 
The cross sections ($\sigma_i$) of the anomalous contributions ($a_i$) relative to the SM value ($\sigma_1$) used to define the fractional cross sections \fai for the Approach 1 and 2 coupling relationships. For the $\kappa_{1}$ and $\kappa_{2}^{\PZ\PGg}$ couplings, the numerical values $\Lambda_1 = \Lambda_1^{\PZ\PGg} = 100\GeV$ are chosen to keep all coefficients of similar order of magnitude. 
}   
\begin{tabular}{cccc}
\fai & $a_i$ & Approach 1 $\sigma_i/\sigma_1$ & Approach 2 $\sigma_i/\sigma_1$ \\
\hline 
\fatwo   & $a_2$ & 0.361 & 6.376 \\ 
\fathree & $a_3$ & 0.153 & 0.153 \\ 
\fL      & $\kappa_1$ & 0.682 & 5.241 \\ 
\fLzg    & $\kappa_2^{\PZ\PGg}$ & 1.746 & \NA \\
\end{tabular} 
\label{tab:ac_crossections} 
\end{table}

\section{The CMS detector}
\label{sec:cmsdet}

{\tolerance=800The CMS apparatus~\cite{Chatrchyan:2008zzk} is a multipurpose, nearly hermetic detector, designed to identify electrons, muons, photons, and (charged and neutral) hadrons~\cite{CMS:2015xaf,CMS:2018rym,CMS:2015myp,CMS:2014pgm}. A global reconstruction ``particle-flow" (PF) algorithm~\cite{CMS:2017yfk} combines the information provided by the all-silicon inner tracker and by the crystal electromagnetic and brass-scintillator hadron calorimeters, operating inside a 3.8\unit{T} superconducting solenoid, with data from gas-ionization muon detectors interleaved with the solenoid return yoke, to build \PGt leptons, jets, missing transverse momentum, and other physics objects~\cite{CMS:2018jrd,CMS:2016lmd,Sirunyan:2019kia}.\par}
Events of interest are selected using a two-tiered trigger system~\cite{Sirunyan:2020zal,Khachatryan:2016bia}. The first level (L1), composed of custom hardware processors, uses information from the calorimeters and muon detectors to select events at a rate of around 100\unit{kHz} within a fixed latency of about 4\mus~\cite{Sirunyan:2020zal}. The second level, known as the high-level trigger (HLT), consists of a farm of processors running a version of the full event reconstruction software optimized for fast processing, and reduces the event rate to around 1\unit{kHz} before data storage~\cite{Khachatryan:2016bia}. A more detailed description of the CMS detector, together with a definition of the coordinate system and kinematic variables, can be found in Ref.~\cite{Chatrchyan:2008zzk}.

\section{Data sets and simulation}
\label{sec:datasets}

The data sets included in this analysis were recorded with the CMS detector in 2016, 2017, and 2018, and correspond to integrated luminosities of 36.3, 41.5, and 59.7\fbinv, respectively~\cite{CMS:LUM-17-003,CMS-PAS-LUM-17-004,CMS-PAS-LUM-18-002}. The collision events must fulfill HLT selection criteria that require the presence of one or two leptons satisfying isolation and identification requirements. For the 2016 data set, the single-electron trigger has a transverse momentum (\pt) threshold of 25\GeV for electrons with pseudorapidity $\abs{\eta} < 2.1$ and 27\GeV for $2.1 < \abs{\eta} < 2.5$, whereas the single-muon trigger has a \pt threshold of 24\GeV for $\abs{\eta} < 2.4$. For the 2017 (2018) data set, the \pt threshold is 35 (32)\GeV for the single-electron trigger (covering $\abs{\eta} < 2.5$) and 27 (24)\GeV for the single-muon trigger ($\abs{\eta}<2.4$). The dilepton $\Pe\PGm$ trigger has \pt thresholds of 23 and 12\GeV for the leading and subleading leptons, respectively, with the same coverage in pseudorapidity for electrons and muons as above. During the first part of data taking in 2016, a lower \pt threshold of 8\GeV for the subleading muon was used.

Monte Carlo event generators are used to model the signal and background processes. For each process, three independent sets of simulated events, corresponding to the three years of data taking, are used. This approach includes year-dependent effects in the CMS detector, data taking, and event reconstruction. All simulated events corresponding to a given data set share the same set of parton distribution functions (PDFs), underlying event (UE) tune, and parton shower (PS) configuration. The PDF sets used are NNPDF 3.0~\cite{Ball:2013hta,Ball:2011uy} for 2016 and NNPDF 3.1~\cite{Ball:2017nwa} for 2017 and 2018. The CUETP8M1~\cite{Khachatryan:2015pea} tune is used to describe the UE in 2016 simulations, whereas the CP5~\cite{Sirunyan:2019dfx} tune is adopted in 2017 and 2018 simulated events. The MC samples are interfaced with \PYTHIA 8.226~\cite{Sjostrand:2014zea} in 2016, and 8.230 in 2017 and 2018, for the modeling of UE, PS, and hadronization. 
{\tolerance=800 Standard Model Higgs boson production through \ggH, VBF, and \VH is simulated at next-to-leading order (NLO) accuracy in quantum chromodynamics (QCD), including finite quark mass effects, using \POWHEG v2~\cite{Nason:2004rx,Frixione:2007vw,Alioli:2010xd,Bagnaschi:2011tu,Nason:2009ai,Luisoni:2013kna,Hartanto:2015uka}. The \textsc{minlo hvj}~\cite{Luisoni:2013kna} extension of \POWHEG v2 is used for the simulation of $\PW\PH$ and quark-induced $\PZ\PH$ production, providing NLO accuracy for the $\VH+0$- and 1-jet processes. For \ggH production, the simulated events are weighted to match the NNLOPS~\cite{Hamilton:2013fea,Hamilton:2015nsa} prediction in the hadronic jet multiplicity (\njet) and Higgs boson \pt distributions. The weighting is based on \pt and \njet as computed in the simplified template cross section scheme 1.0~\cite{Berger:2019wnu}. The \textsc{minlo hjj}~\cite{Frederix:2015fyz} generator, which provides NLO accuracy for $\njet \geq 2$, is also used for \ggH production. The associated production processes with top quarks (\ttH) and bottom quarks (\bbH) are simulated with \POWHEG v2 and \MGvATNLO v2.2.2~\cite{Alwall:2014hca}, respectively, and have a negligible contribution in the analysis phase space. All SM Higgs boson samples are normalized to the cross sections recommended in Ref.~\cite{deFlorian:2227475}. The Higgs boson mass in the event generation is assumed to be 125\GeV, while a value of 125.38\GeV~\cite{HggMass} is used for the calculation of cross sections and branching fractions. The decay to a pair of \PW bosons and subsequently to leptons or hadrons is performed using the \textsc{JHUGen} v5.2.5 generator in 2016, and v7.1.4 in 2017 and 2018, for \ggH, VBF, and quark-induced $\PZ\PH$ samples. The Higgs boson and \PW boson decays are performed using \PYTHIA 8.212 for the other signal simulations.\par}

{The \ggH, VBF, and \VH Higgs boson events with \HVV anomalous couplings are generated with \textsc{JHUGen} at LO accuracy. With respect to the $\kappa_2^{\PZ\PGg}/(\Lambda_1^{\PZ\PGg})^2$ coupling parameter discussed in Section~\ref{sec:phenomenology}, the sign convention of the photon field is determined by the sign in front of the gauge fields in the covariant derivative. In this analysis, we define the covariant derivative $D_\mu = \partial_\mu -\mathrm{i}e \sigma^i W_\mu^i/(2s_w) + \mathrm{i}e B_\mu/(2 c_w)$ following the convention in \textsc{JHUGen}~\cite{Davis:2021tiv}.
The \textsc{JHUGen} and \POWHEG SM Higgs boson simulations were compared after parton showering and no significant differences in the distributions of kinematic observables were found. We adopt the \textsc{JHUGen} simulation to describe the kinematic features in all production modes with \HVV anomalous couplings. The expected yields are scaled to match the SM theoretical predictions~\cite{deFlorian:2227475} for inclusive cross sections and the \POWHEG SM prediction of relative event yields in the event categorization based on associated particles.} 
Simulation of the \ggH + 2 jets process with \Hgg anomalous couplings is done using \textsc{minlo X0jj}~\cite{Nason_2020} at NLO in QCD.
A large number of signal samples with various anomalous couplings were generated. The \textsc{MELA} package~\cite{Gao:2010qx,Bolognesi:2012mm,Anderson:2013afp,Gritsan:2016hjl,Gritsan:2020pib} contains a library of matrix elements from \textsc{JHUGen} for different Higgs boson signal hypotheses. Matrix elements from different coupling signal hypotheses, but with the same production mechanism, are used to reweight the generated signal events. This procedure is used in the construction of the predictions for the different coupling components and their interference, allowing us to cover all points in the signal model phase space with sufficient statistical precision.

Background events are produced using several simulations. The quark-initiated nonresonant \WW process is simulated with \POWHEG v2~\cite{Nason:2013ydw} at NLO accuracy for inclusive production. 
{\tolerance=800 A reweighting is performed to match the diboson \pt spectrum computed at NNLO+NNLL QCD accuracy~\cite{Meade:2014fca,Jaiswal:2014yba}.
The \MCFM v7.0~\cite{Campbell:1999ah,Campbell:2011bn,Campbell:2015qma} generator is used to simulate gluon-induced \WW production at LO accuracy, with the normalization chosen to match the NLO cross section~\cite{Caola:2016trd}. Nonresonant EW production of \WW pairs with two additional jets is simulated at LO accuracy with \MGvATNLO v2.4.2 using the MLM matching and merging scheme~\cite{MLM}. Top quark pair production (\ttbar) and single top quark processes, including $\PQt\PW$, s- and t-channel contributions, are simulated with \POWHEG v2~\cite{Frixione:2007nw,Alioli:2009je,Re:2010bp}. 
A reweighting of the top quark and antiquark \pt spectrum at parton level is performed for the \ttbar simulation in order to match the NNLO and next-to-next-to-leading logarithm (NNLL) QCD predictions, including also the NLO EW contribution~\cite{Czakon:2017wor}.\par}
{\tolerance=5000 The Drell--Yan (DY) production of a charged-lepton pair is simulated with \MGvATNLO v2.4.2 at NLO accuracy with up to two additional partons, using the FxFx matching and merging scheme~\cite{Frederix:2012ps}. Production of a \PW boson associated with an initial state radiation photon (\wg) is simulated with \MGvATNLO v2.4.2 at NLO accuracy with up to 1 additional parton, using the FxFx jet merging. Diboson processes containing at least one \PZ boson or a virtual photon ($\PGg^{*}$) with a mass as low as 100\MeV are generated with \POWHEG v2~\cite{Nason:2013ydw} at NLO accuracy. Production of a \PW boson in association with a $\PGg^{*}$ (\wgs) for masses below 100\MeV is simulated by \PYTHIA 8.212 in the parton showering of \wg events. Triboson processes with inclusive decays are also simulated at NLO accuracy with \MGvATNLO v2.4.2.\par}

For all processes, the detector response is simulated using a detailed description of the CMS detector, based on the \GEANTfour toolkit~\cite{Agostinelli:2002hh}. The distribution of additional $\Pp\Pp$ interactions within the same or nearby bunch crossings (pileup) in the simulation is reweighted to match that observed in data. The efficiency of the trigger system is evaluated in data on a per lepton basis using dilepton events consistent with the \PZ boson decay. The overall efficiencies of the trigger selections used in the analysis are obtained as the average of the per-lepton efficiencies weighted by their probability. The resulting efficiencies are applied directly on simulated events.

\section{Event reconstruction}
\label{sec:evreco}

The identification and measurement of the properties of individual particles (PF candidates) in an event is achieved in the PF algorithm by combining information from various subdetectors.
Electrons are identified and their momenta are measured in the pseudorapidity interval $\abs{\eta} < 2.5$ by combining tracks in the silicon tracker with spatially compatible energy deposits in the electromagnetic calorimeter. Muons are identified and their momenta are measured in the pseudorapidity range $\abs{\eta} < 2.4$ by matching tracks in the muon system and the silicon tracker.
For better rejection of nonprompt leptons, increasing the sensitivity of the analysis, leptons are required to be isolated and well reconstructed using a set of criteria based on the quality of the track reconstruction, shape of calorimetric deposits, and energy flux in the vicinity of the particle's trajectory~\cite{CMS:2015xaf,CMS:2018rym}. In addition, a selection based on a dedicated multivariate analysis (MVA) tagger developed for the CMS \ttH analysis~\cite{cms2023muon} is added in all channels for muon candidates.

Multiple $\Pp\Pp$ interaction vertices are identified from tracking information by use of the adaptive vertex fitting algorithm~\cite{Fruhwirth:2007hz}. The primary $\Pp\Pp$ interaction vertex is taken to be the vertex corresponding to the hardest scattering in the event, evaluated using tracking information alone, as described in Section 9.4.1 of Ref.~\cite{CMS-TDR-15-02}. Leptons are required to be associated to the primary vertex using transverse and longitudinal impact parameter criteria~\cite{CMS:2015xaf,CMS:2018rym}.

Hadronic jets are clustered from PF candidates using the infrared- and collinear-safe anti-\kt algorithm with distance parameters of 0.4 (AK4) and 0.8 (AK8). The jet momentum is determined as the vectorial sum of all particle momenta in the jet. The AK8 jets considered are required to be reconstructed within the silicon tracker acceptance ($\abs{\eta} < 2.4$), whereas AK4 jets are reconstructed in the range $\abs{\eta} < 4.7$. For AK4 jets, contamination from pileup is suppressed using charged-hadron subtraction which removes charged PF candidates originating from vertices other than the primary interaction vertex.
{\tolerance=800 The residual contribution from neutral particles originating from pileup vertices is removed by means of an event-by-event jet-area-based correction to the jet four-momentum~\cite{Sirunyan:2020foa}. For AK8 jets, the pileup-per-particle identification algorithm (PUPPI)~\cite{Bertolini:2014bba} is used to mitigate the effect of pileup at the reconstructed-particle level, making use of local shape information, event pileup properties, and tracking information. Additional selection criteria are applied to remove jets potentially dominated by instrumental effects or reconstruction failures~\cite{Sirunyan:2020foa}.\par}

The AK8 jets are used to reconstruct hadronic \PV boson decays in a single merged jet when the decay products are highly collimated. This approach targets boosted \PW or \PZ bosons originating from the \VH production mode. Such Lorentz-boosted \PV decays are identified using the ratio of the 2- to 1-subjettiness~\cite{Thaler:2010tr}, \tausj, and the groomed jet mass $\mj$. The groomed mass is calculated after applying a modified mass drop algorithm~\cite{Dasgupta:2013ihk,Butterworth:2008iy}, known as the soft-drop algorithm~\cite{Larkoski:2014wba}, with parameters $\beta = 0$, $z_\text{cut} = 0.1$, and $R_0 = 0.8$. The algorithm also identifies two hard subjets within the AK8 jet.

We refer to the identification of jets likely originating from bottom quarks as \PQb tagging~\cite{Sirunyan:2017ezt, CMS-DP-2017-013}. For each AK4 jet in the event, a score is calculated through a multivariate combination of different jet properties, making use of boosted decision trees and deep neural networks. A jet is considered \PQb-tagged if its associated score exceeds a threshold, tuned to achieve a certain tagging efficiency as measured in \ttbar events. The chosen working point corresponds to about 90\% efficiency for bottom quark jets and to a mistagging rate of about 10\% for light-flavor or gluon jets and of about 50\% for charm quark jets.

The missing transverse momentum vector \ptvecmiss is computed as the negative vector sum of the transverse momenta of all the PF candidates in an event, and its magnitude is denoted as \ptmiss~\cite{Sirunyan:2019kia}. The PUPPI algorithm is applied to reduce the pileup dependence of the \ptvecmiss observable by computing the \ptvecmiss from the PF candidates weighted by their probability to originate from the primary interaction vertex~\cite{Sirunyan:2019kia}.

\section{Event selection}
\label{sec:evsel}

The analysis is performed using \HtoWW candidate events in the $\Pe\PGm$ final state. For an event to be selected, the transverse momenta of the leading lepton \ptlone and the subleading lepton \ptltwo must be greater than 25 and 13\GeV, respectively. The \ptltwo threshold in the case of a muon is lowered to 10\GeV for the 2016 data set because of the lower threshold in the corresponding HLT algorithm. Events containing additional leptons with $\pt > 10\GeV$ are discarded. The dilepton system is required to have an invariant mass \mll greater than 12\GeV and transverse momentum \ptll above 30\GeV. A requirement on the missing transverse momentum of $\ptmiss > 20\GeV$ is implemented. We define transverse mass discriminating variables \mTH and \mTl as 
\begin{equation}
\centering
\mTH = \sqrt{2\ptll\ptmiss[1-\cos{\Delta\Phi(\vec{p}_{\text{T}}^{\Pell\Pell},\vec{p}_{\text{T}}^{\text{miss}})}]},\:
\label{eq:mth}
\end{equation}
\begin{equation}
\centering
\mTl = \sqrt{2\ptltwo\ptmiss[1-\cos{\Delta\Phi(\vec{p}_{\text{T}}^{\Pell 2},\vec{p}_{\text{T}}^{\text{miss}})}]},\:
\label{eq:mw2}
\end{equation}
and select events with $\mTH > 60\GeV$ and $\mTl > 30\GeV$.  
The \mTH requirement suppresses the $\text{DY}\to\tautau$ background process and avoids overlap with the \HtoTT analysis~\cite{Sirunyan:2019nbs}. To ensure orthogonality with a future off-shell \HtoWW analysis we require $\mTH < 125\GeV$. In addition, the region $76.2 < \mll < 106.2\GeV$ is excluded  to avoid overlap with the off-shell $\HtoZZ \to 2\Pell 2\PGn$ analysis~\cite{Sirunyan:2019twz}. These requirements will simplify a future combination of Higgs boson decay final states. Finally, events with any b-tagged jets with $\pt > 20\GeV$ are vetoed. These base selection criteria are summarized in Table~\ref{table:selection_base}.

\begin{table}[htb]
\centering
\topcaption{Summary of the base selection criteria.}
\begin{tabular}{lc}
Variable & Selection \\
\hline 
Number of leptons & 2 ($\Pe\PGm$ of opposite charge) \\
$\ptlone$ & $>$25\GeV \\
$\ptltwo$ & $>$13\GeV (10\GeV for 2016 data) \\
$\mll$ & 12--76.2\GeV or $>$106.2\GeV\\
$\ptll$ & $>$30\GeV \\
$\ptmiss$ & $>$20\GeV \\
$\mTl$ & $>$30\GeV \\
$\mTH$ & 60--125\GeV\\
$\text{\njet (\PQb jets)}$ & 0
\end{tabular}
\label{table:selection_base}
\end{table}

For the \HVV coupling analysis, exclusive selection criteria, which are based on the associated jet activity in the event, are applied that target the \ggH, VBF, and \VH production processes.  The AK4 (AK8) jets considered are required to have $\pt > 30~(200)\GeV$. In the \ggH channel, zero or one AK4 jet is required in the event. For the VBF and Resolved \VH channels, we require two AK4 jets with dijet masses of $\mjj > 120\GeV$ and $60 < \mjj < 120\GeV$, respectively. The Boosted \VH channel requires the presence of a \PV-tagged AK8 jet (\PV jet); such jets have a groomed mass in the region $65 < \mj  < 105\GeV$ and satisfy the requirement $\tausj < 0.4$. In the other channels, a \PV jet veto is implemented to ensure orthogonality. These production channels for the \HVV coupling study are summarized in Table~\ref{table:selection_category}. 

\begin{table*}[htb]
\centering
\topcaption{Summary of the \ggH, VBF, and \VH production channels used for the \HVV coupling study.}
\begin{tabular}{lcccc}
Variable & \ggH & VBF & Resolved \VH & Boosted \VH \\
\hline 
$\text{\njet (\PV jets)}$ & 0 & 0 & 0 & $>$0 \\
$\text{\njet (AK4 jets)}$ & 0 \& 1 & 2 & 2 & \NA \\
$\mjj$ & \NA & $>$120\GeV & 60--120\GeV & \NA \\
\end{tabular}
\label{table:selection_category}
\end{table*}

As the production vertex of the \ggH + 2 jets process is sensitive to anomalous \Hgg coupling effects, we use a 2-jet \ggH channel that follows the VBF selection described above for the \Hgg coupling analysis. 
The \HWW decay vertex is not sensitive to anomalous \Hgg effects, and so decay-based variables are not studied in this channel. This permits a relatively tight selection of $\mll < 55\GeV$ which is beneficial for background suppression. The 0- and 1-jet \ggH channels are also included to constrain the \ggH signal strength.  
All channels included for the \Hgg coupling study are summarized in Table~\ref{table:selection_category2}. 

\begin{table}[htb]
\centering
\topcaption{Summary of \ggH channel selections used for the \Hgg coupling study.}
\begin{tabular}{lcc}
Variable & \ggH  & 2-jet \ggH \\
\hline
$\text{\njet (AK4 jets)}$ & 0 \& 1 & 2\\
$\mjj$ & \NA & $>$120\GeV  \\
$\mll$ & \NA & $<$55\GeV  \\
\end{tabular}
\label{table:selection_category2}
\end{table}

Control regions (CRs) are defined using the base selection criteria together with a set of alternative requirements summarized in Table~\ref{table:control_regions}. They are used to validate the background description and to estimate the number of background events in the signal region (SR). A dedicated $\tautau$ CR targets events from the DY process $\PZ \to \tautau$ with $\PGt$ leptons decaying leptonically to produce the $\Pe\PGm$ final state. Also a top quark CR is defined to enhance events with one or more top quarks decaying to a \PW boson and bottom quark. Splitting events according to the number of associated jets, separate $\tautau$ and top quark CRs are defined for the 0-, 1- and 2-jet SRs. 
An additional CR with an enhanced contribution from the nonresonant \WW background is used in the 2-jet SR. All CRs are used in the final data fit to constrain the DY, top quark, and \WW background normalizations. 

\begin{table}[htb]
\centering
\topcaption{Summary of the $\tautau$, top quark, and \WW control region requirements.}
\begin{tabular}{lccc}
Variable & $\tautau$ & top quark & \WW \\
\hline 
$\mll$ & 40--80\GeV & $>$50\GeV & $>$106.2\GeV \\
$\mTH$ & $<$60\GeV & \NA & 60--125\GeV \\
$\mTl$ & \NA & $>$30\GeV & $>$30\GeV \\
$\text{\njet (\PQb jets)}$ & 0 & $>$0 & 0 \\
\end{tabular}
\label{table:control_regions}
\end{table}

Additional $\tautau$, top quark, and \WW CRs are defined requiring a \PV jet. These CRs are used to validate the background description in the Boosted \VH channel. However, they generally do not have a sufficient number of events to significantly constrain the background normalizations in the final fit to the data. As such, we rely on the 2-jet CRs to determine the normalizations to be used in the Boosted \VH channel. Agreement between data and the background prediction in the \PV jet CRs is observed when using normalizations determined in the 2-jet CRs.

\section{Background estimation}
\label{sec:background}

{\tolerance=800 The nonprompt-lepton backgrounds originating from leptonic decays of heavy quarks, hadrons misidentified as leptons, and electrons from photon conversions are suppressed by  identification and isolation requirements imposed on electrons and muons. In this analysis, the nonprompt-lepton background primarily originates from $\PW+$jets events and is estimated from data, as described in detail in Ref.~\cite{Sirunyan:2018egh}. The procedure involves measuring the rate at which a nonprompt lepton passing a loose selection further passes a tight selection (misidentification rate) and the corresponding rate for a prompt lepton to pass this selection (prompt rate). The misidentification rate is measured in a data sample enriched in multijet events, whereas the prompt rate is measured using a tag-and-probe method~\cite{CMS:2010svw} in a data sample enriched in DY events. The nonprompt-lepton background estimation is validated with data in a CR enriched with $\PW+$jets events, in which a pair of same-sign leptons is required.\par}
The backgrounds from top quark processes and nonresonant \WW production are estimated using a combination of MC simulations and the dedicated CRs described in the previous section. The normalisations of these backgrounds are left as free parameters in the fit, keeping different parameters for each jet multiplicity region. The top quark background normalization is measured from the observed data in the top quark enriched CRs. A separate normalization parameter is included for the quark-induced and gluon-induced \WW backgrounds.
For the 2-jet regions, the \WW enriched CR is used to constrain the \WW background normalisation parameters.
In the 0- and 1-jet channels, these parameters are constrained directly in the signal regions, which span the high \mll phase space enriched in \WW events.

The $\text{DY}\to\tautau$ background process is estimated with a data-embedding technique~\cite{tau_embedding}. As for the top quark and \WW backgrounds, the DY normalization is left unconstrained in the data fit. The $\text{DY}\to\tautau$ enriched CR described in Section~\ref{sec:evsel} is used to constrain the free normalization parameters in the 0-, 1-, 2-jet regions. The data-embedded samples cover the events that pass the $\Pe\PGm$ triggers, which represent the vast majority of the selected events. The remaining $\text{DY}\to\tautau$ events, which enter the analysis through the single-lepton triggers (${\approx}5\%$ of the total), are estimated using MC simulation.

{\tolerance=800 The $\PW\PZ$ and \wgs background contributions are simulated as described in Section~\ref{sec:datasets}, and a data-to-simulation scale factor is derived in a three-lepton CR, as described in Ref.~\cite{Sirunyan:2018egh}. The contribution of the \wg process may also be a background because of photon conversions in the detector material. This process is estimated using MC simulation and validated using data in a CR requiring events with a leading \PGm and a trailing \Pe with same sign and a separation in $\Delta R=\sqrt{\smash[b]{\Delta\phi^2 + \Delta\eta^2}}$ (where $\phi$ is the azimuthal angle in radians) smaller than 0.5. Triple vector boson production is a minor background in all channels and is estimated using MC simulation.\par}

\section{Observables and kinematic discriminants}
\label{sec:KD}

In this paper, we search for anomalous \HVV and \Hgg coupling effects by studying:

\begin{enumerate}
\item the two quark jets from VBF and \VH production (\HVV coupling);
\item the \HtoWW decay products (\HVV coupling); and
\item the two quark jets from \ggH + 2 jets production (\Hgg coupling).
\end{enumerate}

\begin{figure*}[!htb]
\centering 
\includegraphics[width = 2in]{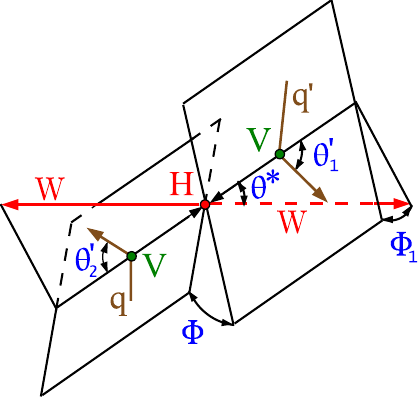}
\includegraphics[width = 2in]{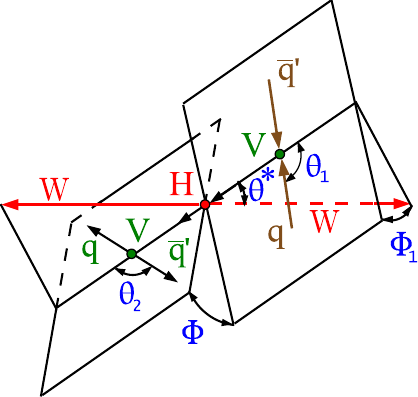}
\includegraphics[width = 2in]{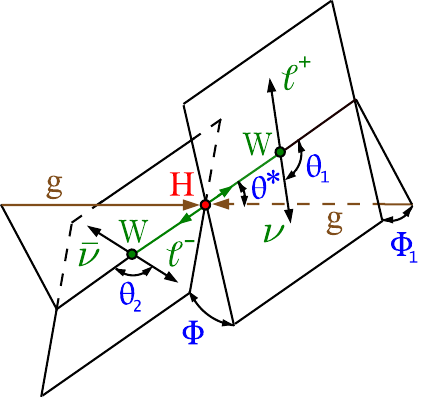}\\
\caption{
Topologies of the Higgs boson production and decay for vector boson fusion $\PQq{\PQq^\prime}\to \PQq{\PQq^\prime} \PH$ (left), $\PQq\bar{\PQq}^\prime\to \VH$ (center), and gluon fusion with decay $\Pg\Pg \to \PH \to 2 \ell 2\nu$ (right). For the electroweak production topologies, the intermediate vector bosons and their decays are shown in green and the $\PH \to \PW\PW$ decay is marked in red. For the $\Pg\Pg \to \PH \to 2 \ell 2\nu$ topology, the $\PW$ boson leptonic decays are shown in green. In all cases, the incoming particles are depicted in brown and the angles characterizing kinematic distributions are marked in blue. Five angles fully characterize the orientation of the production and decay chain and are defined in the suitable rest frames.
\label{fig:kinematics}
}
\end{figure*}

{\tolerance=800 The VBF, \VH, and \ggH production and decay topologies relevant for the \HVV coupling are illustrated in Fig.~\ref{fig:kinematics}.\par}

When combined with the momentum transfer of the vector bosons, the five angles illustrated for VBF/\VH production provide complete kinematic information for production and decay of the Higgs boson. The illustration for Higgs boson production via \ggH in association with two jets is identical to the VBF diagram, except for replacing the intermediate vector bosons by gluons. Full production kinematic information is extracted for VBF, \VH, and \ggH + 2 jets candidate events using discriminants built from the matrix element calculations of the \textsc{MELA} package. The MELA approach is designed to reduce the number of observables to a minimum, while retaining all essential information. To form the production-based MELA kinematic discriminants, we use jets to reconstruct the four-momentum of the associated production particles. The presence of two neutrinos in the final state means it is not possible to reconstruct the four-momentum of all the Higgs boson decay products. Therefore, decay-based kinematic discriminants built from matrix elements are not used in this analysis. Instead, we rely on kinematic variables related to the measured final state of the Higgs boson decay. The strategies used for each of the topologies listed above are now discussed in more detail. 

\subsection{Kinematic features of two quark jets in VBF and \texorpdfstring{\VH}{VH} channels}
\label{subsec:vbf}

Kinematic distributions of associated particles in VBF and \VH production are sensitive to the anomalous \HVV couplings of the Higgs boson. 
{\tolerance=800 As illustrated in Fig.~\ref{fig:kinematics}, a set of seven observables can be defined for the VBF and \VH production topologies : $\Omega = \{\theta_{1}^{(\prime)}, \theta_{2}^{(\prime)}, \theta^*, \Phi, \Phi_1, q_1^{2}, q_2^{2} \}$, with $q_1^2$ and $q_2^2$ the squared four-momenta of the vector bosons~\cite{Anderson:2013afp}. Three types of discriminants are defined using the full kinematic description characterized by $\Omega$. The first type of discriminant is designed to separate signal and background Higgs boson production processes:\par}
\begin{equation}
\mathcal{D}_\text{sig} = \frac{\mathcal{P}_\text{sig}(\Omega) }{\mathcal{P}_\text{sig}(\Omega) +\mathcal{P}_\text{bkg}(\Omega)}, 
\label{eq:melaDbkg} 
\end{equation}
where the probability density $\mathcal{P}$ for a specific process is calculated from the matrix elements provided by the \textsc{MELA} package.
The second type of discriminant separates the anomalous coupling BSM process from that of the SM:
\begin{equation}
\mathcal{D}_\text{BSM} = \frac{\mathcal{P}_\text{BSM}(\Omega) }{\mathcal{P}_\text{BSM}(\Omega) + \mathcal{P}_\text{SM}(\Omega)}. 
\label{eq:melaDbsm} 
\end{equation}
Throughout this document the generic BSM label is generally replaced by the specific anomalous coupling state targeted. For the $a_{3}$ $CP$-odd and $a_{2}$ $CP$-even coupling parameters, we use, respectively, $\mathcal{D}_{0-}$ and $\mathcal{D}_{0+}$, whereas for the $\Lambda_1$ coupling parameters we use $\mathcal{D}_{\Lambda_1}$ and $\mathcal{D}_{\Lambda_1}^{\PZ\gamma}$. The third type of discriminant isolates the interference contribution:
\begin{equation}
\mathcal{D}_\text{int} = \frac{ \mathcal{P}_{\text{SM-BSM}}^\text{int}(\Omega)} {\mathcal{P}_\text{SM}(\Omega) +\mathcal{P}_\text{BSM}(\Omega)}, 
\label{eq:melaDint} 
\end{equation}
where $\mathcal{P}_{\text{SM-BSM}}^\text{int}$ is the interference part of the probability distribution for a process with a mixture of the SM and BSM contributions. The $CP$ label is generally used for the $a_3$ coupling parameter, as the BSM signal in this case is a pseudoscalar and the interference discriminant is a $CP$-sensitive observable. The $\mathcal{P}$ values are normalized to give the same integrated cross sections in the relevant phase space of each process. Such normalization leads to a balanced distribution of events in the range between 0 and 1 for $\mathcal{D}_\text{sig}$ and $\mathcal{D}_\text{BSM}$, and between $-1$ and $+1$ for $\mathcal{D}_\text{int}$. 

\begin{figure}[!htb]
\centering 
\includegraphics[width=0.475\textwidth]{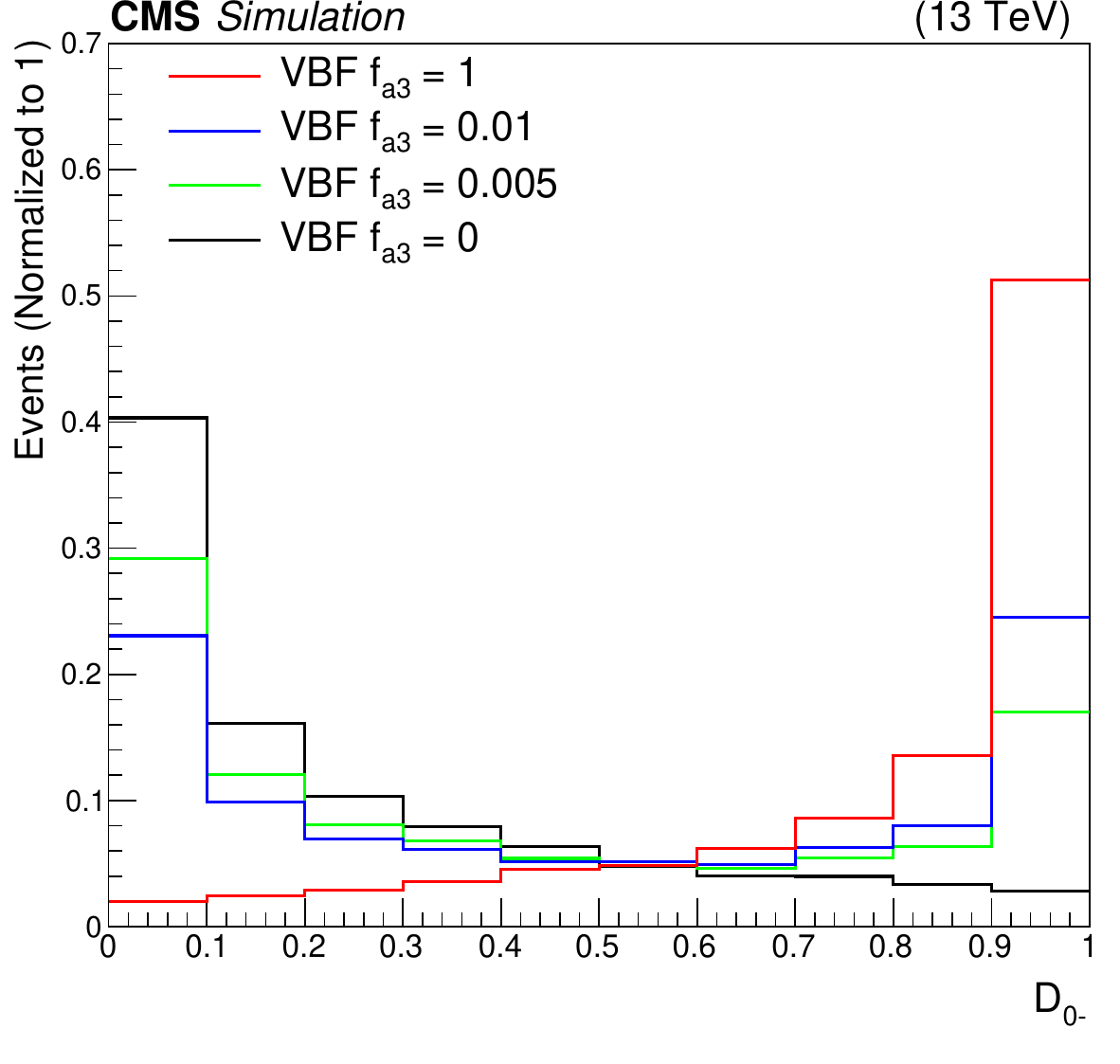}
\includegraphics[width=0.475\textwidth]{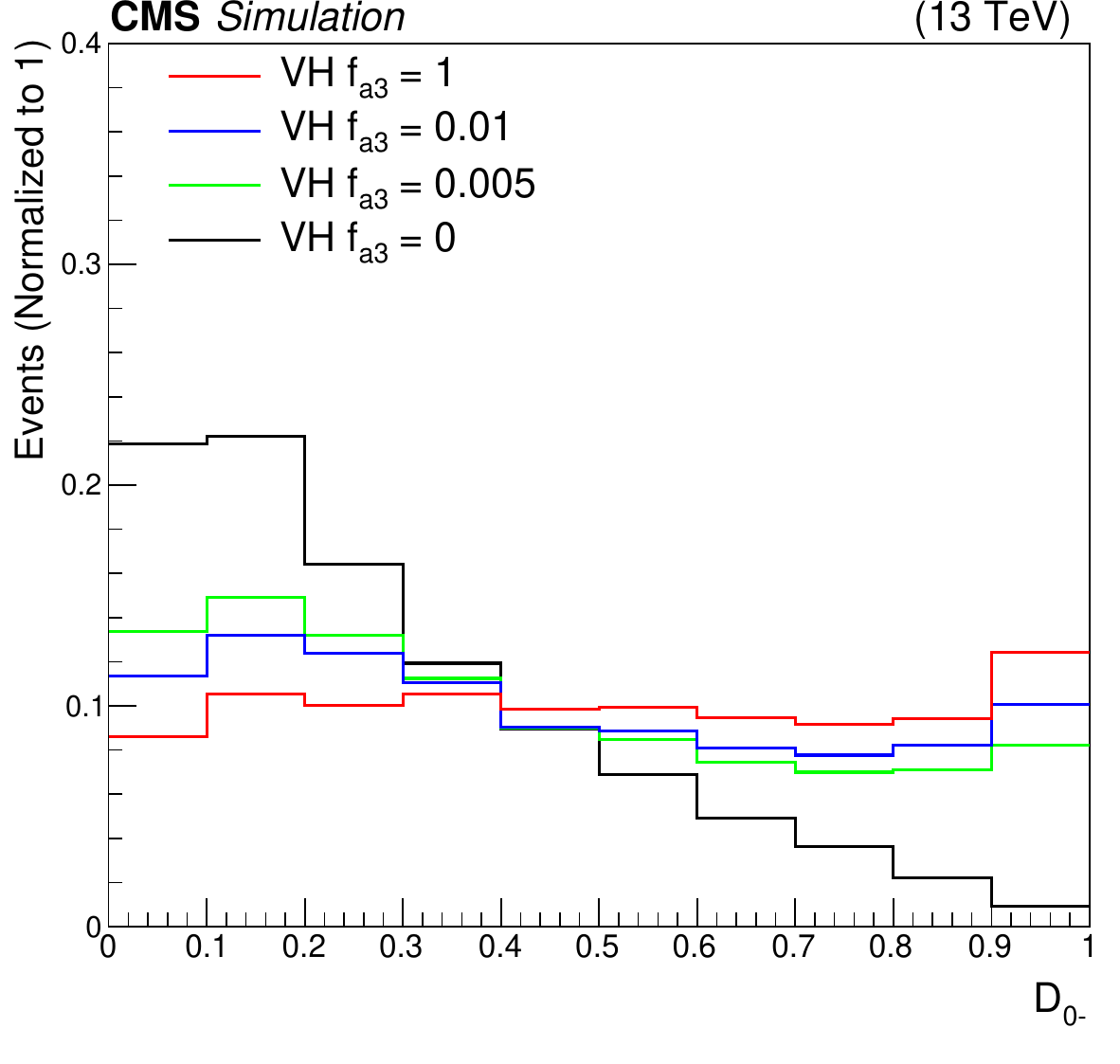}
\caption
{
	The $\mathcal{D}_{0-}$ discriminant in the VBF (\cmsLeft) and Resolved \VH (\cmsRight) production channels for a number of VBF (\cmsLeft) and \VH (\cmsRight) signal hypotheses.
Pure $a_1$ ($\fathree = 0$) and $a_3$ ($\fathree = 1$) \HVV signal hypotheses are shown along with two mixed coupling hypotheses ($\fathree = 0.005$ and $\fathree = 0.01$). 
All distributions are normalized to unity.
\label{fig:DExample}
}
\end{figure}

The selected events are split into three main production channels: VBF, Resolved \VH, and Boosted \VH. In the first two channels, the four-momenta of the two AK4 jets assigned as the associated particles are used in the MELA probability calculation. For the Boosted \VH category, we use the four-momentum of the two subjets of the \PV-tagged AK8 jet. An estimate of the Higgs boson four-momentum is also required for the probability calculation. This can not be measured directly since the final state contains two neutrinos. As such, we construct a proxy Higgs boson four-momentum in the following manner. The $p_{\text{x}}$ and $p_{\text{y}}$ of the dineutrino system are estimated from the \ptvecmiss in a given event. The corresponding $p_{\text{z}}$ is then set to equal that of the dilepton system, which is based on the observed correlation between these variables at the generator level for simulated signals. Finally, the mass of the dineutrino system is set equal to the mean value of the generator-level dineutrino mass. The resulting four-momentum can then be combined with that of the measured dilepton system to create a proxy Higgs boson four-momentum. We note that the MELA probability calculation for the production vertices is largely based on the kinematic features of the associated particles, so the reconstruction of the proxy Higgs boson has a relatively small effect on the final discriminants. As an illustrative example of the MELA based discriminants used in this analysis, Fig.~\ref{fig:DExample} shows the $\mathcal{D}_{0-}$ discriminant in the VBF and Resolved \VH production channels for a number of different signal hypotheses. The discriminants are designed to target the dominant signal production process in a given channel.

In the VBF channel, a $\mathcal{D}_\text{VBF}$ discriminant is constructed, following Eq.~(\ref{eq:melaDbkg}), where $\mathcal{P}_\text{sig}$ corresponds to the probability for the VBF production hypothesis, and $\mathcal{P}_\text{bkg}$ corresponds to that of gluon fusion production in association with two jets. The discriminant is also suitable for separating SM backgrounds from the VBF signal process. In the Resolved and Boosted \VH channels, the corresponding discriminants do not give a significant level of separation with respect to \ggH production or SM backgrounds. This is due to the relatively tight selection criteria, which limit the phase space to \VH-like events. Hence, these discriminants are not included in the \VH channels.

The $\mathcal{D}_{CP}$ discriminant is sensitive to the sign of the interference between the $CP$-even SM and $CP$-odd BSM states. An asymmetry between the number of events detected with positive and negative $\mathcal{D}_{CP}$ values is expected for mixed $CP$ states. Therefore, a forward-backward categorization (forward defined as $\mathcal{D}_{CP} > 0$ and backward as $\mathcal{D}_{CP} < 0$) is used to analyze the $CP$-odd couplings. Similarly, $\mathcal{D}_\text{int}$ gives sensitivity to the sign of the interference between the SM and $a_2$ \HVV BSM states. A forward-backward $\mathcal{D}_\text{int}$ categorization is also included. The value of $\mathcal{D}_\text{int}$ used to define the categories is chosen to symmetrize the SM Higgs boson expectation. In the case of the $\Lambda_1$ measurements, the interference discriminants were shown to be highly correlated with the $\mathcal{D}_\text{BSM}$ discriminants and so are not considered. 
 
We now discuss the categorization and construction of the final multidimensional discriminants used for the two \HVV coupling approaches defined in Section~\ref{sec:phenomenology}. The binning of the final discriminants was optimized to ensure sufficient statistical precision in the predictions of all bins, while retaining the kinematic information required to discriminate between the SM and anomalous coupling signal hypotheses.

\subsubsection{VBF/\texorpdfstring{\VH}{VH} analysis strategy for Approach 1}

In Approach 1, each of the four anomalous \HVV coupling parameters ($a_2$, $a_3$, $\kappa_1/(\Lambda_1)^2$, and $\kappa_2^{\PZ\PGg}/(\Lambda_1^{\PZ\PGg})^2$) are analyzed separately. For this purpose, we construct a multidimensional discriminant for each of the four anomalous couplings in the VBF, Resolved \VH, and Boosted \VH channels. 

In the VBF channel, we use two bins of the production discriminant $\mathcal{D}_\text{VBF}$, corresponding to low and high purity, using a bin boundary of 0.75. The $\mll$ variable, which is sensitive to anomalous effects at the \HtoWW decay vertex, is included with two bins in the range 12--76.2\GeV. A bin boundary of 45\GeV is chosen based on the expected signal shape changes induced by anomalous effects. Finally, one of the $\mathcal{D}_{\text{BSM}}$ discriminants is included with ten equally sized bins. Depending on the anomalous coupling under study this discriminant may be $\mathcal{D}_{0+}$, $\mathcal{D}_{0-}$, $\mathcal{D}_{\Lambda_1}$ or $\mathcal{D}_{\Lambda_1}^{\PZ\gamma}$. 

For the \VH channels, the $\mll$ and $\mathcal{D}_\text{BSM}$ observables are used to build 2D kinematic discriminants. The $\mll$ bins are the same as for the VBF channel. In the Resolved \VH channel, we use four $\mathcal{D}_\text{BSM}$ bins of equal size. For the Boosted \VH case, three variable bins with boundaries of 0.6 and 0.8 are used, a large first bin is chosen because relatively little signal is expected at low values of $\mathcal{D}_\text{BSM}$. A distinct multidimensional discriminant is constructed for each anomalous coupling hypothesis in the \VH channels. 

\begin{figure*}[!htbp]
\centering 
\includegraphics[width=0.475\textwidth]{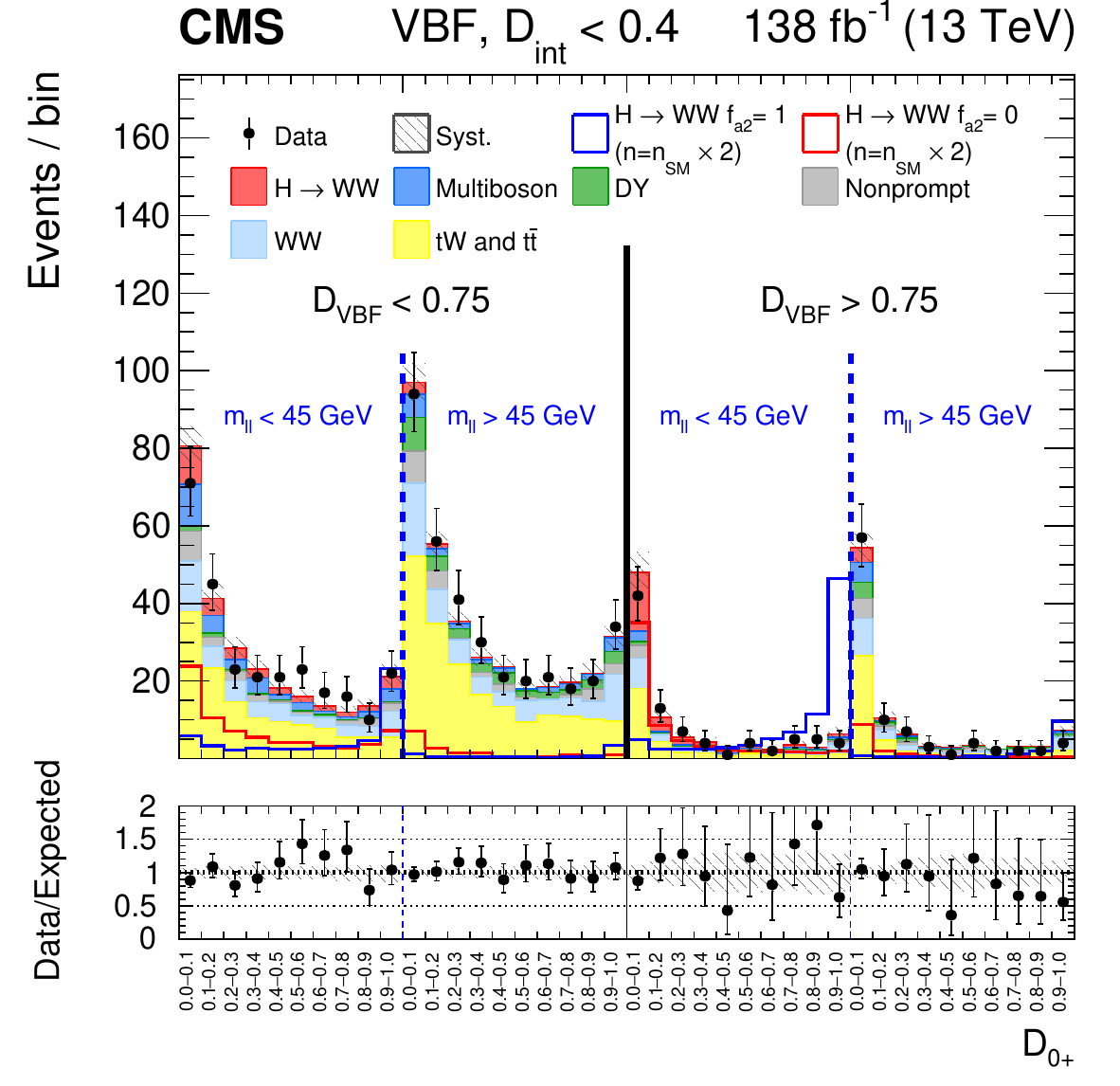}
\includegraphics[width=0.475\textwidth]{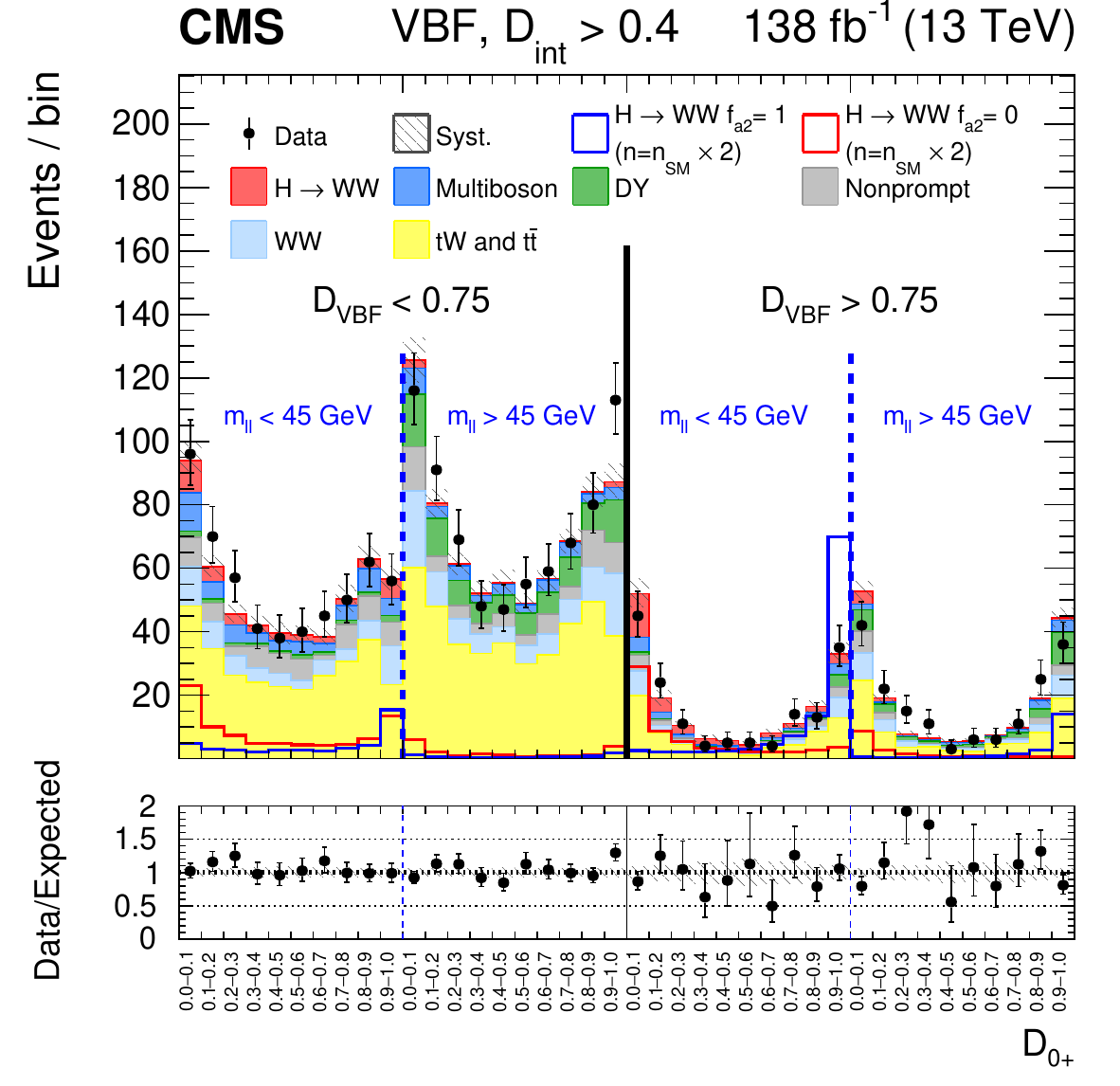} \\
\includegraphics[width=0.475\textwidth]{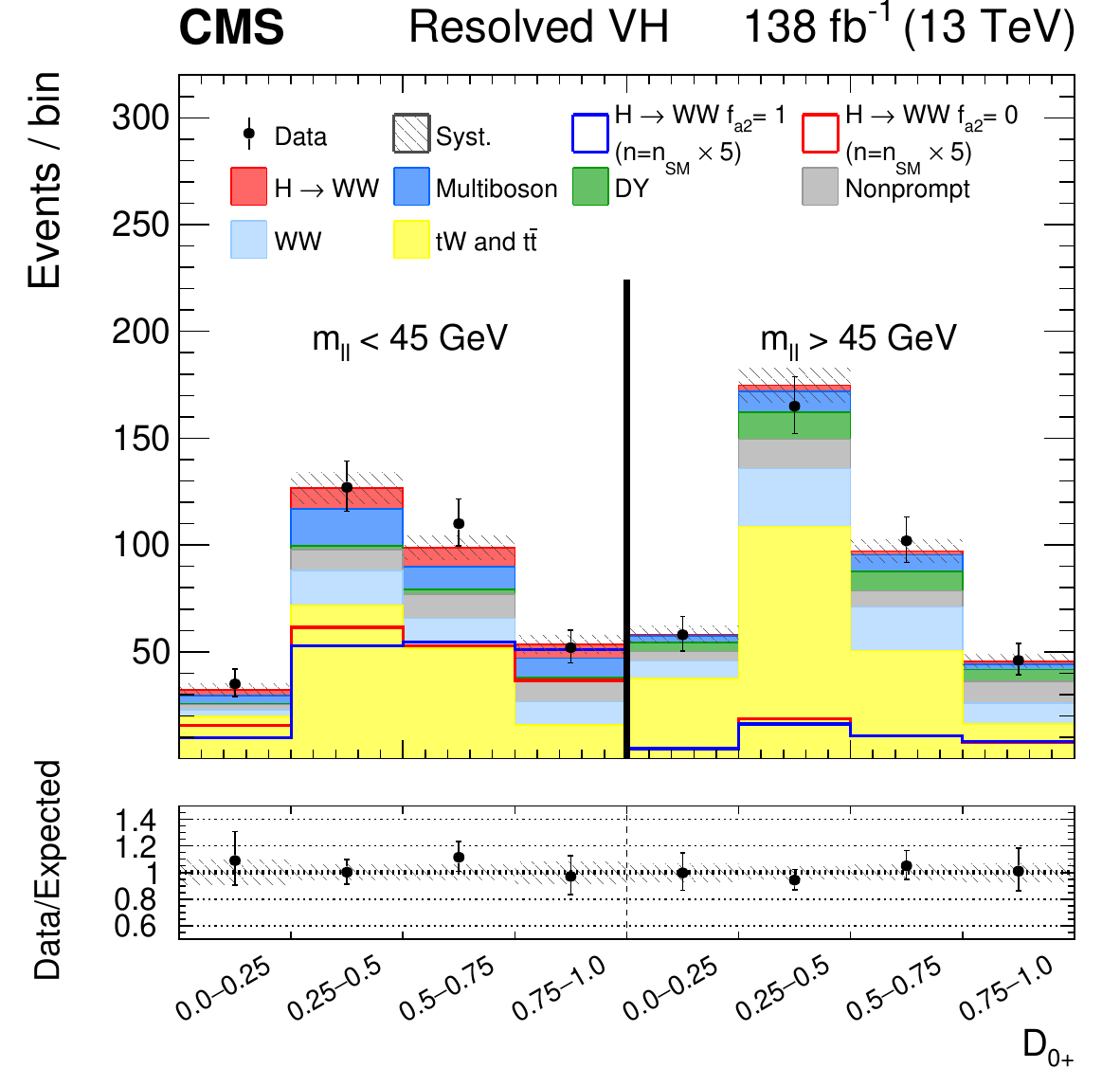}
\includegraphics[width=0.475\textwidth]{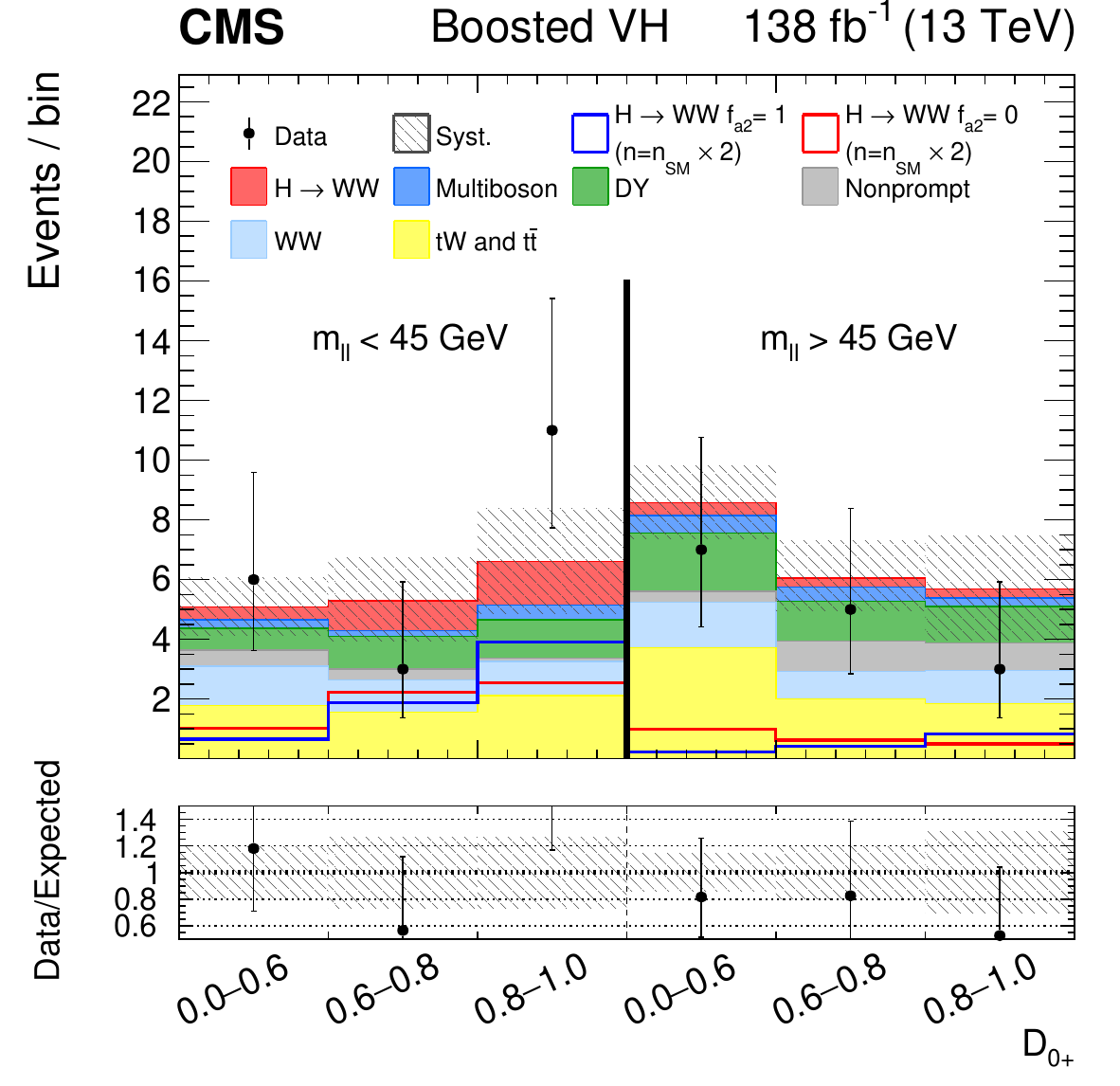}\\
\caption{
Observed and predicted distributions after fitting the data for $[\mathcal{D}_\text{VBF}, \mll, \mathcal{D}_{0+}]$ in the VBF channel (upper), and for $[\mll, \mathcal{D}_{0+}]$ in the Resolved \VH (lower left) and Boosted \VH (lower right) channels. For the VBF channel, the $\mathcal{D}_{\text{int}} < 0.4$ (left) and $\mathcal{D}_{\text{int}} > 0.4$ (right) categories are shown.
The predicted Higgs boson signal is shown stacked on top of the background distributions. For the fit, the $a_{1}$ and $a_2$ \HVV coupling contributions are included. 
The corresponding pure $a_{1}$ ($\fatwo = 0$) and $a_2$ ($\fatwo = 1$) signal hypotheses are also shown superimposed, their yields correspond to the predicted number of SM signal events scaled by an arbitrary factor to improve visibility.
The uncertainty band corresponds to the total systematic uncertainty. The lower panel in each figure shows the ratio of the number of events observed to the total prediction.
}

\label{fig:HVVH0PHpostfitRun2}
\end{figure*}

\begin{figure*}[!htbp]
\centering 
\includegraphics[width=0.46\textwidth]{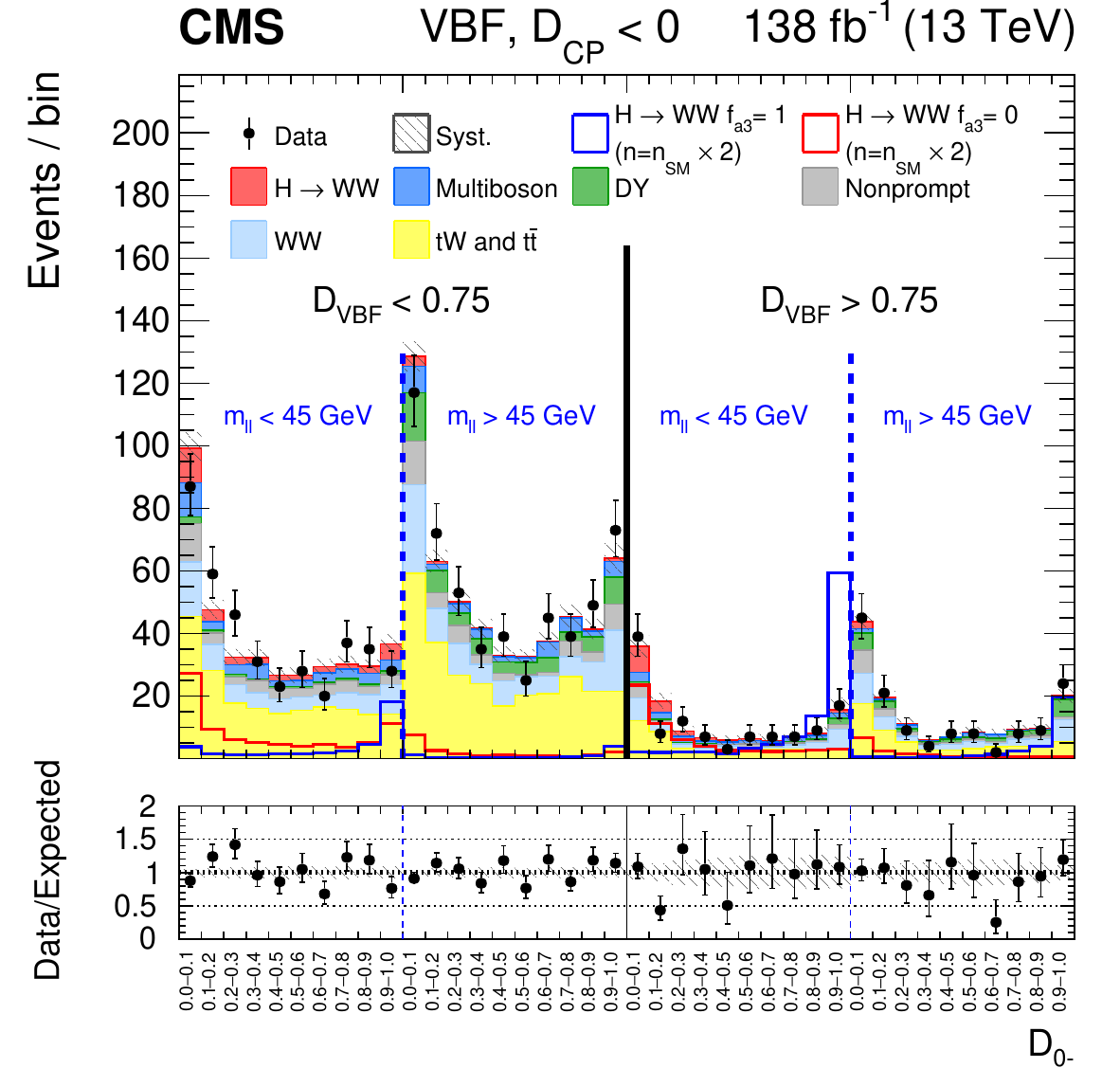}
\includegraphics[width=0.46\textwidth]{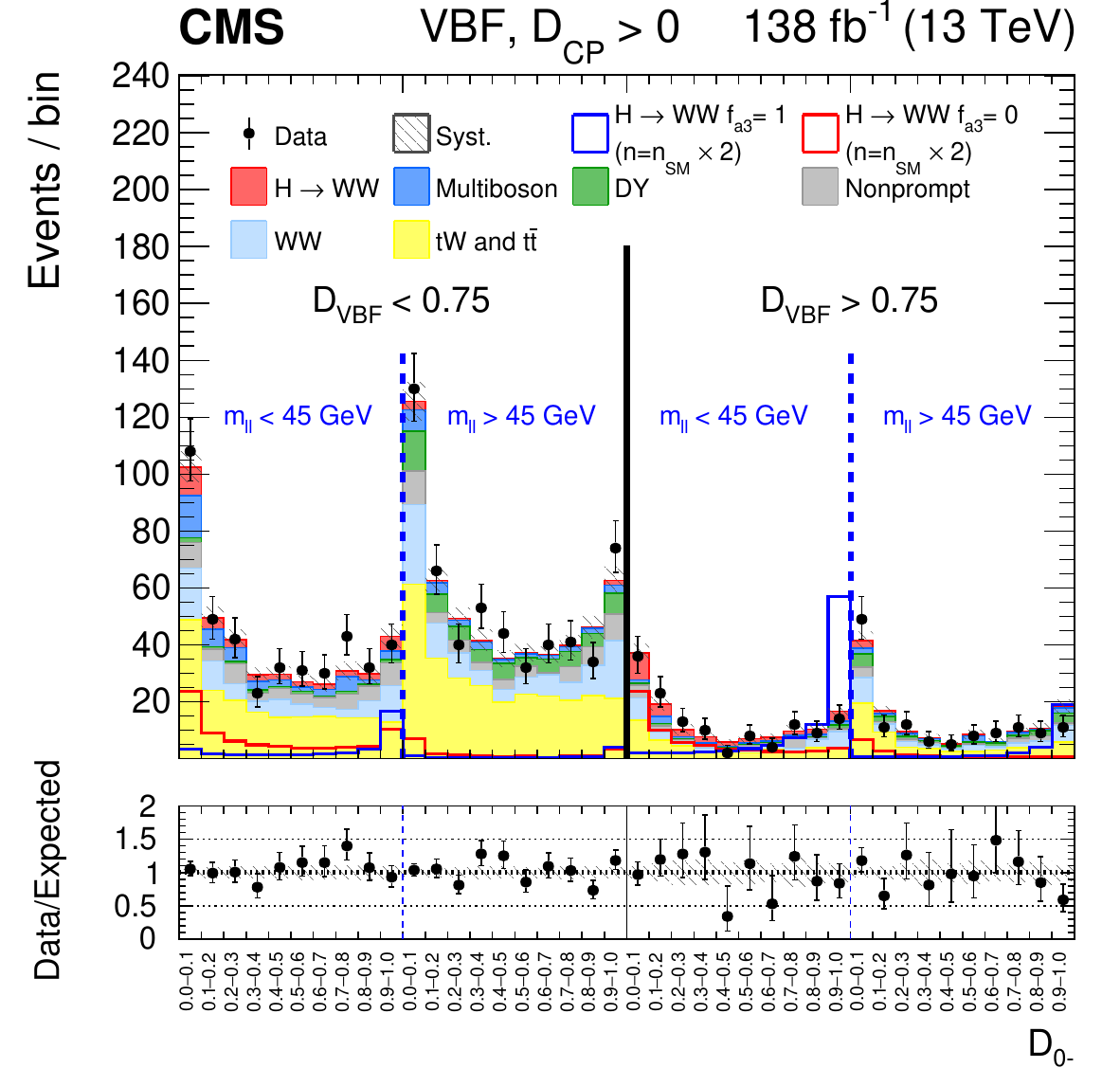} \\
\includegraphics[width=0.46\textwidth]{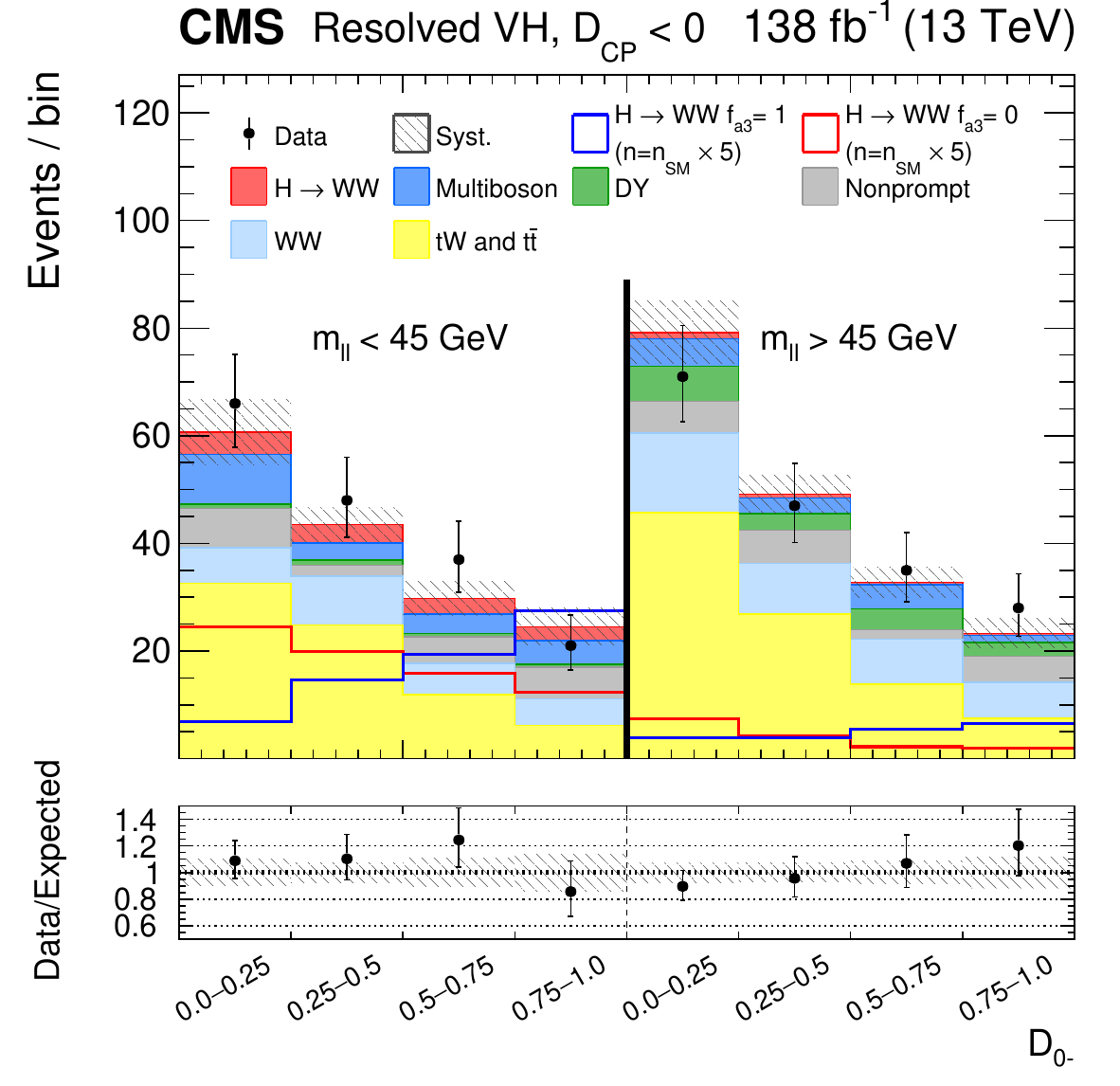}
\includegraphics[width=0.46\textwidth]{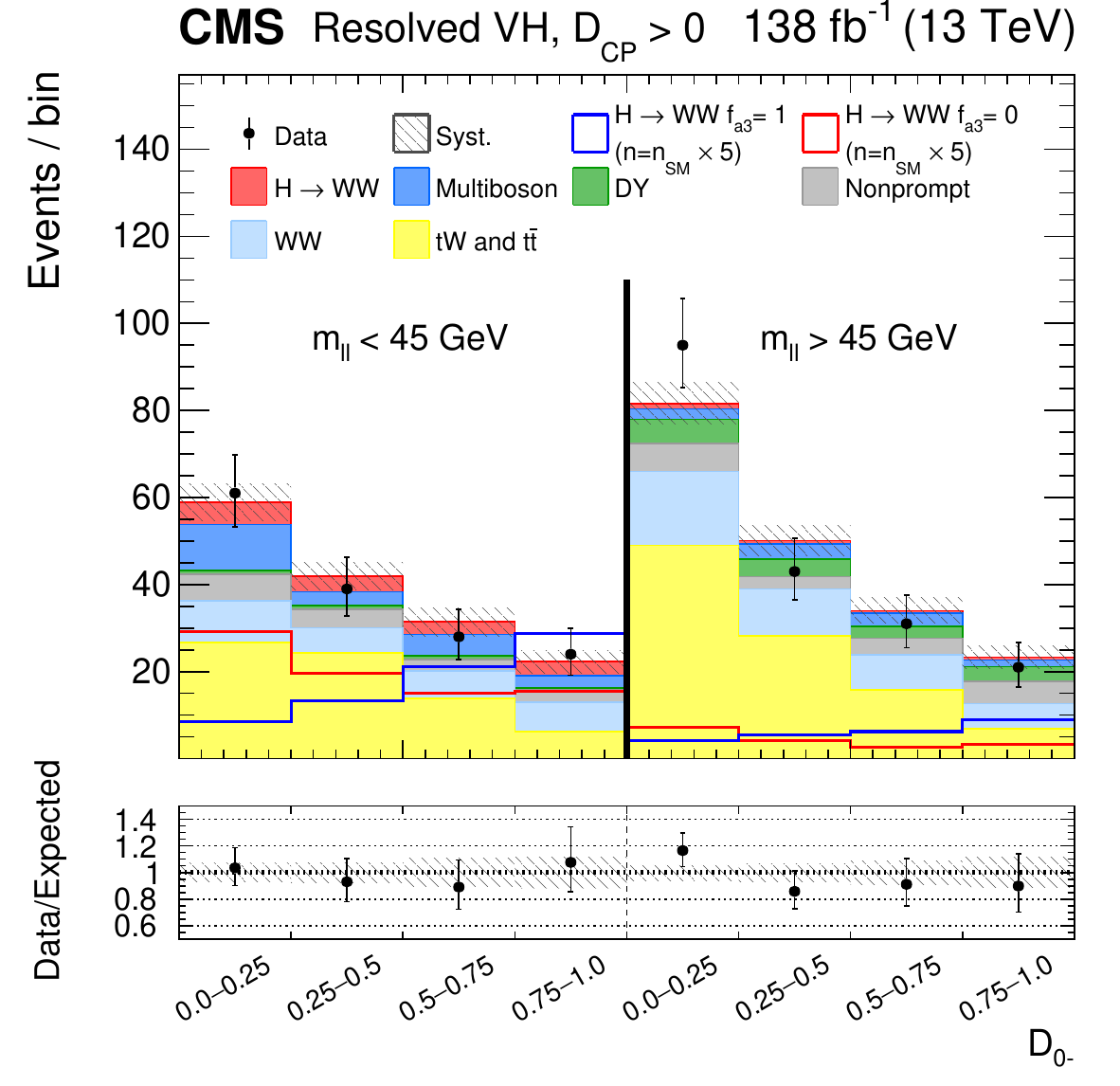} \\
\includegraphics[width=0.46\textwidth]{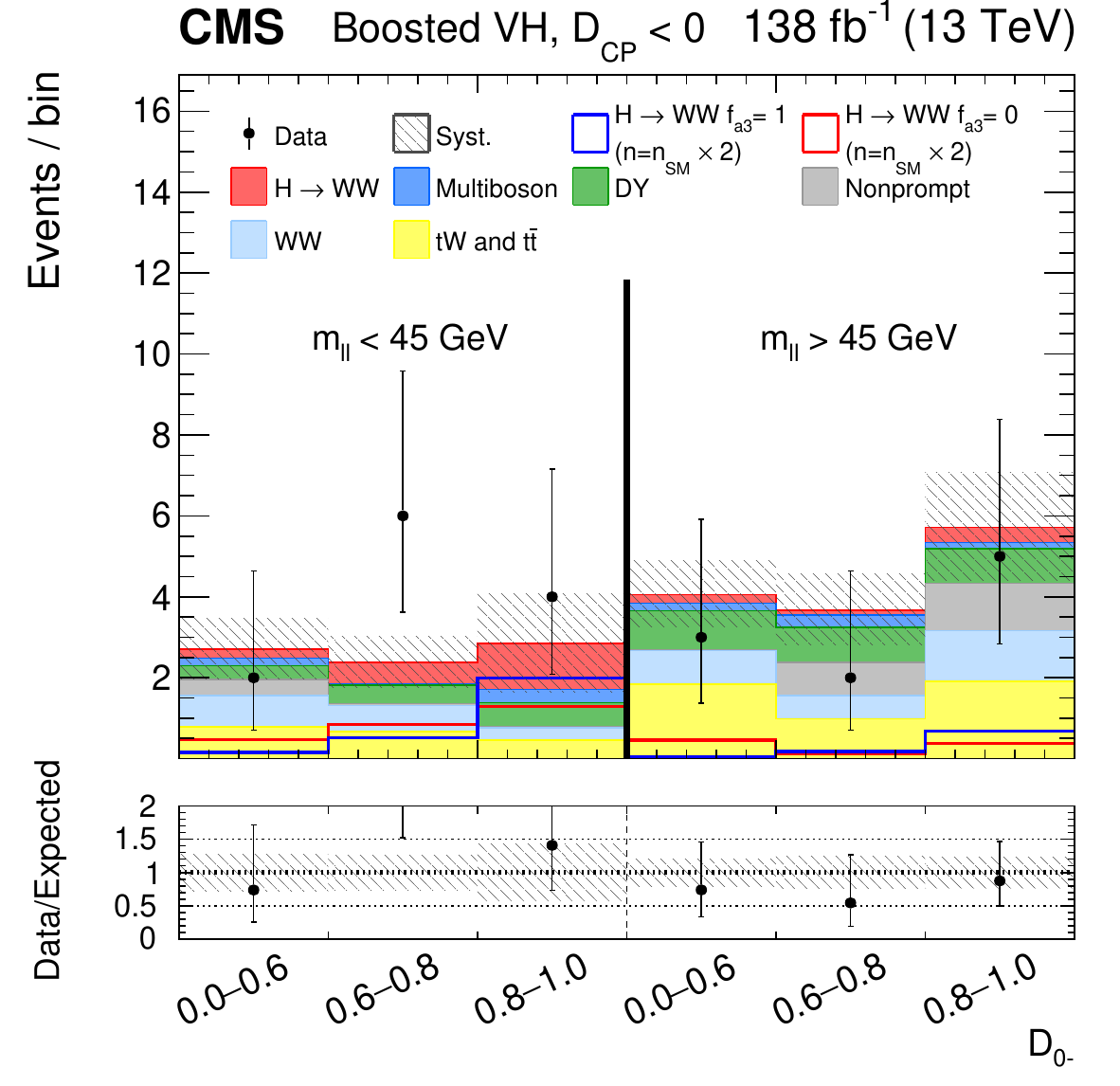}
\includegraphics[width=0.46\textwidth]{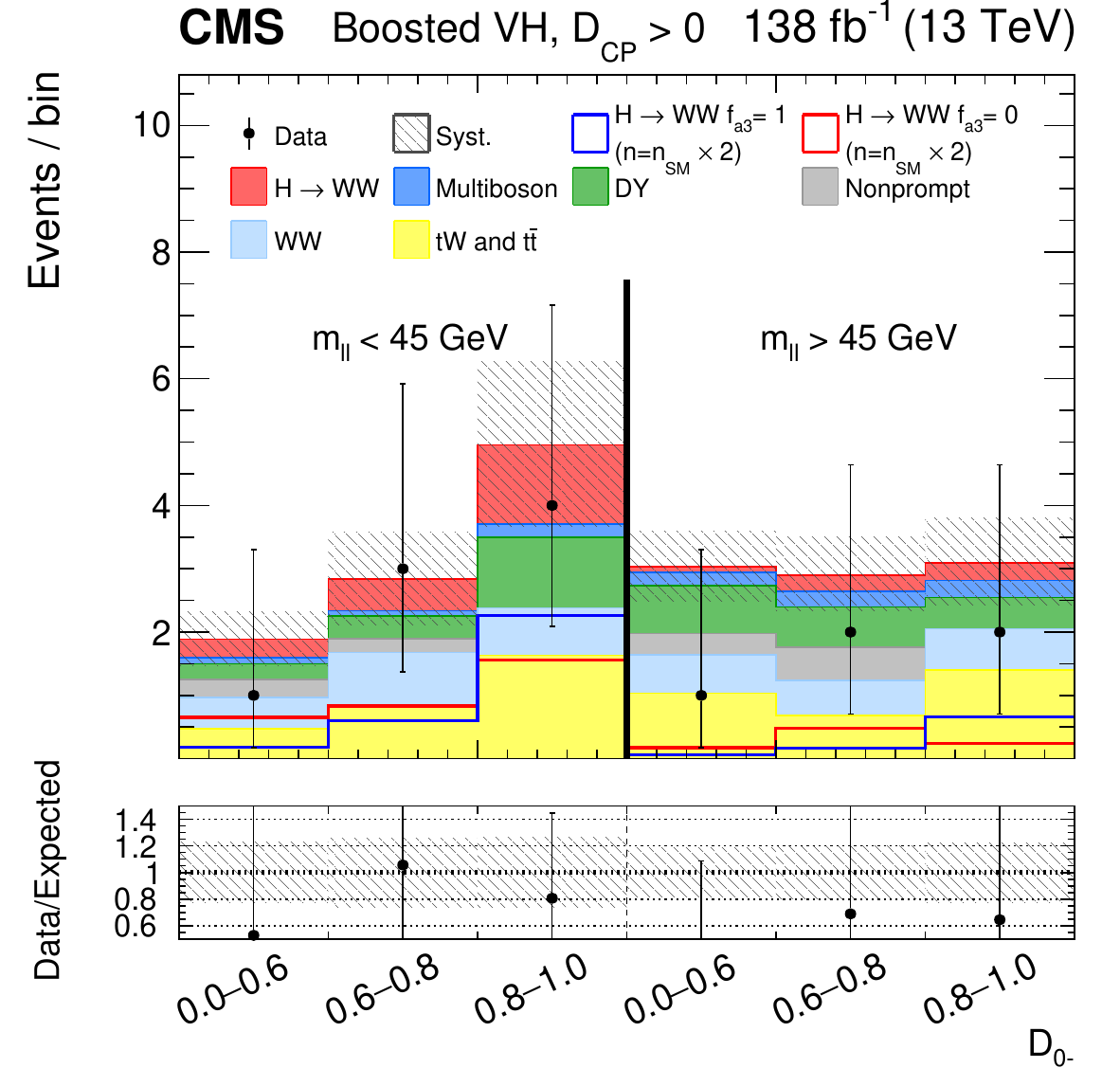}\\
\caption{
Observed and predicted distributions after fitting the data for $[\mathcal{D}_\text{VBF}, \mll, \mathcal{D}_{0-}]$ in the VBF channel (upper), and for $[\mll, \mathcal{D}_{0-}]$ in the Resolved \VH (middle) and Boosted \VH (lower) channels. For each channel, the $\mathcal{D}_{CP} < 0$ (left) and $\mathcal{D}_{CP} > 0$ (right) categories are shown.
For the fit, the $a_{1}$ and $a_3$ \HVV coupling contributions are included. More details are given in the caption of Fig.~\ref{fig:HVVH0PHpostfitRun2}.
}
\label{fig:HVVH0MpostfitRun2}
\end{figure*}

\begin{figure*}[!htbp]
\centering 
\includegraphics[width=0.46\textwidth]{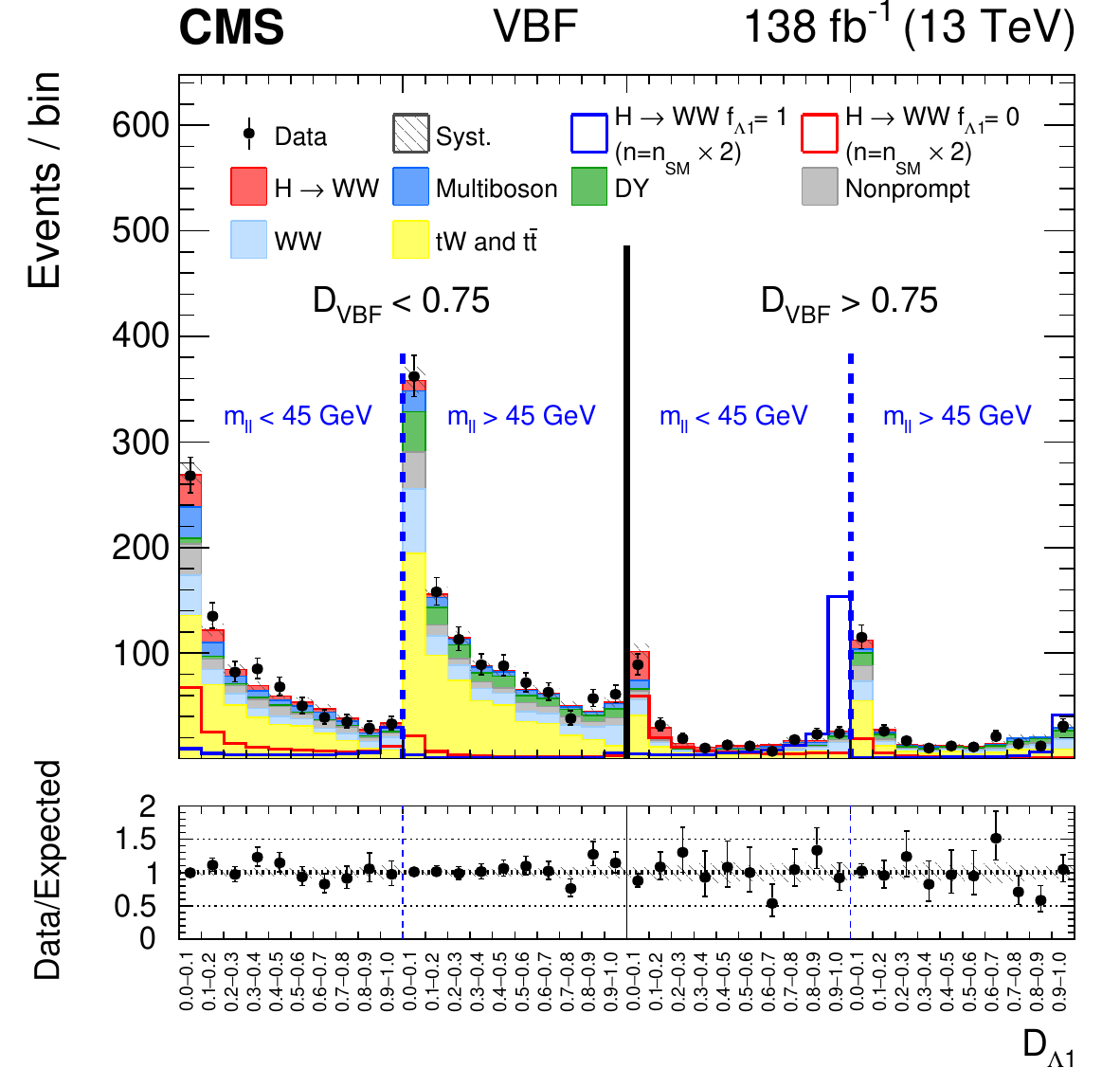}
\includegraphics[width=0.46\textwidth]{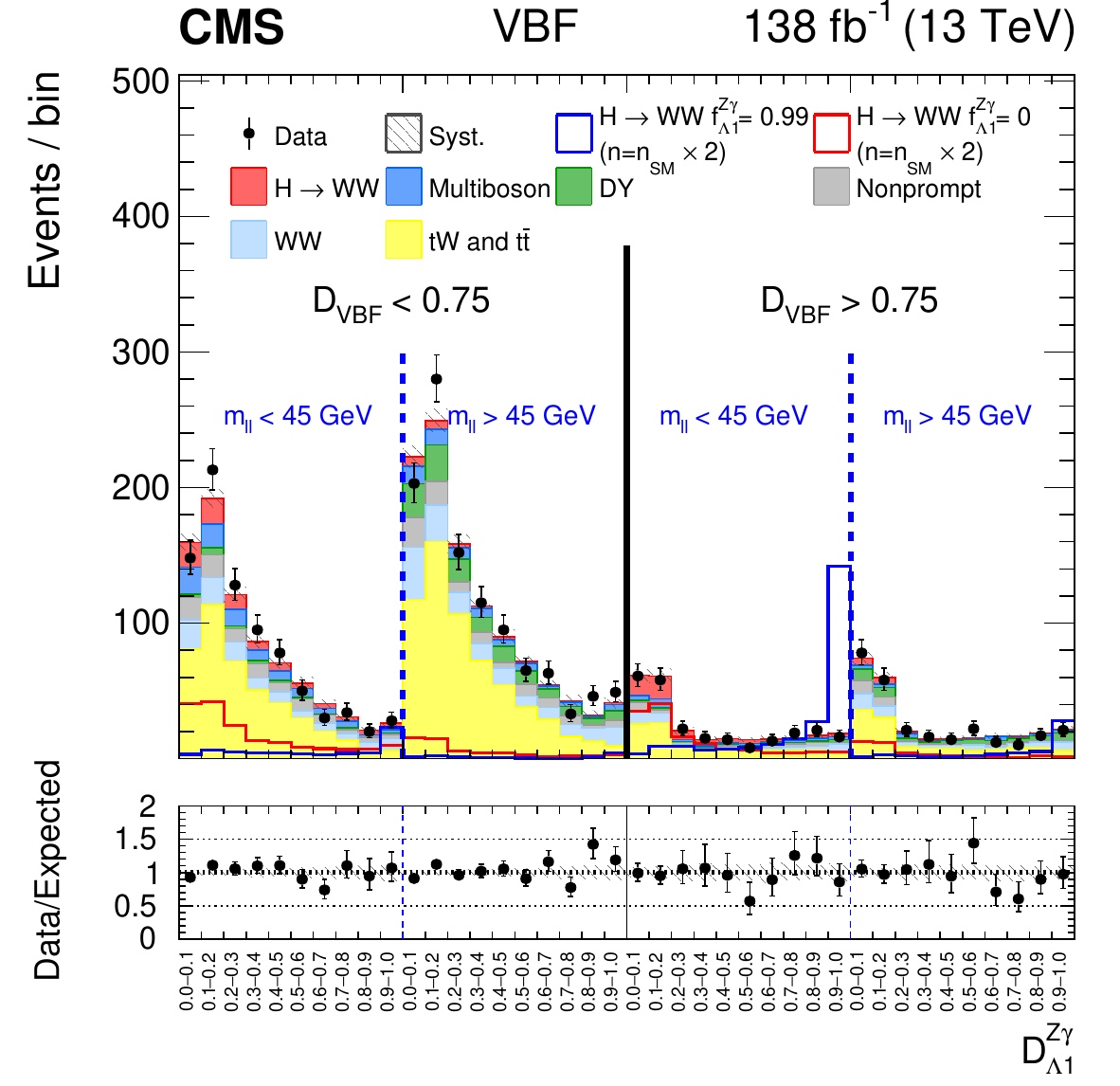} \\
\includegraphics[width=0.46\textwidth]{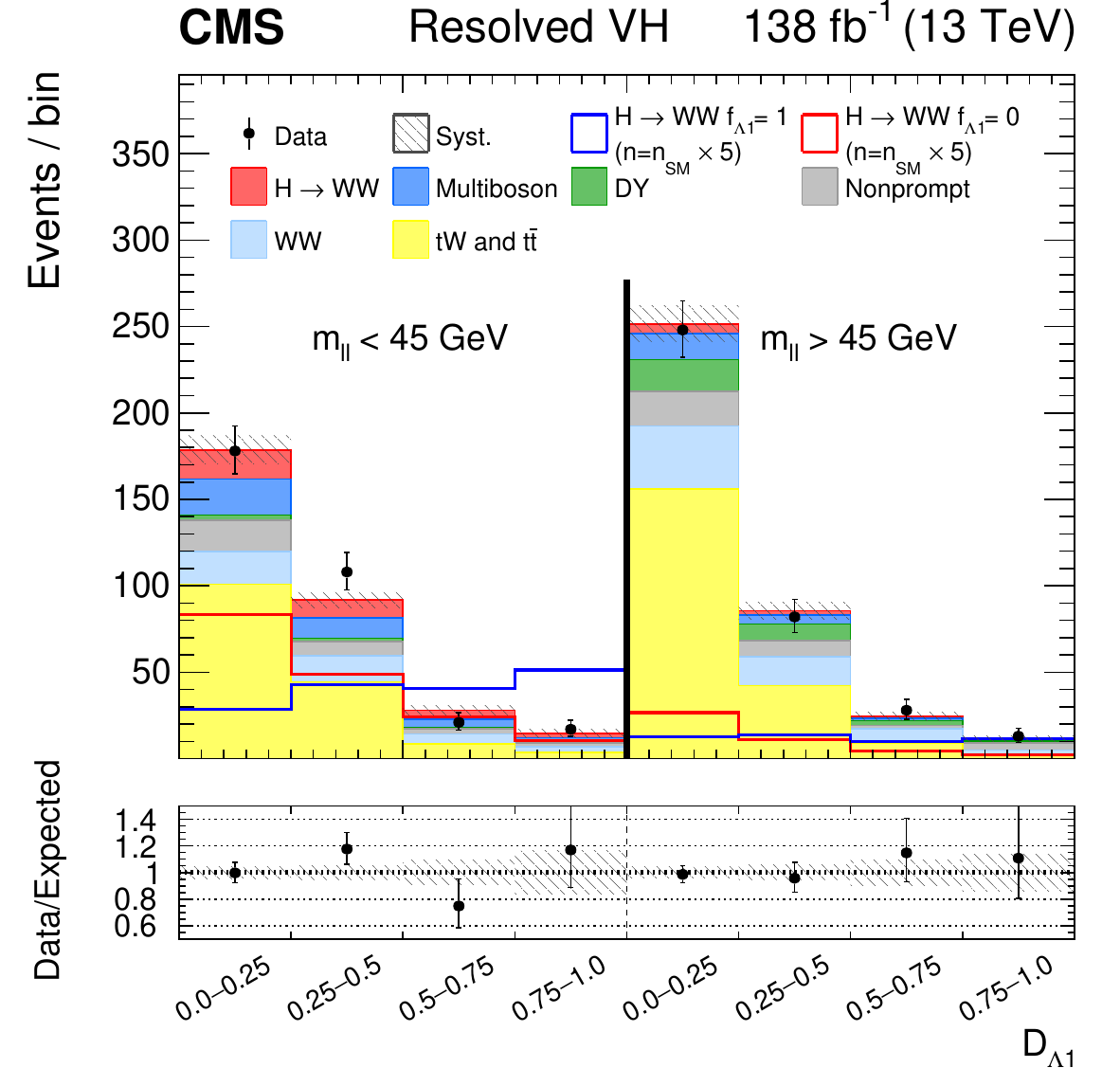}
\includegraphics[width=0.46\textwidth]{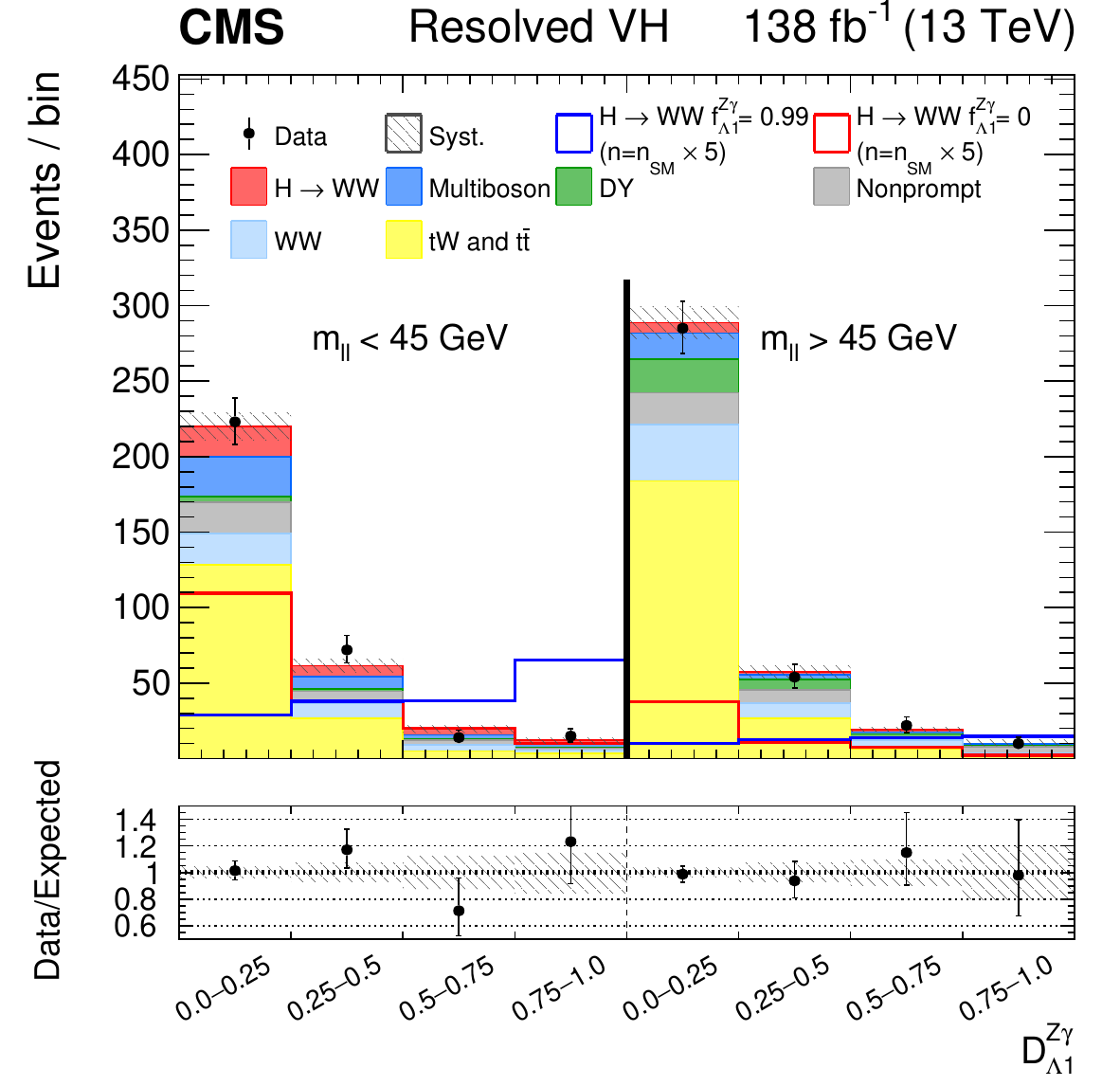} \\
\includegraphics[width=0.46\textwidth]{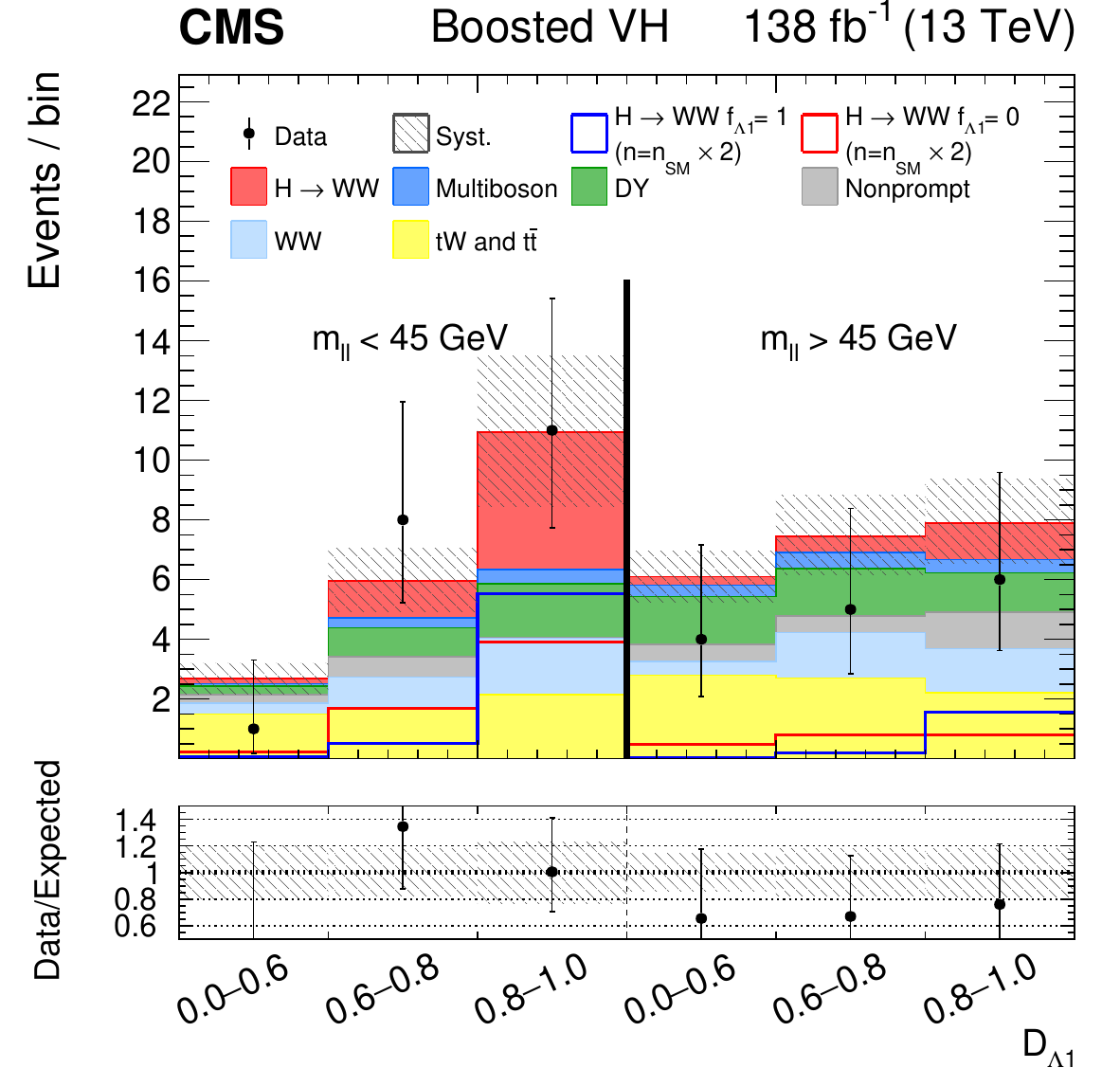}
\includegraphics[width=0.46\textwidth]{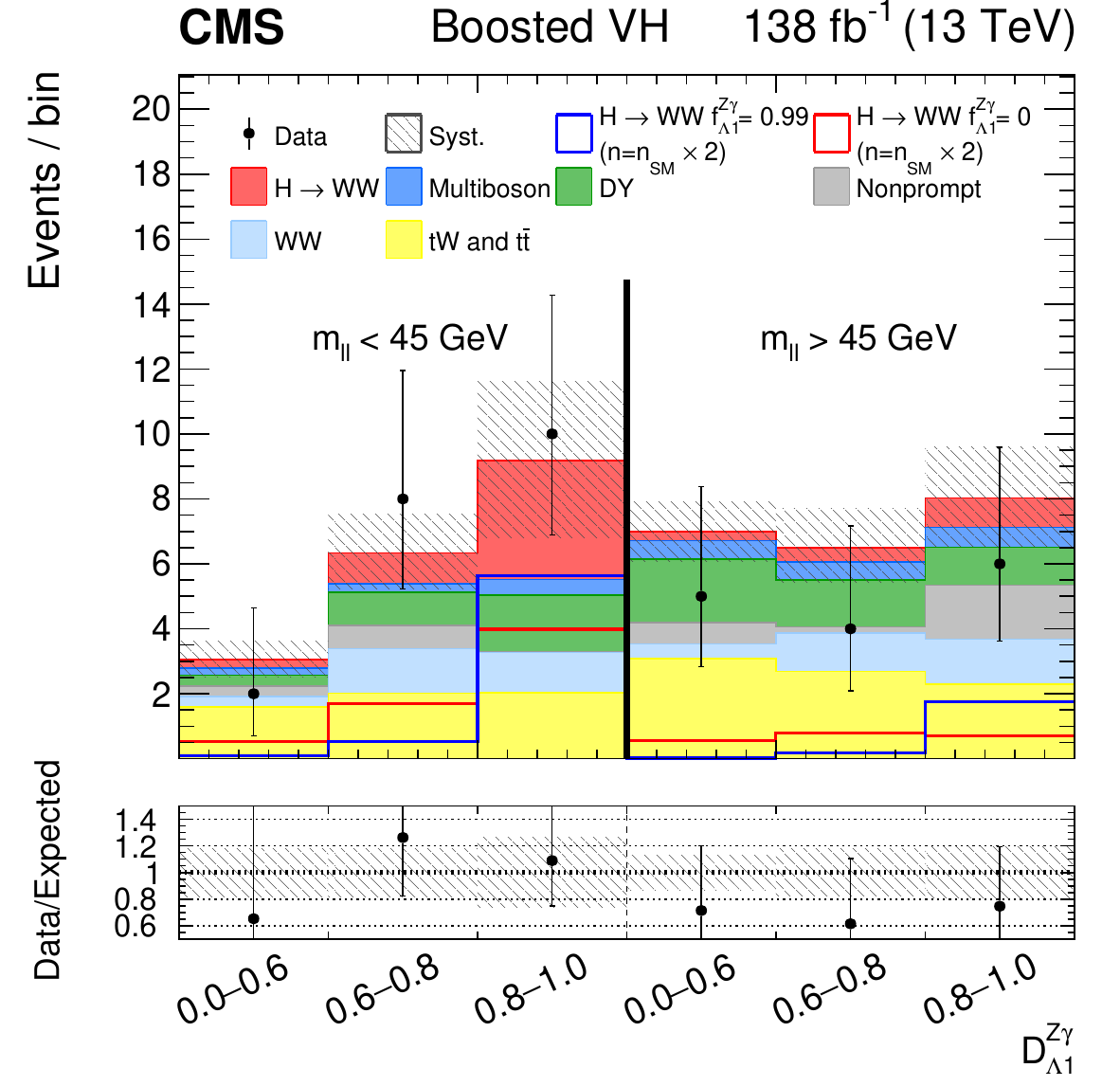}\\
\caption{
Observed and predicted distributions after fitting the data for $[\mathcal{D}_\text{VBF}, \mll, \mathcal{D}_{\Lambda 1}]$ (upper left) and $[\mathcal{D}_\text{VBF}, \mll, \mathcal{D}_{\Lambda 1}^{\PZ\gamma}]$ (upper right) in the VBF channel, and for $[\mll, \mathcal{D}_{\Lambda 1}]$ (left) and $[\mll, \mathcal{D}_{\Lambda 1}^{\PZ\gamma}]$ (right) in the Resolved \VH (middle) and Boosted \VH (lower) channels.
For the fits, the $a_{1}$ and $\kappa_1/(\Lambda_1)^2$ (left) or $a_{1}$ and $\kappa_2^{\PZ\PGg}/(\Lambda_1^{\PZ\PGg})^2$ (right) \HVV coupling contributions are included. More details are given in the caption of Fig.~\ref{fig:HVVH0PHpostfitRun2}. 
} 
\label{fig:HVVH0L1postfitRun2}
\end{figure*}

For the $a_{3}$ coupling parameter, a forward-backward categorization of events based on $\mathcal{D}_{CP}$ is implemented. In the case of the $a_{2}$ coupling parameter, $\mathcal{D}_\text{int}$ is largely correlated with $\mathcal{D}_{0+}$ in the \VH channels. Therefore, a forward-backward $\mathcal{D}_\text{int}$ categorization is implemented only in the VBF channel. Figures~\ref{fig:HVVH0PHpostfitRun2}--\ref{fig:HVVH0L1postfitRun2} show the discriminants used in the final fit to the data for the $a_{2}$, $a_{3}$, $\kappa_1/(\Lambda_1)^2$, and $\kappa_2^{\PZ\PGg}/(\Lambda_1^{\PZ\PGg})^2$ Approach 1 coupling studies in the VBF and \VH channels. A summary of the observables used in the \HVV Approach 1 analysis may be found in Table~\ref{tab:Dobservables}.

\subsubsection{VBF/\texorpdfstring{\VH}{VH} analysis strategy for Approach 2}

In Approach 2, we use one categorization strategy and build one multidimensional discriminant in each channel to target all the \HVV coupling parameters ($a_2$, $a_3$, $\kappa_1/(\Lambda_1)^2$) simultaneously. In the VBF channel, the $\mathcal{D}_{CP}$ and $\mathcal{D}_\text{int}$ discriminants are used to create four interference categories. Both $\mathcal{D}_\text{VBF}$ and $\mll$ are used as for Approach 1. All three $\mathcal{D}_\text{BSM}$ discriminants that target the $a_{2}, a_{3}$ and $\kappa_1/(\Lambda_1)^2$ coupling parameters are included. However, the number of bins we  implement is limited by the number of simulated events. Also the $\mathcal{D}_\text{BSM}$ discriminants are significantly correlated and so have similar performance for all couplings. Therefore, we use the $CP$-odd discriminant $\mathcal{D}_{0-}$ and just one of the $CP$-even discriminants, $\mathcal{D}_{0+}$, both with three bins and bin boundaries of 0.1 and 0.9. A dedicated rebinning strategy is applied to the $[\mathcal{D}_{0-}, \mathcal{D}_{0+}]$ distribution merging bins dominated by the SM Higgs boson prediction or with low precision in the background prediction. In the \VH channels, just two categories using $\mathcal{D}_{CP}$ are defined and the discriminant is built using \mll as for Approach 1. Again, both $\mathcal{D}_{0-}$ and $\mathcal{D}_{0+}$ are chosen for the final discriminant. For the Resolved \VH channel, we use three bins with boundaries of 0.25 and 0.75, whereas for the Boosted \VH case we use two bins with a boundary of 0.8. The same rebinning strategy described for the VBF channel is applied to both Resolved and Boosted \VH multidimensional discriminants. Table~\ref{tab:Dobservables} includes a summary of the observables used in the \HVV Approach 2 analysis.

\subsection{Kinematic features of \texorpdfstring{\HtoWW}{H to WW} decay products in 0- and 1-jet \texorpdfstring{\ggH}{ggH} channels}
\label{subsec:ggh}

\begin{figure}[!htb]
\centering 
\includegraphics[width=0.475\textwidth]{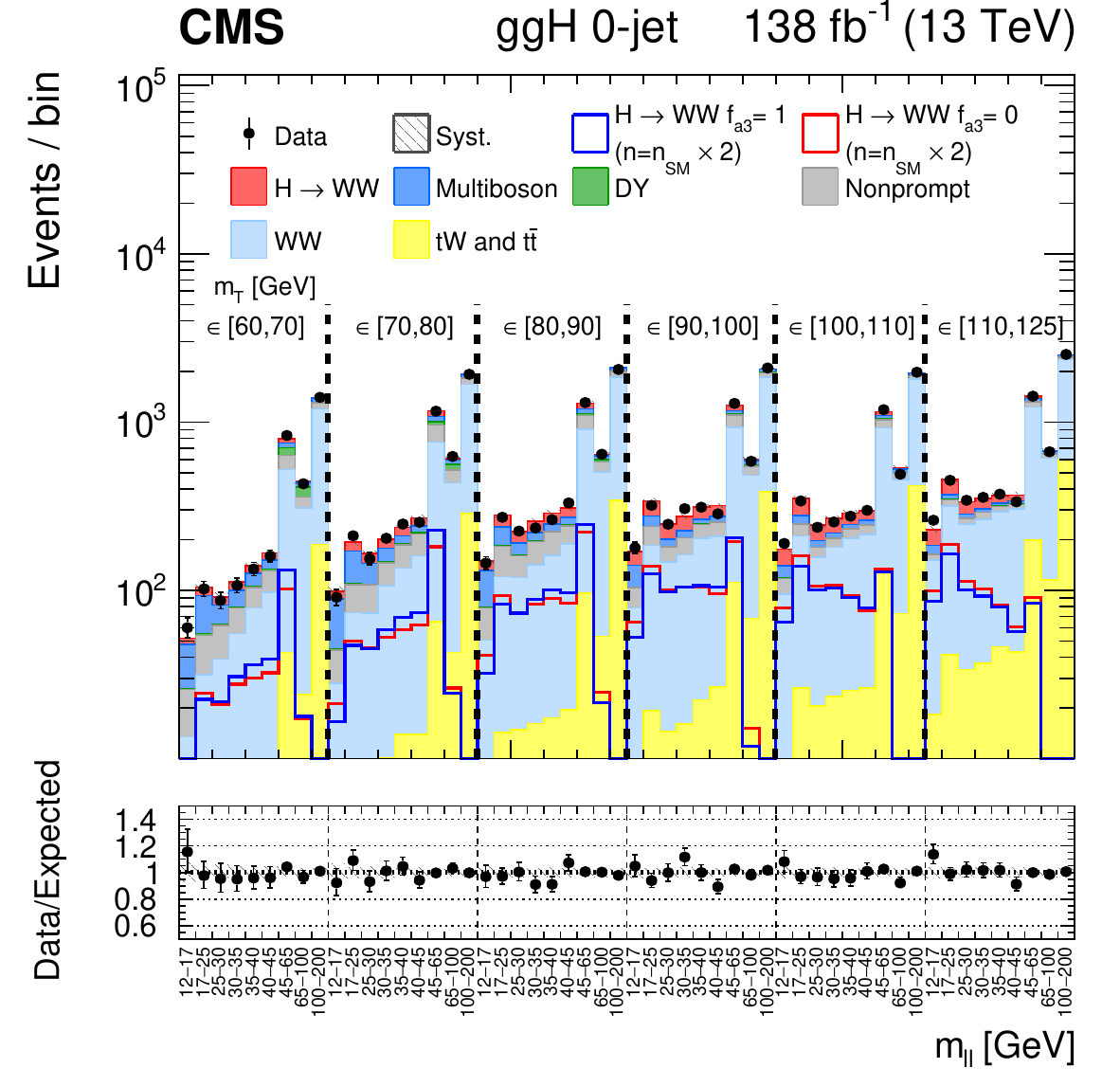}
\includegraphics[width=0.475\textwidth]{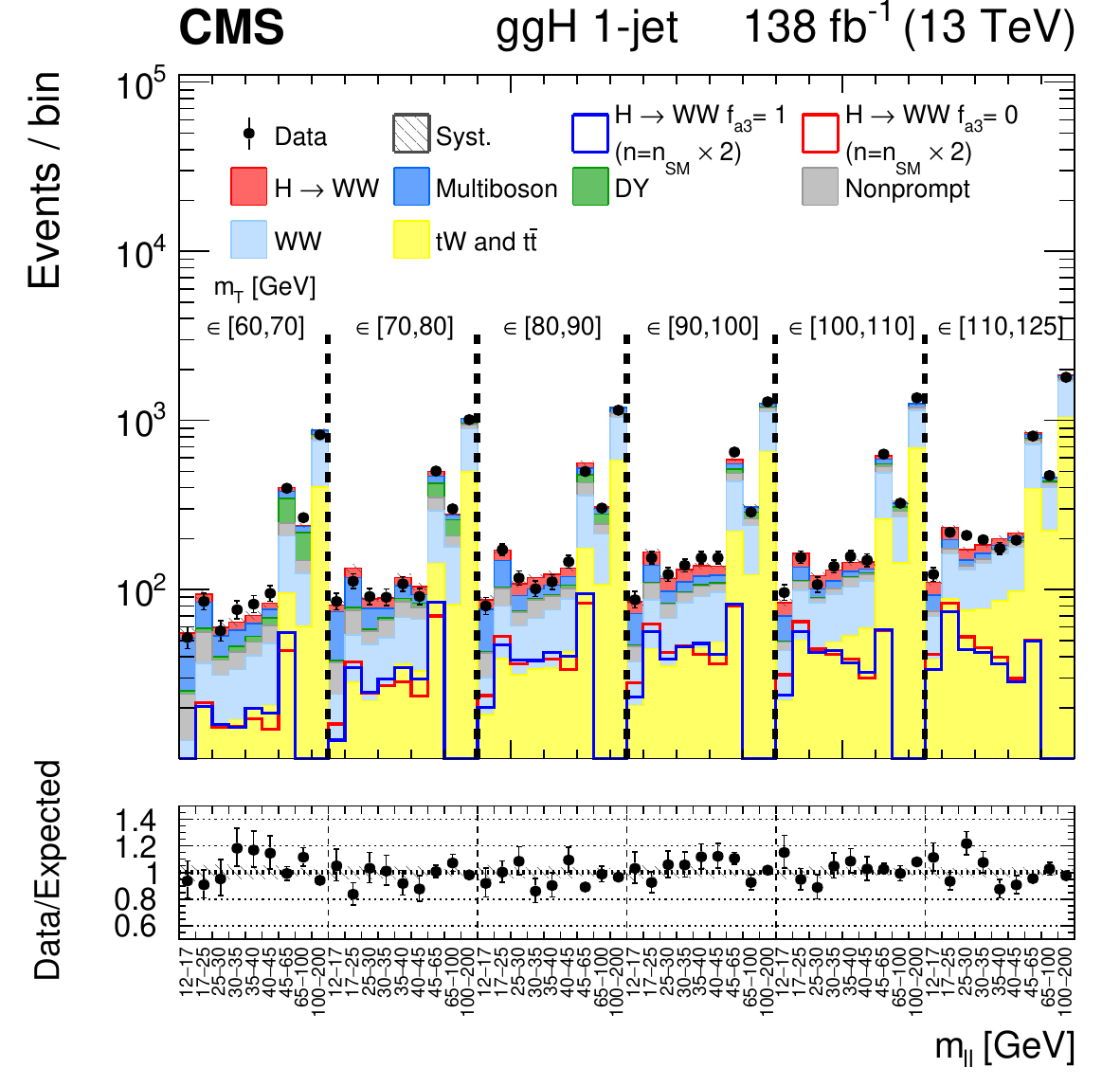}\\
\caption
{
Observed and predicted distributions after fitting the data for $[\mT, \mll]$ in the 0- (\cmsLeft) and 1-jet (\cmsRight) \ggH channels. 
For the fit, the $a_{1}$ and $a_3$ \HVV coupling contributions are included.
More details are given in the caption of Fig.~\ref{fig:HVVH0PHpostfitRun2}. 
\label{fig:HVVggHpostfitRun2}
}
\end{figure}

Similar to the SM \HtoWW analysis~\cite{CMS:2022uhn}, we use $\mll$ and $\mT$ to build 2D discriminants in the 0- and 1-jet \ggH channels. The distributions have nine bins for \mll in the range 12--200\GeV and six bins for \mT in the range 60--125\GeV. The bin widths vary and are optimized to achieve good separation between the SM Higgs boson signal and backgrounds, as well as between the different anomalous coupling signal hypotheses. In particular, a finer binning with respect to the SM \HtoWW analysis is implemented in regions where anomalous effects are most significant. Figure~\ref{fig:HVVggHpostfitRun2} shows the $[\mT, \mll]$ distributions in the 0- and 1-jet \ggH channels. The same $[\mT, \mll]$ discriminant is used to study all \HVV anomalous couplings for both Approach 1 and 2.  

\subsection{Kinematic features of two quark jets in 2-jet \texorpdfstring{\ggH}{ggH} channel}
\label{subsec:ggh2j}
 
\begin{figure}[!htb]
\centering 
\includegraphics[width=0.475\textwidth]{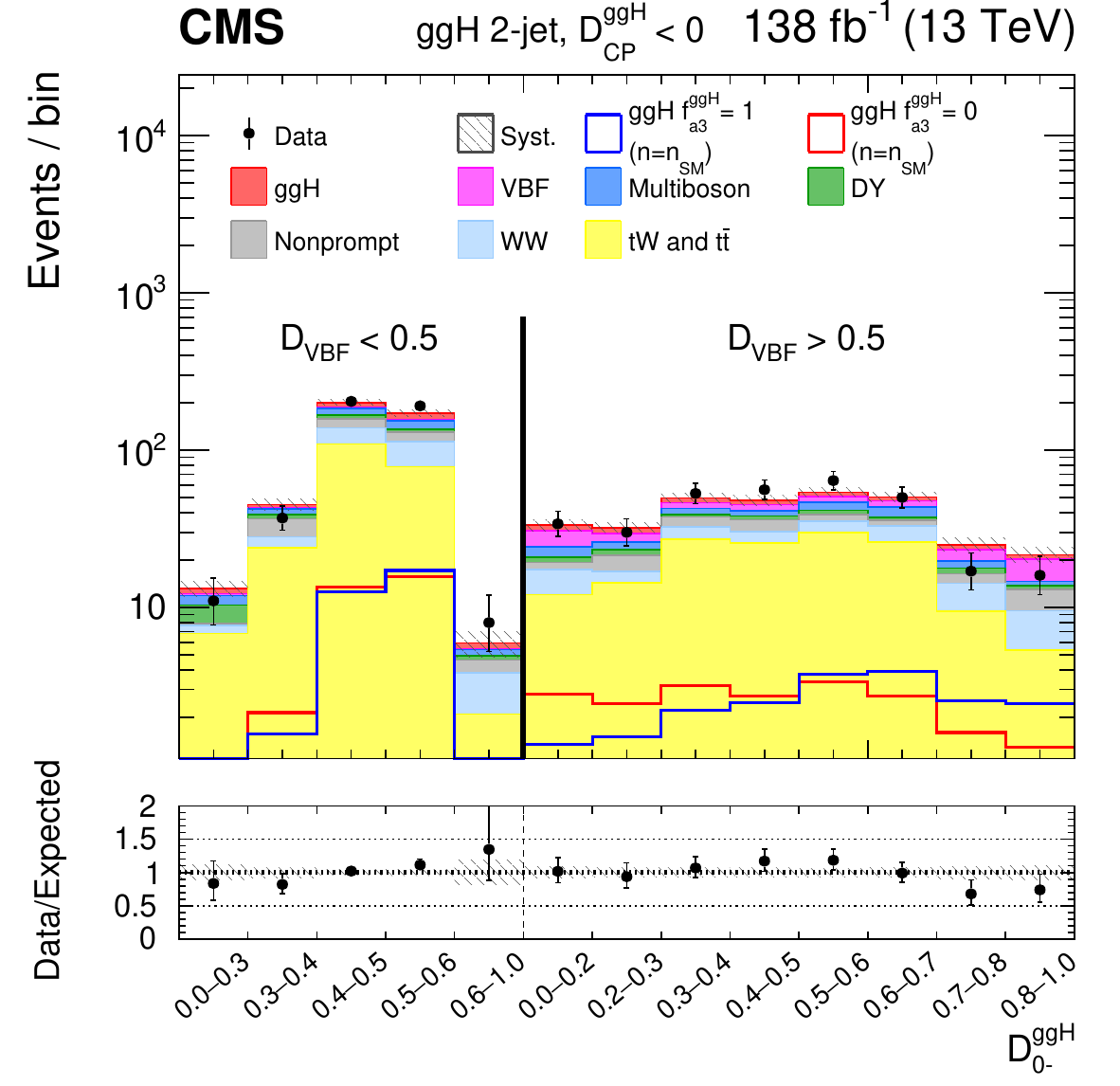}
\includegraphics[width=0.475\textwidth]{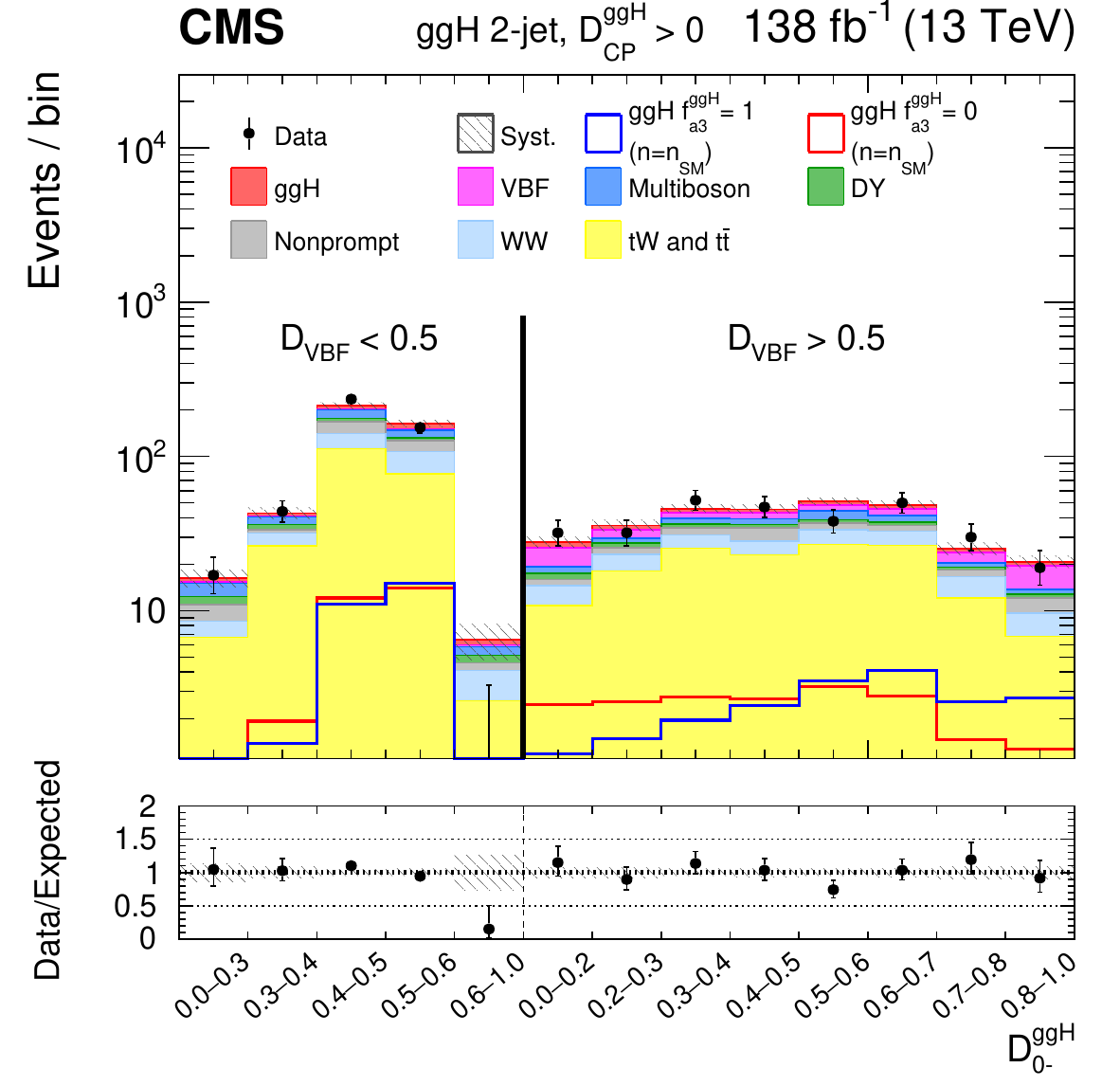}\\
\caption{
Observed and predicted distributions after fitting the data for $[\mathcal{D}_\text{VBF}$, $\mathcal{D}^{\ggH}_{0-}]$ in the 2-jet \ggH channel.
Both the $\mathcal{D}^{\ggH}_{\text{CP}} < 0$ (\cmsLeft) and $\mathcal{D}^{\ggH}_{\text{CP}} > 0$ (\cmsRight) categories are shown.
In this case, the VBF and \ggH signals are shown separately. 
For the fit, the $a_2^{\glgl}$ and $a_3^{\glgl}$ coupling contributions are included.
The corresponding pure $a_2^{\glgl}$ ($\fagg = 0$) and $a_3^{\glgl}$ ($\fagg = 1$) signal hypotheses are also shown superimposed, their yields correspond to the predicted number of SM signal events.
More details are given in the caption of Fig.~\ref{fig:HVVH0PHpostfitRun2}. 
}
\label{fig:HGGpostfitRun2}
\end{figure}

For the \Hgg coupling, we adopt a similar approach to the VBF $CP$ study, where the $CP$-odd $a_{3}$ \HVV coupling parameter is included. In this case, the optimal observables are $\mathcal{D}^{\ggH}_{0-}$ and $\mathcal{D}^{\ggH}_{CP}$, targeting the $CP$-odd $a_{3}$ \Hgg coupling parameter. A forward-backward categorization is implemented using $\mathcal{D}^{\ggH}_{CP}$, and the $\mathcal{D}_\text{VBF}$ and $\mathcal{D}^{\ggH}_{0-}$ observables are used to build 2D discriminants. The \mll variable is not considered in this case because it is not sensitive to anomalous \Hgg effects. For $\mathcal{D}_\text{VBF}$, the bin boundary is relaxed to 0.5 to ensure sufficient \ggH events are accepted in the more VBF-like bin. For $\mathcal{D}_{0-}$, eight (five) bins are used in the more (less) VBF-like bin with larger bin sizes at the extremes of the distribution to ensure sufficient precision in the background and signal predictions. {The 0- and 1-jet channels discussed previously are also included in this study to constrain the \ggH signal strength. The $[\mathcal{D}_\text{VBF}$, $\mathcal{D}^{\ggH}_{0-}]$ distributions used to analyze the \Hgg $a_{3}$ anomalous coupling in the 2-jet \ggH channel are shown in Fig.~\ref{fig:HGGpostfitRun2}. A summary of the observables used in the \Hgg analysis is given in Table~\ref{tab:Dobservables}.}

\begin{table*}[!htb]
\centering
\topcaption{
The kinematic observables used for the interference based categorization and for the final discriminants used in the fits to data to study the \HVV and \Hgg couplings. For each of the anomalous \HVV couplings in Approach 1, we have a dedicated analysis in the VBF and \VH channels. In Approach 2, we use one analysis to target all anomalous \HVV couplings simultaneously.   
}
\begin{tabular}{llll}
Analysis & Channel & Categorization & Final discriminant \\
\hline  
\multirow{8}{*}{\HVV } & VBF ($a_{2}$) & $\mathcal{D}_\text{int}$ & $[\mathcal{D}_\text{VBF}$, $\mll$, $\mathcal{D}_{0+}]$ \\ 
\multirow{8}{*}{Approach 1} & VBF ($a_{3}$) & $\mathcal{D}_{CP}$ & $[\mathcal{D}_\text{VBF}$, $\mll$, $\mathcal{D}_{0-}]$ \\ 
& VBF ($\kappa_1$) & \NA & $[\mathcal{D}_\text{VBF}$, $\mll$, $\mathcal{D}_{\Lambda 1}]$ \\
& VBF ($\kappa_2^{\PZ\PGg}$) & \NA & $[\mathcal{D}_\text{VBF}$, $\mll$, $\mathcal{D}_{\Lambda_1}^{\PZ\gamma}]$ \\
& \VH ($a_{2}$) & \NA & $[\mll$, $\mathcal{D}_{0+}]$ \\
& \VH ($a_{3}$) & $\mathcal{D}_{CP}$ & $[\mll$, $\mathcal{D}_{0-}]$ \\ 
& \VH ($\kappa_1$) & \NA & $[\mll$, $\mathcal{D}_{\Lambda 1}]$ \\
& \VH ($\kappa_2^{\PZ\PGg}$) & \NA & $[\mll$, $\mathcal{D}_{\Lambda_1}^{\PZ\gamma}]$ \\
& 0- $\&$ 1-jet \ggH & \NA & $[\mT$, $\mll]$ \\[\cmsTabSkip]
\multirow{2}{*}{\HVV} & VBF & $\mathcal{D}_{CP}$, $\mathcal{D}_\text{int}$ & $[\mathcal{D}_\text{VBF}$, $\mll$, $\mathcal{D}_{0-}$, $\mathcal{D}_{0+}]$ \\ 
\multirow{2}{*}{Approach 2} & VH & $\mathcal{D}_{CP}$ & $[\mll$, $\mathcal{D}_{0-}$, $\mathcal{D}_{0+}]$ \\
& 0- $\&$ 1-jet \ggH & \NA & $[\mT$, $\mll]$ \\[\cmsTabSkip]
\multirow{2}{*}{\Hgg} 
& 2-jet \ggH & $\mathcal{D}^{\ggH}_{CP}$ & $[\mathcal{D}_\text{VBF}$, $\mathcal{D}^{\ggH}_{0-}]$ \\
& 0- $\&$ 1-jet \ggH & \NA & $[\mT$, $\mll]$ \\
\end{tabular}
\label{tab:Dobservables}
\end{table*}

\section{Systematic uncertainties}
\label{sec:systematics}

The signal extraction is performed using binned templates to describe the various signal and background processes. Systematic uncertainties that change the normalization or shape of the templates are included. All the uncertainties are modeled as nuisance parameters that are profiled in the maximum likelihood fit described in Section~\ref{sec:results}. The systematic uncertainties arise from both experimental or theoretical sources.
         
\subsection{Experimental uncertainties}

The following experimental systematic uncertainties are included in the final fit to data: 

\begin{itemize}
\item The total uncertainty associated with the measurement of the integrated luminosity for 2016, 2017, and 2018 is $1.2\%\text{~\cite{CMS:LUM-17-003}, }2.3\%\text{~\cite{CMS-PAS-LUM-17-004}, and }2.5\%$~\cite{CMS-PAS-LUM-18-002}, respectively. This uncertainty is partially correlated among the three data sets, resulting in an overall uncertainty of 1.6\%.
\item The systematic uncertainty in the trigger efficiency is determined by varying the tag lepton selection criteria and the \PZ boson mass window used in the tag-and-probe method. It affects both the normalization and the shape of the signal and background distributions, and is kept uncorrelated among data sets. The total normalization uncertainty is less than 1\%. 
\item The tag-and-probe method is also used to determine the lepton identification and isolation efficiency. Corrections are applied to account for any discrepancy in the efficiencies measured in data and simulation. The corresponding systematic uncertainty is about 1\% for electrons and 2\% for muons. 
\item The uncertainties in the determination of the lepton momentum scale mainly arise from the limited data sample used for their estimation. The impact on the normalization of the signal and background templates ranges between 0.6--1.0\% for the electron momentum scale and is about 0.2\% for the muon momentum scale. 
They are treated as uncorrelated among the three data-taking years.     
\item {\tolerance=800 The jet energy scale uncertainty is modeled by implementing eleven independent nuisance parameters corresponding to different jet energy correction sources, six of which are correlated among the three data sets. Their effects vary in the range of 1--10\%, mainly depending on the jet multiplicity in the analysis phase space. Another source of uncertainty arises from the jet energy resolution smearing applied to simulated samples to match the \pt resolution measured in data. The effect varies in a range of 1--5\%, depending on the jet multiplicity and is uncorrelated among the data sets. These uncertainties are included for both AK4 and AK8 jets. In addition, the \mj scale and resolution, and \PV tagging corrections with their corresponding uncertainties are included for \PV-tagged AK8 jets. These variables are calibrated in a top quark-antiquark sample enriched in hadronically decaying \PW bosons~\cite{Khachatryan:2014vla}.\par}  
\item The effects of the unclustered energy scale, jet energy scale, and lepton \pt scales are included for the calculation of the missing transverse momentum. The resulting normalization systematic uncertainty is 1--10\% and is treated as uncorrelated among the years. 
\item Both the normalization and shape of the signal and background templates are affected by the jet pileup identification uncertainty. The effect is below 1\%.
\item The uncertainty associated with the \PQb tagging efficiency is modeled by seventeen nuisance parameters out of which five are of a theoretical origin and are correlated among the three data sets. The remaining set of four parameters per data set are treated as uncorrelated as they arise from the statistical accuracy of the efficiency measurement~\cite{Sirunyan:2017ezt}.  
\item Estimation of the nonprompt-lepton background is affected by the limited size of the data sets used for the misidentification rate measurements. It is also affected by the difference in the flavor composition of jets misidentified as leptons between the misidentification rate measurement region (enriched in multijet events) and the signal phase space. The effects on the nonprompt-lepton background estimation range between a few percent to about 10\% depending on the SR and are treated as nuisance parameters uncorrelated between electrons and muons and among the three data sets. A normalization uncertainty of 30\%~\cite{Sirunyan:2018egh} is assigned to fully cover for any discrepancies with respect to data in a $\PW+$jets CR and is treated as uncorrelated among data sets. 
\item The statistical uncertainties due to the limited number of simulated events are also included for all bins of the background distributions used to extract the results~\cite{Barlow:249779}.      

\end{itemize}
\label{sec:exp_uncty}
\subsection{Theoretical uncertainties}

Multiple theoretical uncertainties are considered and are correlated among data sets, unless stated otherwise: 

\begin{itemize}
\item The uncertainties related to the choice of PDF and \alpS have a minor effect on the shape of the distributions. Therefore, only normalization effects related to the event acceptance and to the cross section are included. However, these uncertainties are not considered for the backgrounds that have their normalization constrained through data in dedicated CRs. For the Higgs boson signal processes, these uncertainties are calculated by the LHC Higgs cross section working group~\cite{deFlorian:2227475}. 
\item The theoretical uncertainties arising from missing higher-order corrections in the cross section calculations are also included. Background simulations are reweighted to the alternative scenarios corresponding to renormalization $\mu_{\text{R}}$ and factorization $\mu_{\text{F}}$ scales varied by factors 0.5 or 2 and the envelopes of the varied templates are taken as the one standard deviations. For background processes that have their normalization constrained through data in dedicated CRs, we consider only the shape effect of the uncertainties coming from the missing higher-order corrections. The \WW nonresonant background has the uncertainties derived by varying $\mu_{\text{R}}$, $\mu_{\text{F}}$, and the resummation scale. For the \ggH and VBF signal processes, the effects of the missing higher-order corrections on the overall cross section are decoupled into multiple sources according to the recipes described in Ref.~\cite{deFlorian:2227475}.
\item The uncertainty due to the pileup modeling was included for the main simulated background processes (DY, \WW, top quark) as well as the \ggH and VBF signals. The effect is determined by varying the total inelastic $\Pp\Pp$ cross section (69.2\unit{mb}~\cite{PhysRevLett.117.182002, Sirunyan:2018nqx}) within the assigned 5\% uncertainty. 
\item The PS modeling mainly affects the jet multiplicity, causing migration of events between categories that results in template shape changes. Associated uncertainties are evaluated by reweighting events with varied PS weights computed with \PYTHIA 8.212. The effect on the signal strength is found to be below 1\%.
\item Uncertainties associated with UE modeling are evaluated by varying the UE tune parameters used in the MC sample generation. Systematic uncertainties are correlated between the 2017 and 2018 data sets since they share the same UE tunes, whereas for 2016 the uncertainty is considered uncorrelated. The UE uncertainty has a minimal effect on the template shapes and affects the normalization by about 1.5\%.
\item A 15\% uncertainty is applied to the relative fraction of the \glgl-induced component in nonresonant \WW production~\cite{HiggsCAT}. The relative fraction between single top quark and \ttbar processes is assigned a systematic uncertainty of 8\%~\cite{Chatrchyan:2013iaa}. Additional process-specific (DY, $\PV\PZ$, $\PV\PGg$, $\PV\PGg^*$) uncertainties, related to corrections to account for possible discrepancies between data and simulation, are assigned and are correlated among data sets.              
\end{itemize}
\label{sec:theory_uncty}

\section{Results}

\label{sec:results}
The optimization and validation of the analysis were performed using simulation and data in CRs. The data in the SRs were examined once all details of the analysis were finalized.
For the final results, we perform a binned maximum likelihood fit to the data combining all channels and data-taking periods. The statistical approach was developed by the ATLAS and CMS Collaborations in the context of the LHC Higgs Combination Group~\cite{LHC-HCG}. The likelihood function is defined for candidate events as:
\begin{linenomath}
\ifthenelse{\boolean{cms@external}}
{\begin{multline}
    \mathcal{L}(\text{data} | \mu_{\ggH}, \mu_{\text{EW}}, \fai, \theta) =\\
    \prod_{\mathrm{j}}\text{Poisson}(n_{\mathrm{j}}|s_{\mathrm{j}}(\mu_{\ggH}, \mu_{\text{EW}}, \fai, \theta)+b_{\mathrm{j}}(\theta)) p(\tilde{\theta} | \theta)\text{,}
    \label{eq:likelihoodDensity}
\end{multline}}
{\begin{equation}
    \mathcal{L}(\text{data} | \mu_{\ggH}, \mu_{\text{EW}}, \fai, \theta) = \prod_{\mathrm{j}}\text{Poisson}(n_{\mathrm{j}}|s_{\mathrm{j}}(\mu_{\ggH}, \mu_{\text{EW}}, \fai, \theta)+b_{\mathrm{j}}(\theta)) p(\tilde{\theta} | \theta)\text{,}
    \label{eq:likelihoodDensity}
 \end{equation}
}
\end{linenomath}
where j runs over all bins and $n_{\mathrm{j}}$ is the observed number of data events in each bin.
Total signal and background expectations in each bin are represented by $s_{\mathrm{j}}$ and $b_{\mathrm{j}}$, respectively.
The individual signal and background processes considered in each category are described using binned templates of multidimensional discriminants as described in Section~\ref{sec:KD}. Each signal process is parametrized as a linear combination of terms originating from the SM, and anomalous couplings and their interference. The signal expectation depends on the parameters $\mu_{\ggH}$, $\mu_{\text{EW}}$, and \fai, and is constrained by the data fit. Both the signal and background expectations are functions of $\theta$, which represents the full set of nuisance parameters corresponding to the systematic uncertainties. The CRs described in Section~\ref{sec:evsel} are included in the fit in the form of single bins, representing the number of events in each CR.

The $\mu_{\ggH}$ and $\mu_{\text{EW}}$ parameters correspond to the Higgs boson signal strength modifiers for the \ggH and VBF/\VH signals, respectively. Signal yields for the VBF and \VH processes are related to each other because the same \HVV couplings enter both in production and decay of the Higgs boson. The \ggH signal is initiated predominantly by the top fermion couplings and is unrelated to the VBF and \VH production mechanisms. As the signal strength modifiers are free parameters in the fit, the overall signal event yield is not used to discriminate between alternative signal hypotheses. The \fai parameter corresponds to the anomalous coupling cross section fraction and determines the shape of the signal expectation. The cross section fraction for the SM coupling is simply taken as $1 - \abs{\fai}$. In Approach 1, the SM and just one anomalous \HVV coupling are included, and each \fai is thus studied independently. Depending on the particular anomalous coupling under investigation, \fai may represent \fatwo, \fathree, \fL, or \fLzg. For Approach 2, the SM and three anomalous \HVV couplings are included. In this case, \fai represents \fatwo, \fathree and \fL, which are studied simultaneously. It is explicitly required that $\abs{\fatwo} + \abs{\fathree} + \abs{\fL} \le 1$ to avoid probing an unphysical parameter space. Finally, there is just one anomalous coupling corresponding to \fagg to consider for the \Hgg vertex. For this study, we also include the effect of the $CP$-odd \HVV anomalous coupling on the VBF process. This is achieved by including \fathree as a free parameter in the fit. The $p(\tilde{\theta}|\theta)$ are the probability density functions (PDFs) for the observed values of the nuisance parameters, $\tilde{\theta}$, obtained from calibration measurements. The systematic uncertainties that affect only the normalizations of the signal and background processes are treated as PDFs following a log-normal distribution, whereas shape-altering systematic uncertainties are treated as Gaussian PDFs~\cite{LHC-HCG}.

Additional interpretations in terms of the SMEFT Higgs and Warsaw basis coupling parameters are also considered using Eqs.~(\ref{eq:EFTpar1}--\ref{eq:EFTpar4}) and Eqs.~(\ref{eq:EFTWpar1}--\ref{eq:EFTWpar4}), respectively. In each case, four independent couplings are studied simultaneously and the effect of the couplings on the total width of the Higgs boson is taken into account. For the \fai measurements, this effect is absorbed by the signal strength modifiers. A parameterization of the partial widths of the main Higgs boson decay modes as a function of the couplings is used to determine the effect on the Higgs boson width~\cite{hzz_ac_2021, Gritsan:2020pib}. 

The likelihood is maximized with respect to the signal modifier parameters and with respect to the nuisance parameters. Confidence level (\CL) intervals are determined from profile likelihood scans of the respective parameters. The allowed 68\% and 95\% \CL intervals are defined using the set of parameter values at which the profile likelihood function $-2\Delta\ln{\mathcal{L}}$ = 1.00 and 3.84~\cite{Sirunyan_2019}, respectively, for which exact coverage is expected in the asymptotic limit~\cite{10.1214/aoms/1177732360}. The likelihood value at a given \fai is determined by the shape of the signal hypothesis and the relative signal event yields between categories. Expected results are obtained using the Asimov data set~\cite{Cowan_2011} constructed using the SM values of the signal modifier parameters.

\begin{figure*}[!htbp]
\centering
\includegraphics[width=0.475\textwidth]{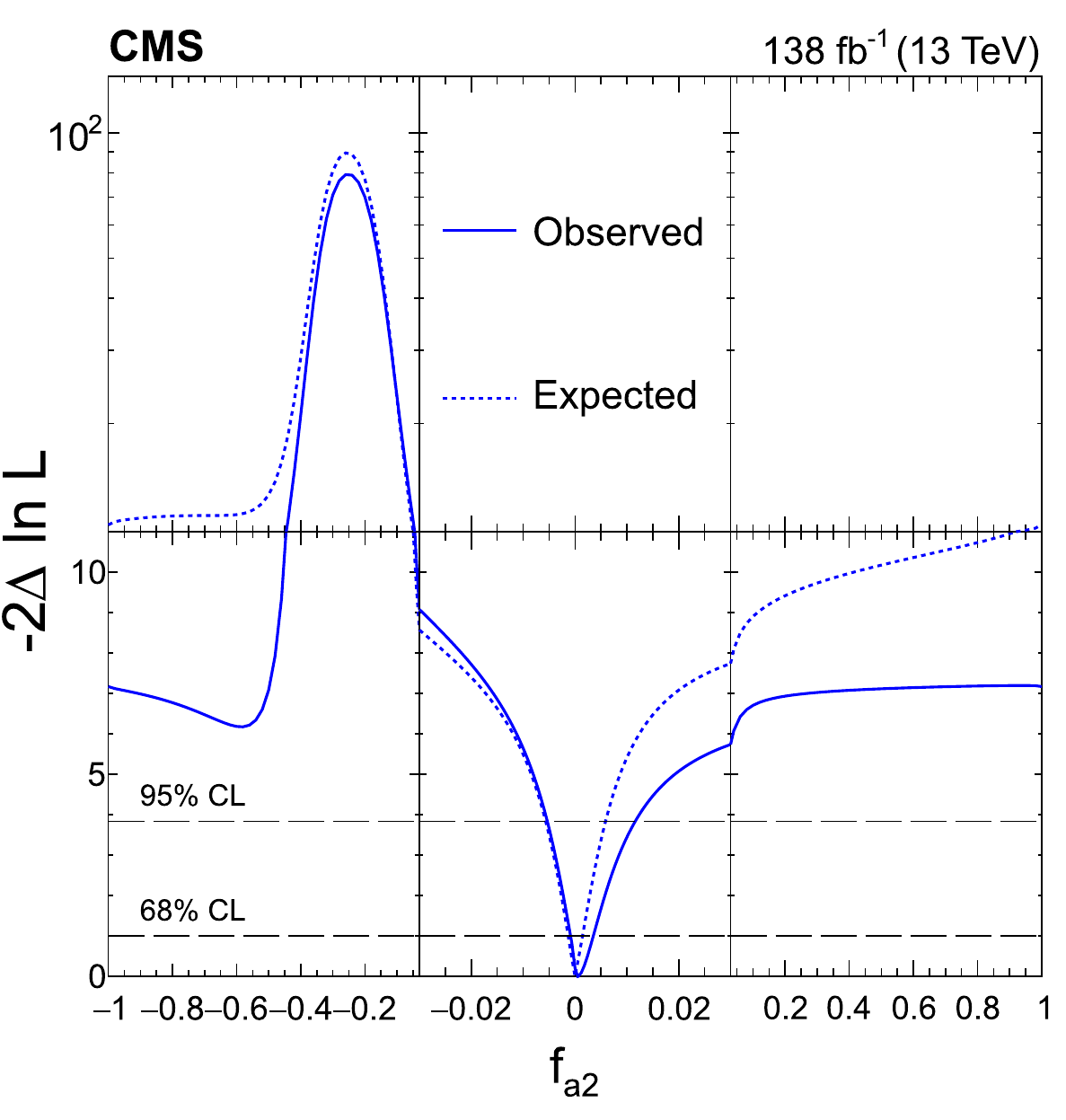}
\includegraphics[width=0.475\textwidth]{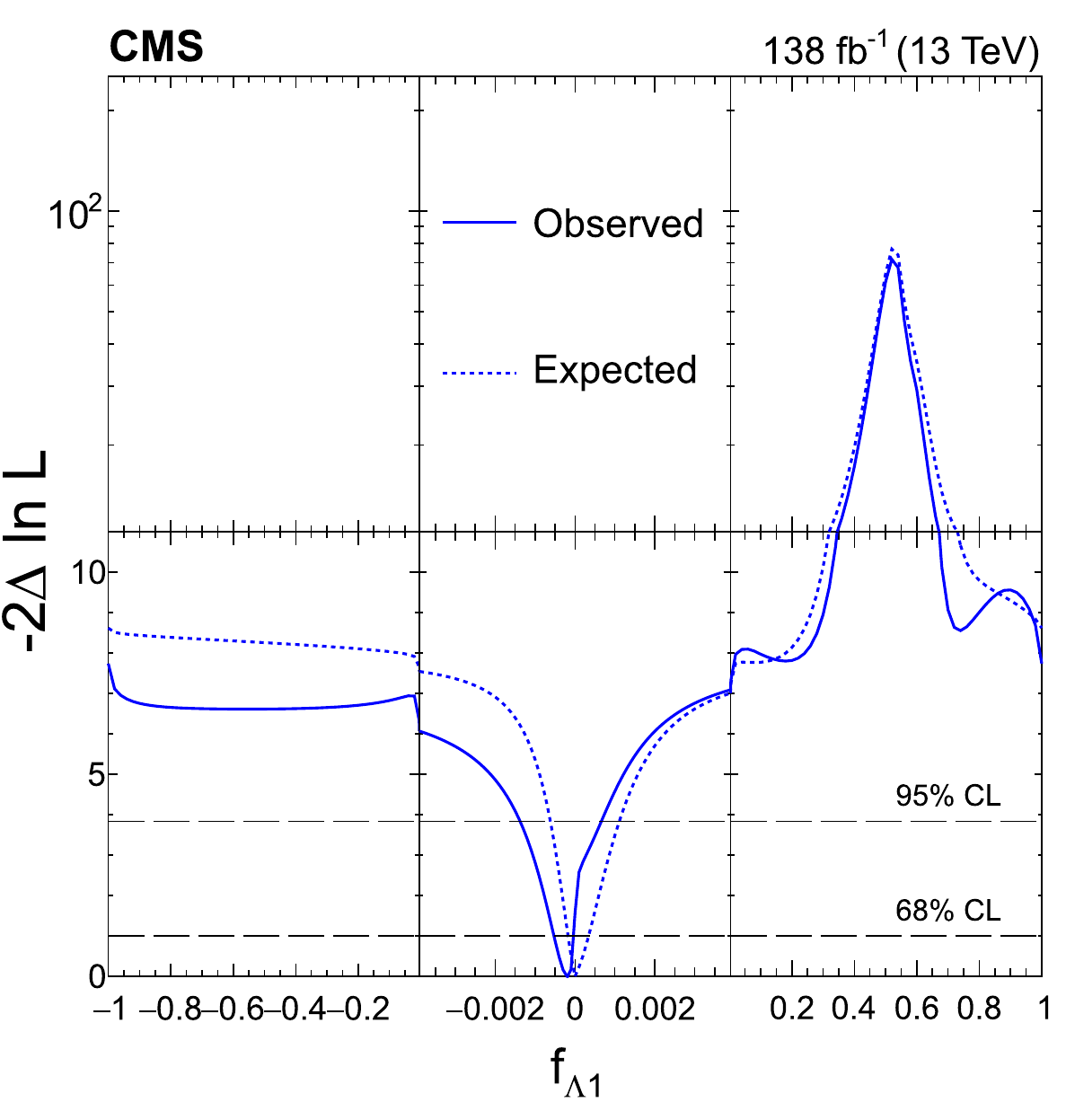} \\
\includegraphics[width=0.475\textwidth]{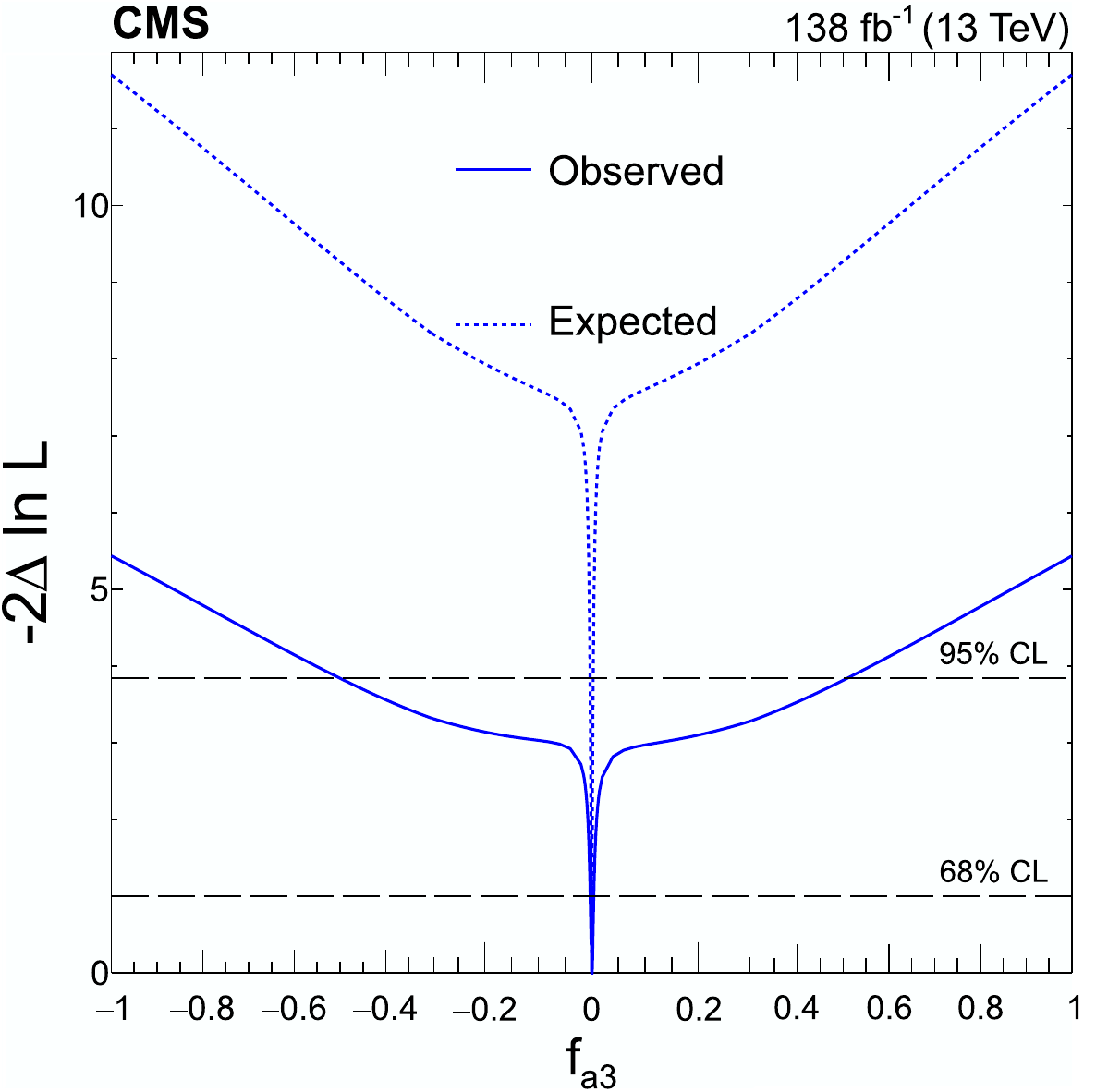}
\includegraphics[width=0.475\textwidth]{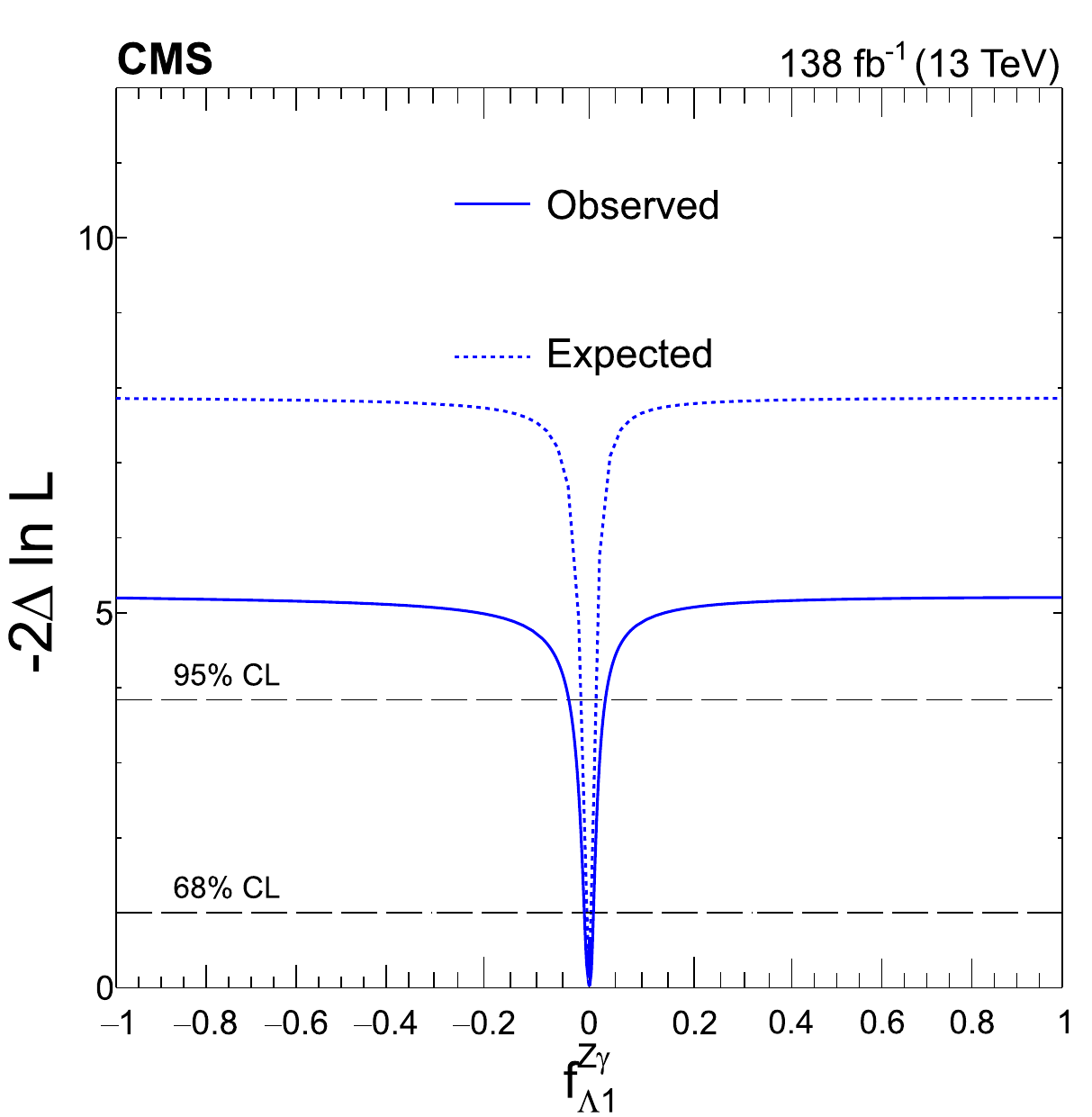}\\
\caption{
Expected (dashed) and observed (solid) profiled likelihood on \fatwo (upper left), \fL (upper right), \fathree (lower left), and \fLzg (lower right) using Approach 1. In each case, the signal strength modifiers are treated as free parameters. The dashed horizontal lines show the 68 and 95\% \CL regions. Axis scales are varied for \fatwo and \fL to improve the visibility of important features.
}
\label{fig:Comb1}
\end{figure*}

\begin{figure*}[!htb]
\centering
\includegraphics[width=0.475\textwidth]{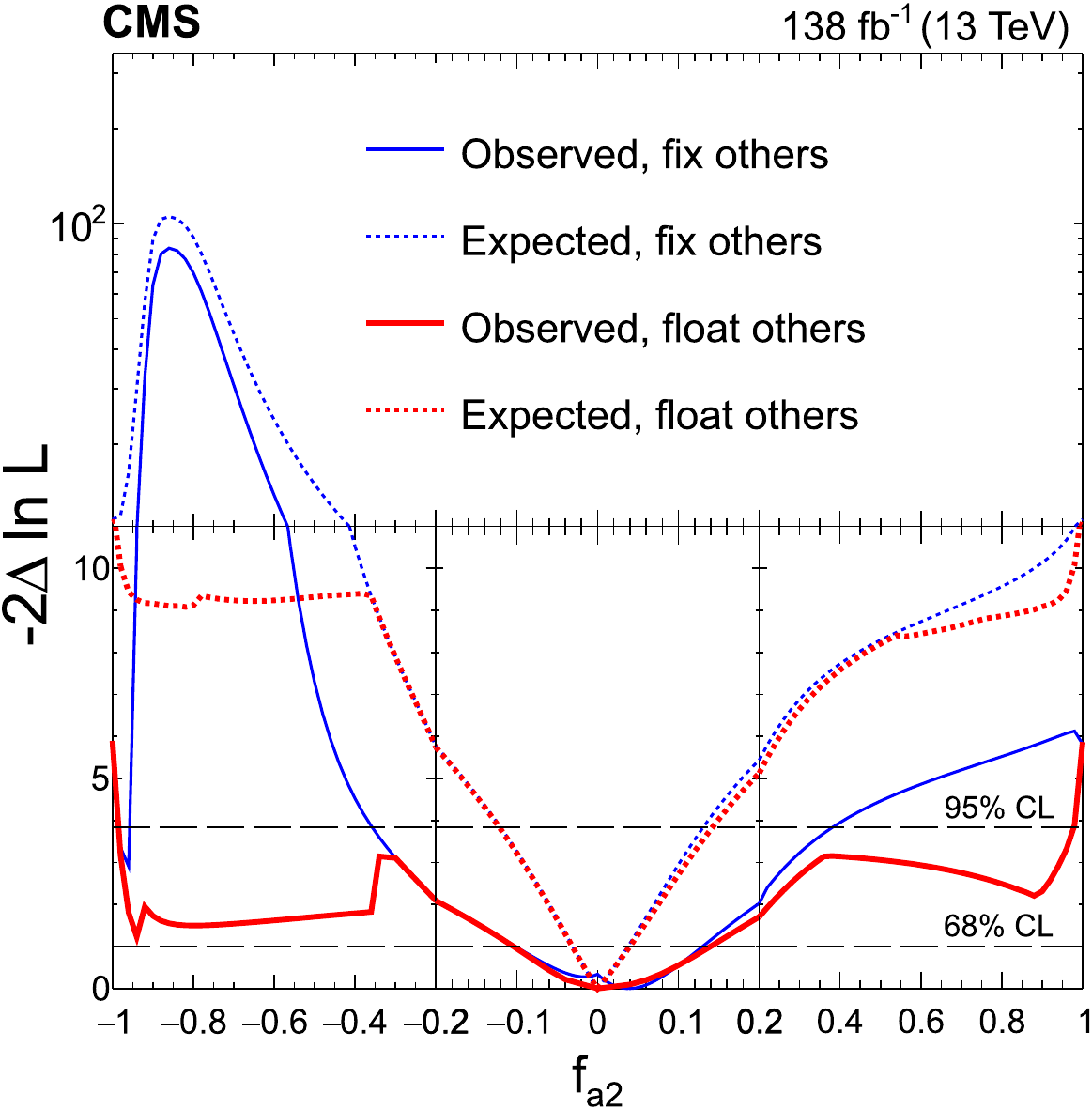}
\includegraphics[width=0.475\textwidth]{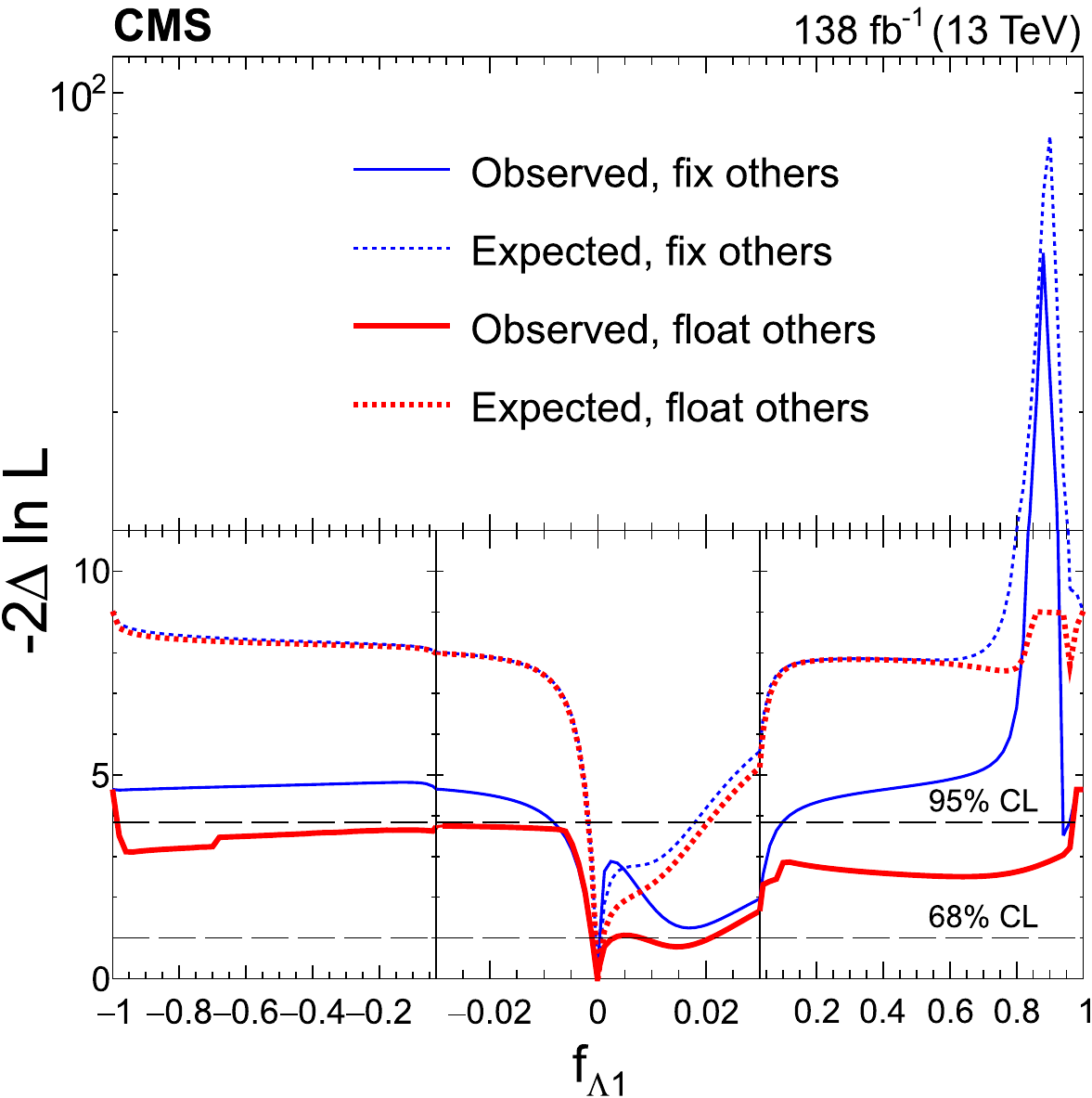} \\
\includegraphics[width=0.475\textwidth]{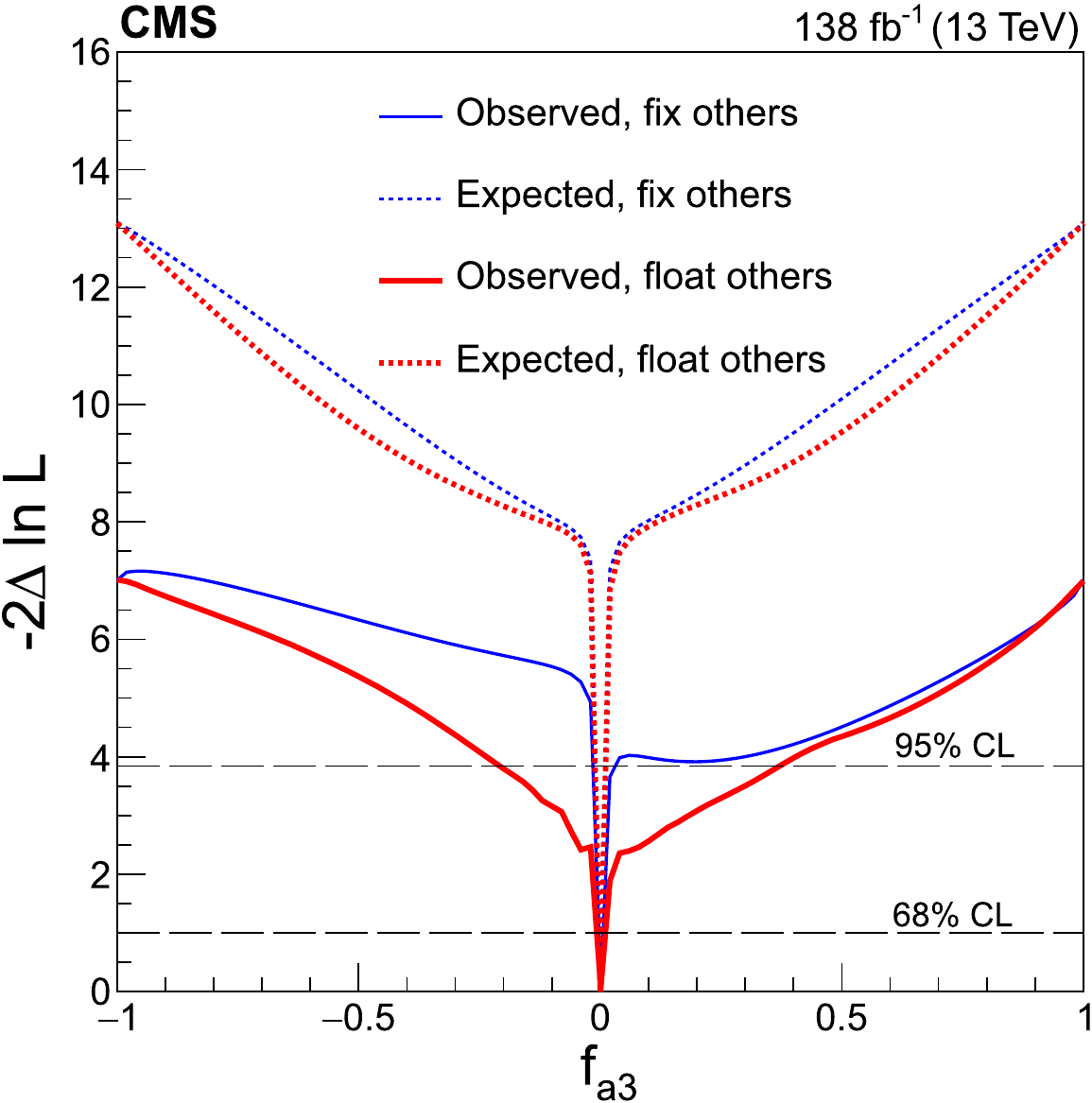}
\caption{
Expected (dashed) and observed (solid) profiled likelihood on \fatwo (upper left), \fL (upper right) and \fathree (bottom) using Approach 2. The other two anomalous coupling cross section fractions are either fixed to zero (blue) or left floating in the fit (red). In each case, the signal strength modifiers are treated as free parameters. The dashed horizontal lines show the 68 and 95\% \CL regions. Axis scales are varied for \fatwo and \fL to improve the visibility of important features.
}
\label{fig:Comb2}
\end{figure*}

\begin{figure*}[!h]
\centering
\includegraphics[width=0.475\textwidth]{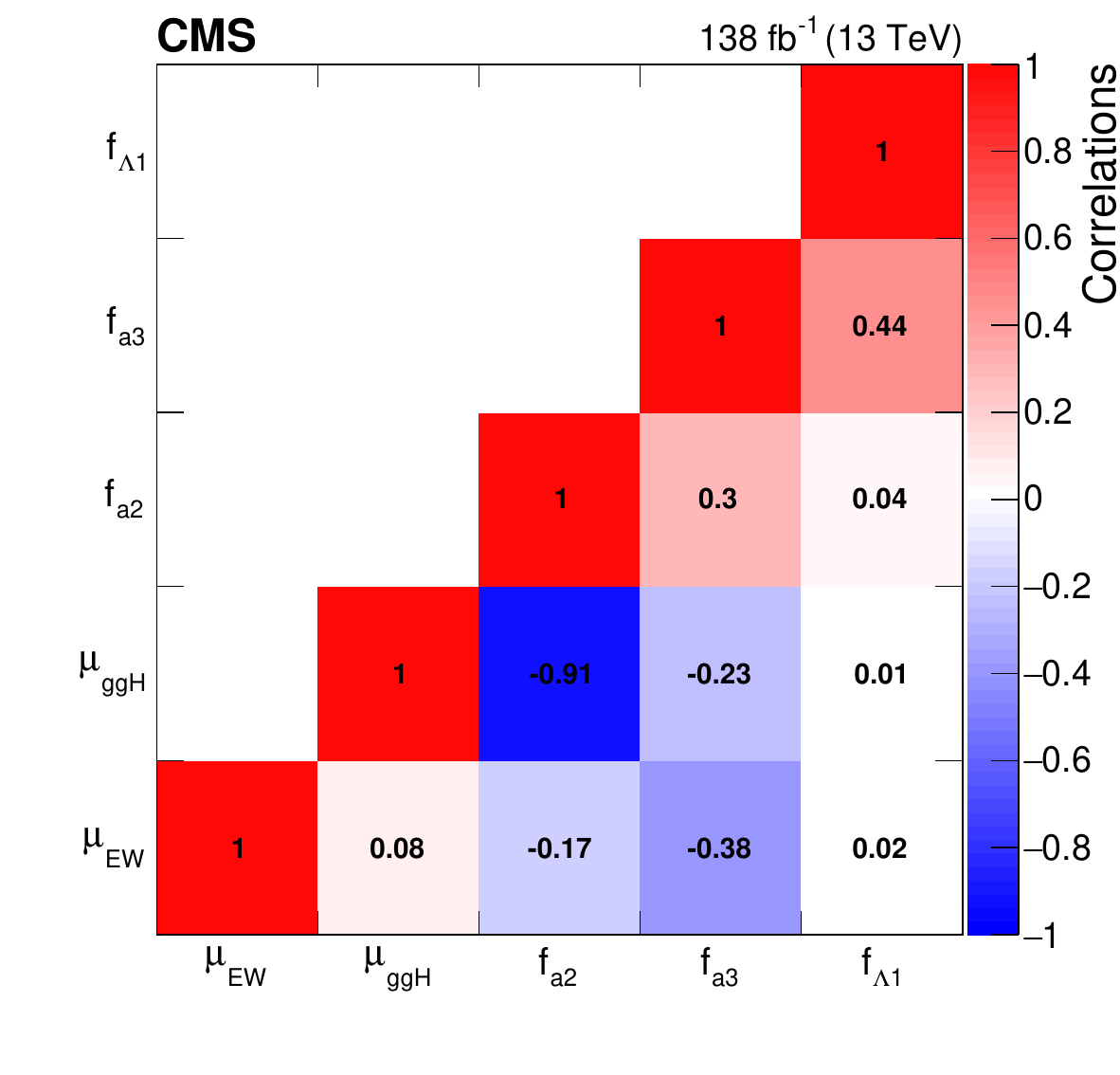}
\includegraphics[width=0.475\textwidth]{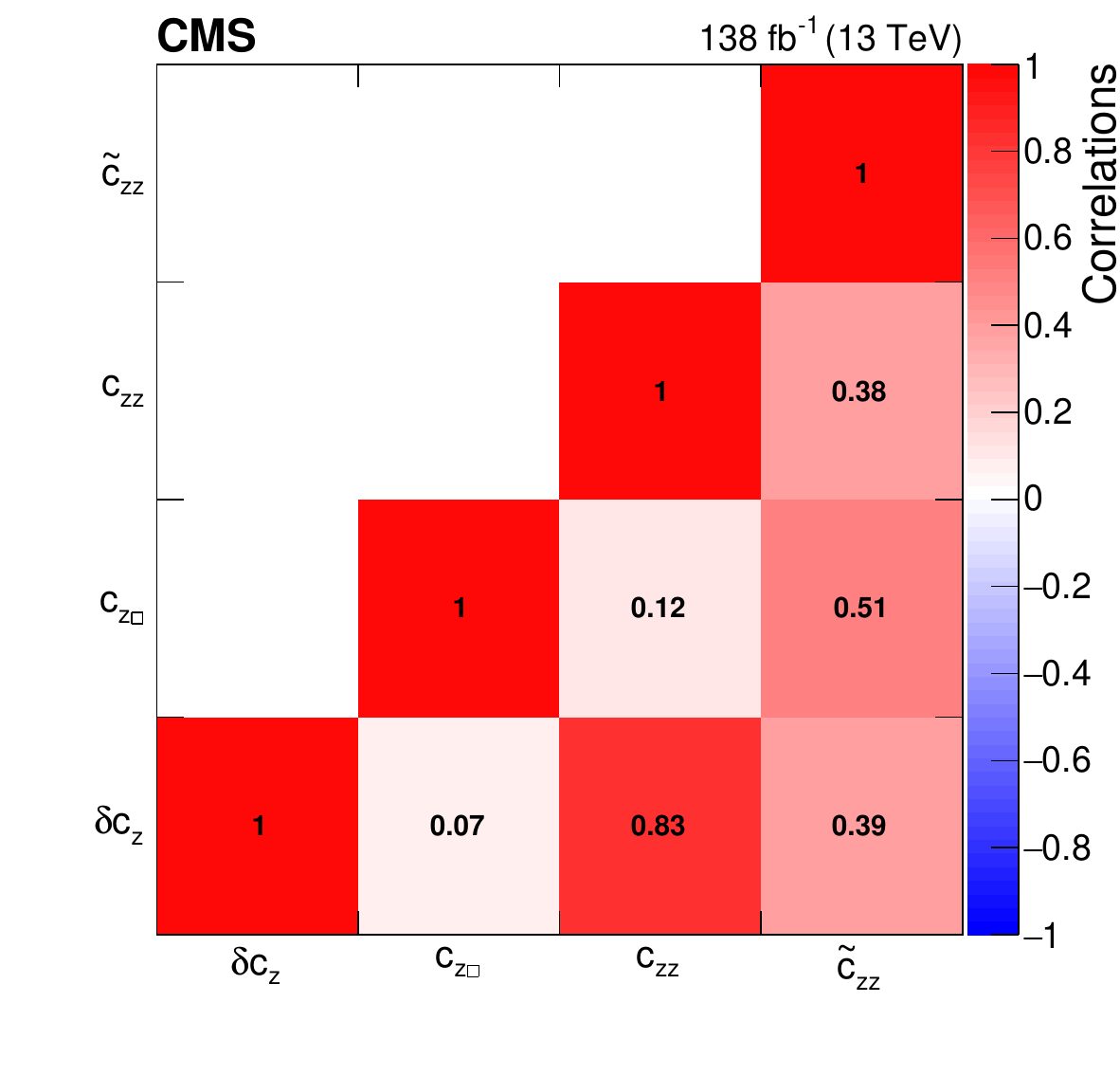}
\caption{
The observed correlation coefficients between \HVV anomalous coupling cross section fractions and signal strength modifiers (left) and between SMEFT Higgs basis coupling parameters (right). 
}
\label{fig:correlations}
\end{figure*}

For Approach 1, where we assume $a_{i}^{\PZ\PZ}=a_{i}^{\PW\PW}$, the expected and observed \fatwo, \fathree, \fL, and \fLzg likelihood scans are shown in Fig.~\ref{fig:Comb1}. Significant interference effects for negative values of \fatwo, around $-0.25$, and positive values of \fL, around 0.5, are evident. Relatively large changes in the signal shape with respect to the SM are predicted at these values. Also evident are narrow minima around \fai = 0. The anomalous coupling terms in Eq.~(\ref{eq:formfactfullampl}) have a $q_{i}^{2}$ dependence, which can be larger at the VBF/\VH production vertex than at the Higgs decay vertex. This causes the cross section and the shape of the VBF/\VH signal hypothesis to change rapidly with \fai. For \fLzg, there are no anomalous effects at the Higgs decay vertex and so the only structure present is the narrow minimum related to the VBF/\VH production vertex. The axis scales are varied to improve the visibility of important features for \fatwo and \fL. For Approach 2, where the SU(2) x U(1) coupling relationships from Eqs.~(\ref{eq:EFT1}--\ref{eq:EFT5}) are adopted, the expected and observed \fatwo, \fathree and \fL likelihood scans are shown in Fig.~\ref{fig:Comb2}. 
The results are shown for each \fai separately with the other two \fai either fixed to zero or left floating in the fit. The measured values of the signal strength parameters correspond to $\mu_{\text{EW}}=0.9^{+0.19}_{-0.24}$ and $\mu_{\ggH}=0.9^{+0.38}_{-0.20}$ when all parameters float simultaneously. 
It is notable that the observed $-2\Delta\ln{\mathcal{L}}$ profile values are generally lower than expected. 
This is consistent with a downward statistical fluctuation in the number of VBF and \VH events. 
The lowest $\mu_{\text{EW}}$ value measured is 0.82 for the Approach 1 \fathree fit which can be compared with the highest value of 0.97 for the corresponding \fL fit. 
In each case, the uncertainty in $\mu_{\text{EW}}$ is about 20\% and as such all fitted values are consistent with both the SM and each other. 
More generally, all anomalous \HVV coupling parameter measurements are consistent with the expectations for the SM Higgs boson. The p-value compatibility of the full Approach 2 fit, where all signal parameters float simultaneously, with the SM is 91\%. A summary of constraints on the anomalous \HVV coupling parameters with the best fit values and allowed 68\% and 95\% \CL intervals are shown in Table~\ref{tab:HVVsum}. The most stringent constraints on the \HVV anomalous coupling cross section fractions are at the per mille level. 
Some constraints are less stringent than expected due to the fitted values of $\mu_{\text{EW}}$ being lower than the SM expectation.
The observed correlation coefficients between \HVV anomalous coupling cross section fractions and signal strength modifiers are displayed in Fig.~\ref{fig:correlations}.

\begin{figure*}[!h]
\centering
\includegraphics[width=0.475\textwidth]{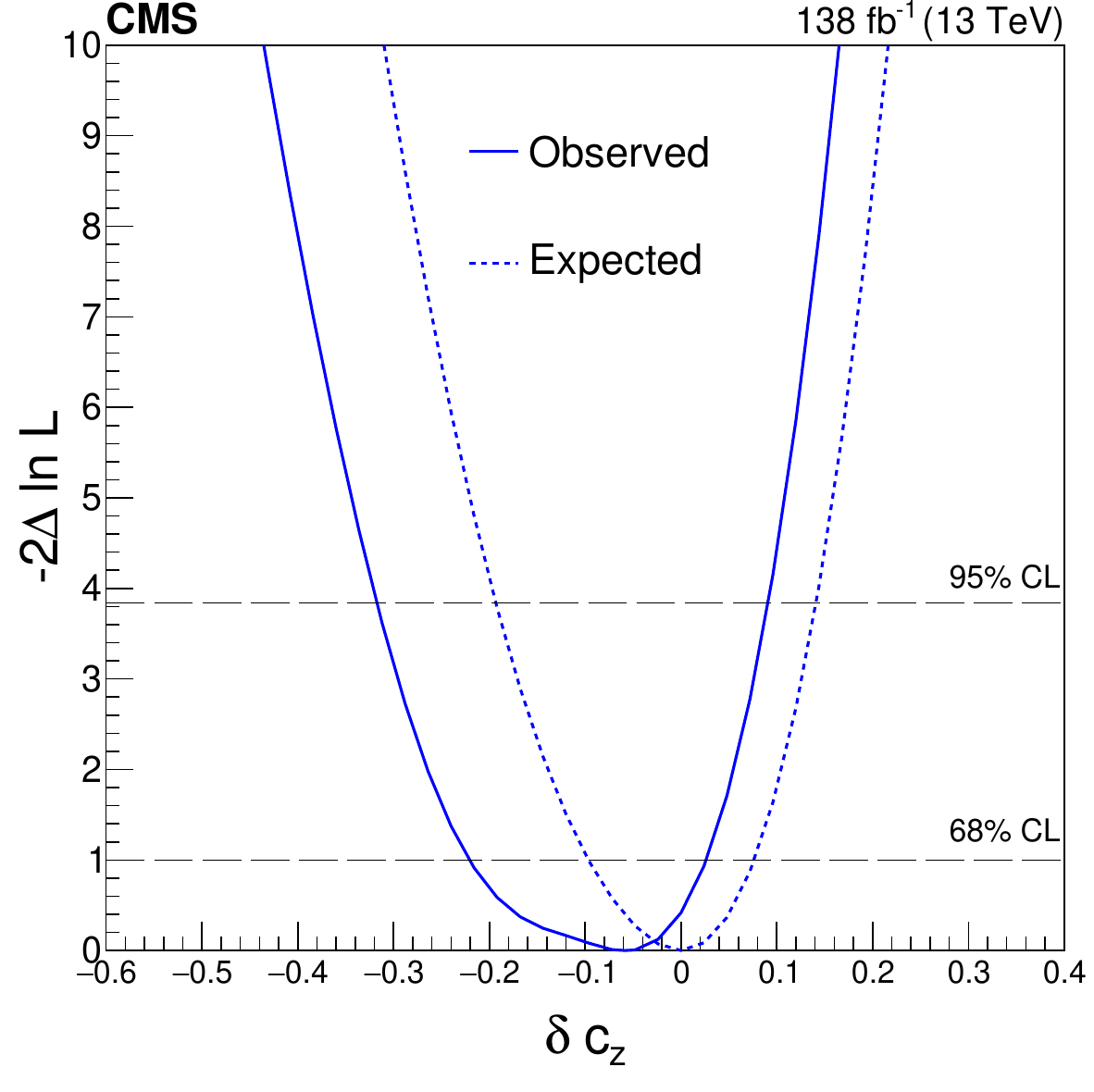}
\includegraphics[width=0.475\textwidth]{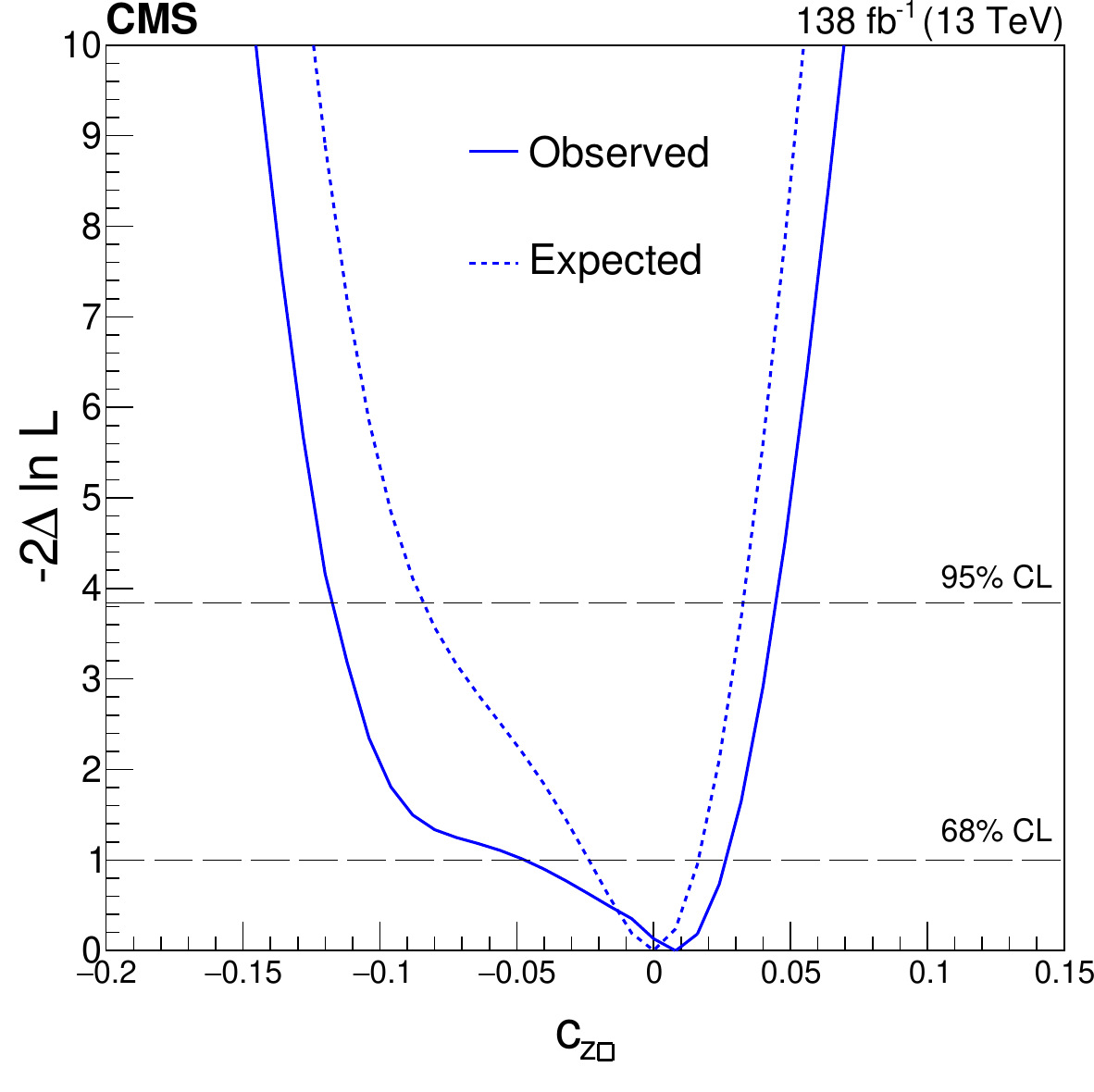}\\
\includegraphics[width=0.475\textwidth]{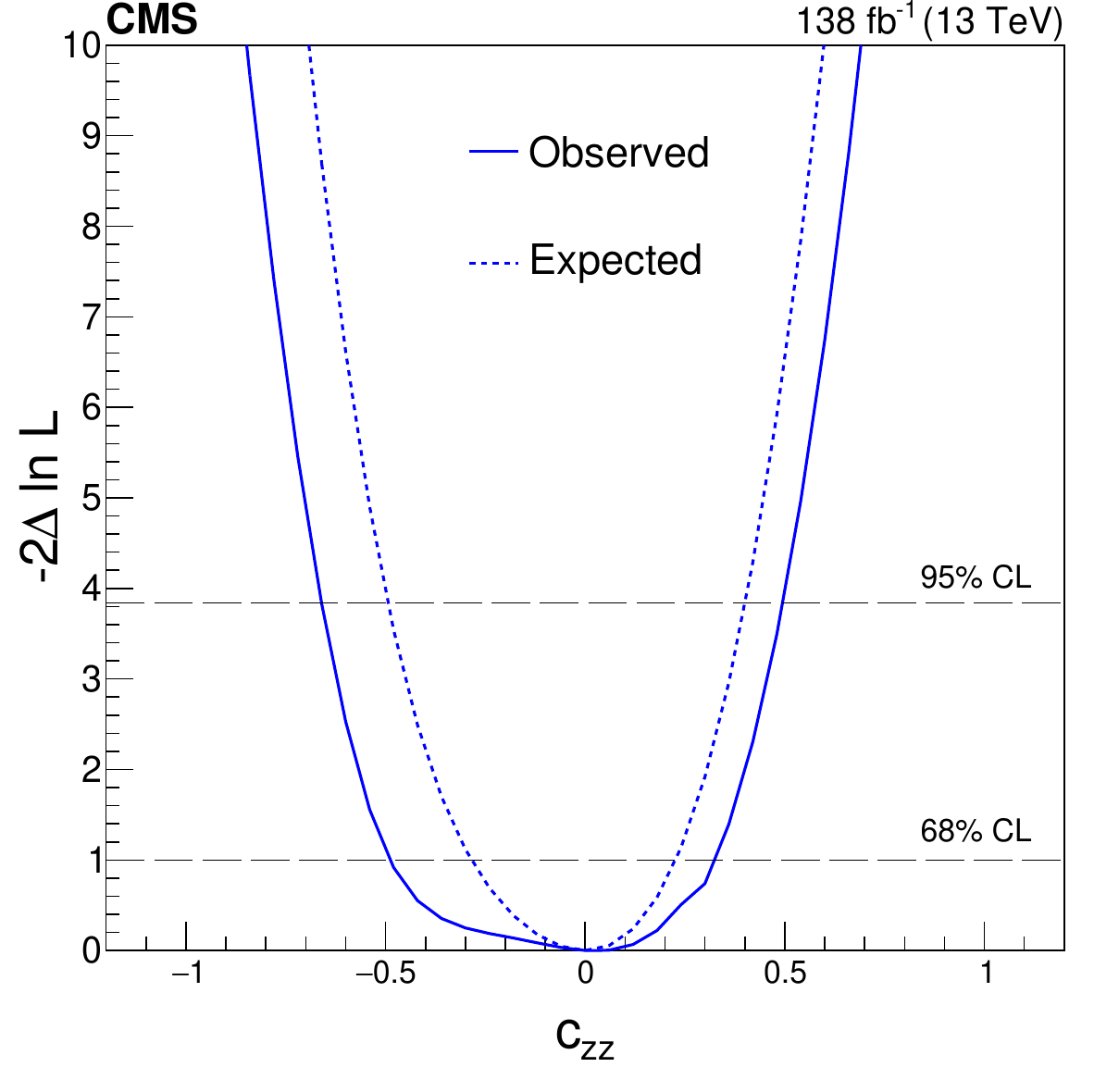}
\includegraphics[width=0.475\textwidth]{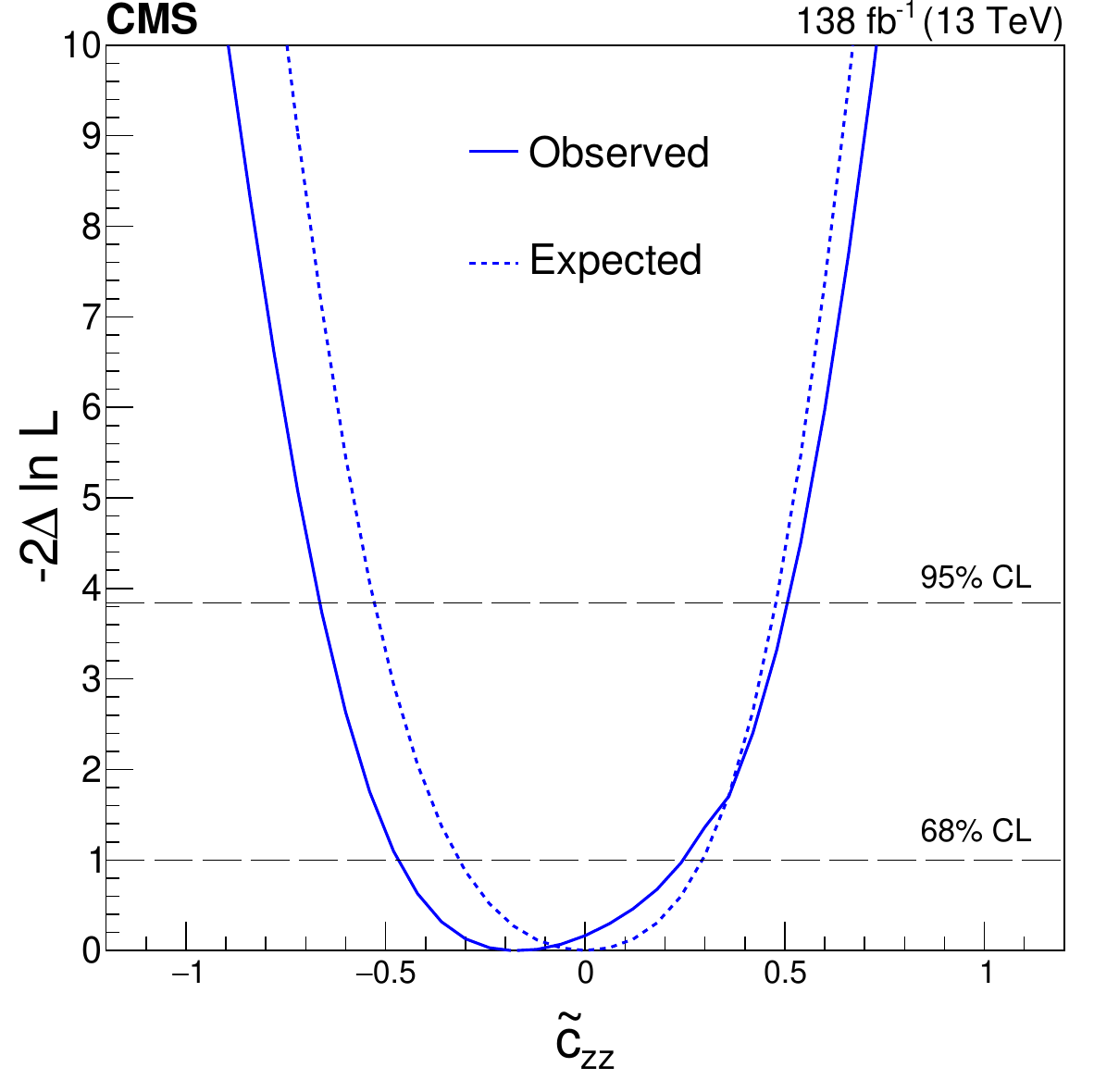}\\
\caption{
Expected (dashed) and observed (solid) profiled likelihood on the $\delta c_\text{z}$ (upper left), $c_{\text{z}\Box}$ (upper right), $c_\text{zz}$ (lower left), and $\tilde{c}_\text{zz}$ (lower right) couplings of the SMEFT Higgs basis. All four couplings are studied simultaneously. The dashed horizontal lines show the 68 and 95\% \CL regions. 
}
\label{fig:HiggsB}
\end{figure*}

\begin{figure}[!htb]
\centering
\includegraphics[width=0.48\textwidth]{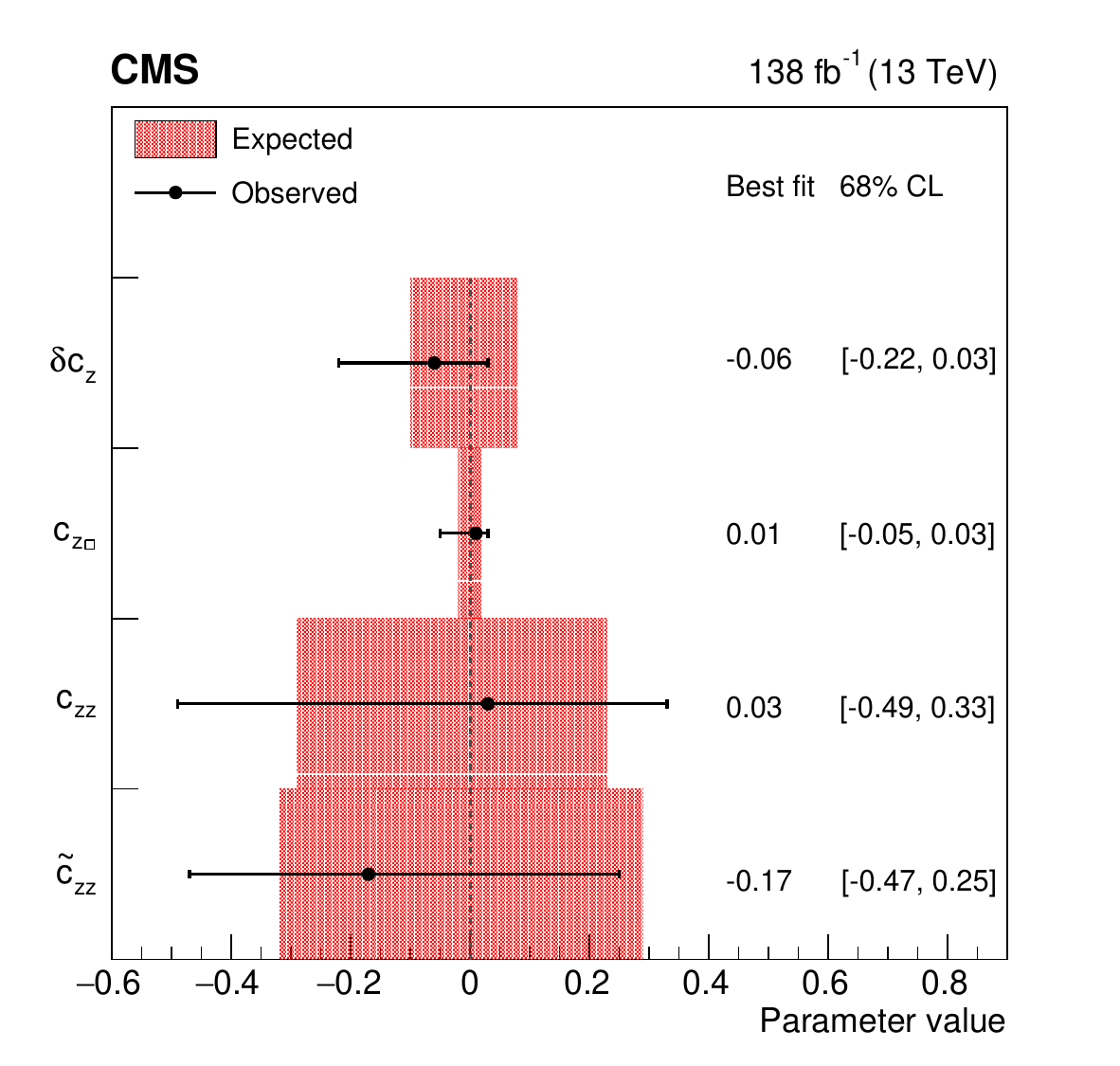}
\includegraphics[width=0.48\textwidth]{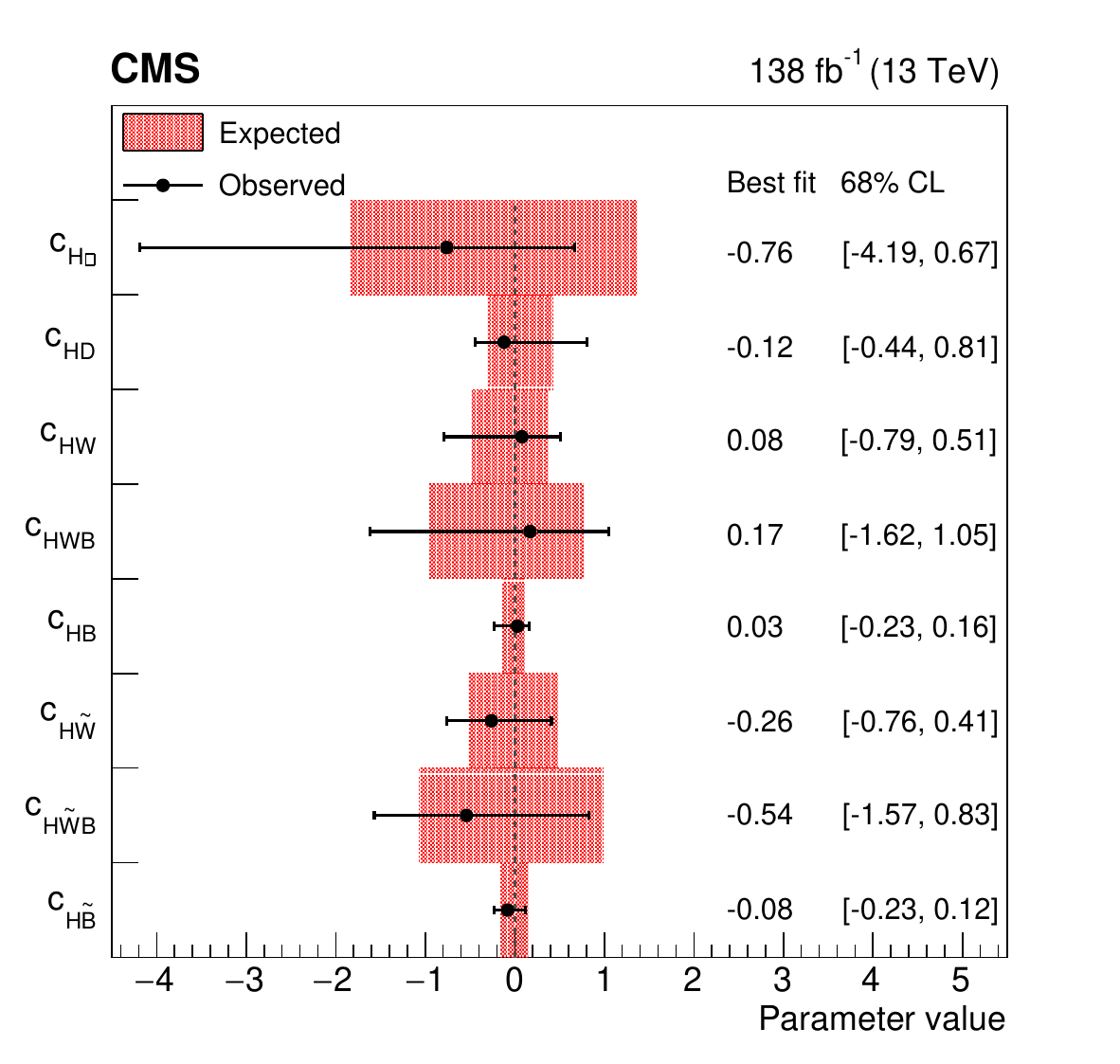}\\
\caption{
Summary of constraints on the SMEFT Higgs (\cmsLeft) and Warsaw (\cmsRight) basis coupling parameters with the best fit values and 68\% \CL uncertainties. For the Warsaw basis, only one of $c_\text{HW}$ , $c_\text{HWB}$, and $c_\text{HB}$ is independent, the same is also true for $c_{\text{H}\tilde{\text{W}}}$, $c_{\text{H}\tilde{\text{W}}\text{B}}$, and $c_{\text{H}\tilde{\text{B}}}$.
}
\label{fig:HVVsumSMEFT}
\end{figure}

For the SMEFT Higgs basis interpretation, the expected and observed constraints on the $\delta c_\text{z}$, $c_{\text{z}\Box}$, $c_\text{zz}$, and $\tilde{c}_\text{zz}$ coupling parameters are shown in Fig.~\ref{fig:HiggsB}. Table~\ref{tab:HVVsumHB} presents a summary of the constraints on the couplings whereas Fig.~\ref{fig:correlations} reports the observed correlation coefficients between them. For the Warsaw basis interpretation, the expected and observed constraints on the $c_{\text{H}\Box}$, $c_\text{HD}$, $c_\text{HW}$, $c_\text{HWB}$, $c_\text{HB}$, $c_{\text{H}\tilde{\text{W}}}$, $c_{\text{H}\tilde{\text{W}}\text{B}}$, and $c_{\text{H}\tilde{\text{B}}}$ coupling parameters are presented in Table~\ref{tab:HVVsumWB}. To cover all the Warsaw basis coupling parameters, three independent fits to the data were performed with a different choice of four independent couplings in each. A summary of the constraints on the SMEFT Higgs and Warsaw basis coupling parameters is presented in Fig.~\ref{fig:HVVsumSMEFT}.

\begin{figure}[!htb]
\centering
\includegraphics[width=0.475\textwidth]{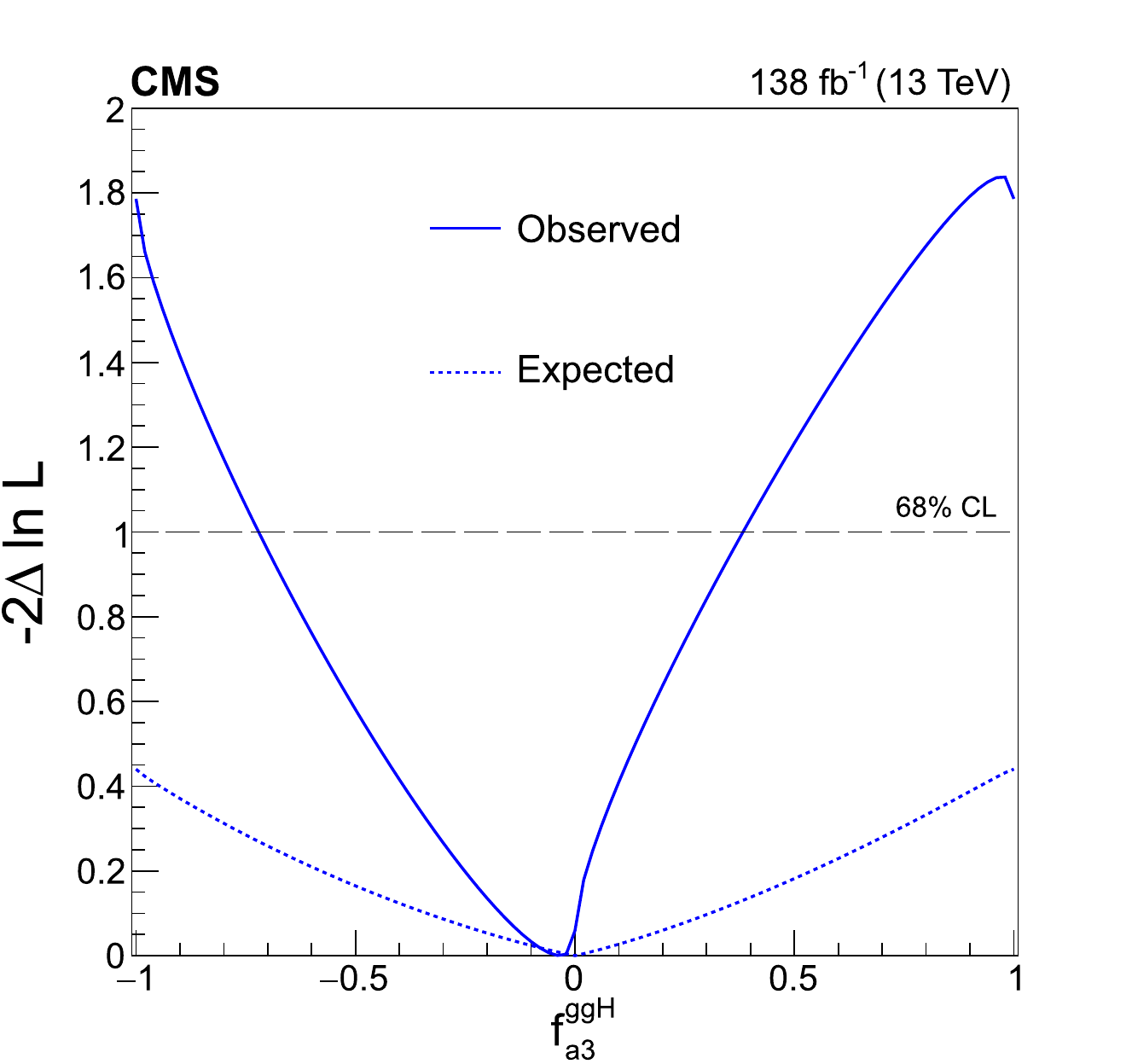}
\caption{
Expected (dashed) and observed (solid) profiled likelihood on \fagg. The signal strength modifiers and the $CP$-odd \HVV anomalous coupling cross section fraction are treated as free parameters. The crossing of the observed likelihood with the dashed horizontal line shows the observed 68\% \CL region.
}
\label{fig:Comb3}
\end{figure}

Finally, the expected and observed \fagg likelihood scans are shown in Fig.~\ref{fig:Comb3}. The result is consistent with the expectation for a SM Higgs boson. 
Excluding the effect of the $CP$-odd \HVV anomalous coupling, by fixing \fathree to zero, has a negligible effect. 
For \abs{\fagg} approaching unity, the observed $-2\Delta\ln{\mathcal{L}}$ profile values are larger than expected. 
This is consistent with downward statistical fluctuations in the data for a couple of bins where sensitivity to the $a_{3}$ \Hgg coupling contribution is enhanced (Fig.~\ref{fig:HGGpostfitRun2} \cmsLeft).  
The constraint on the anomalous \Hgg coupling parameter with the best fit value and allowed 68\% \CL interval is shown in Table~\ref{tab:HVVsum}.

\clearpage

\begin{table*}
\centering
\topcaption{
Summary of constraints on the anomalous \HVV and \Hgg coupling parameters with the best fit values and allowed 68 and 95\% \CL (in square brackets) intervals. For Approach 1, each \fai is studied independently. For Approach 2, each \fai is shown separately with the other two cross section fractions either fixed to zero or left floating in the fit. In each case, the signal strength modifiers are treated as free parameters. 
}
\begin{tabular}{lllll}
Analysis & \fai & & Observed ($\times 10^{-3}$) & Expected ($\times 10^{-3}$)\\
\hline 
\multirow{12}{*}{\HVV $~~~~~~~~~~~ \begin{cases} \\ \\ \\ \\ \\ \\ \\ \\ \\ \\ \end{cases}$ } 
                                     &  \multirow{3}{*}{\fatwo} & best fit & 0.5  & 0.0  \\
\multirow{12}{*}{Approach 1}  &                          & 68\% \CL & [$-$0.8, 3.5] & [$-$1.4, 1.3] \\
                                     &                          & 95\% \CL & [$-$5.7, 12.0]  & [$-$5.2, 6.1] \\

 &  \multirow{3}{*}{\fathree} & best fit   & 0.9 & 0.0 \\
 &                            & 68\% \CL   & [$-$2.7, 4.1] & [$-$0.7, 0.7]  \\
                              & & 95\% \CL & [$-$553.0, 561.0] & [$-$2.8, 2.9] \\

 &  \multirow{3}{*}{\fL}      & best fit   & $-$0.2 & 0.0 \\
 &                            & 68\% \CL   & [$-$0.5, 0.0] & [$-$0.2, 0.5] \\
                              & & 95\% \CL & [$-$1.4, 0.7] & [$-$0.6,1.4]  \\

 &  \multirow{3}{*}{\fLzg}    & best fit   & 3.0 & 0.0 \\
 &                            & 68\% \CL   & [$-$11.0, 9.1] & [$-$5.0, 3.8] \\
                              & & 95\% \CL & [$-$55.0, 42.0]  & [$-$14.0, 11.0]  \\

\multirow{9}{*}{\HVV $~~~~~~~~~~~ \begin{cases} \\ \\ \\ \\ \\ \\ \\ \end{cases}$ } 
                                    &  \multirow{3}{*}{\fatwo} & best fit & 38.0 & 0.0  \\
\multirow{9}{*}{Approach 2}  &                          & 68\% \CL & [$-$112.2, 129.3] & [$-$30.9,37.5] \\
\multirow{9}{*}{(Fix others)}  &                        & 95\% \CL & [$-$376.6, 430.0]$\cup$[$-$989.2, $-$826.3] & [$-$126.1,136.8] \\

 &  \multirow{3}{*}{\fathree} & best fit   & 0.8 & 0.0 \\
 &                            & 68\% \CL   & [$-$0.8, 3.5] & [$-$0.8,1.1] \\
                              & & 95\% \CL & [$-$7.6, 58.8] & [$-$3.4,4.3] \\

 &  \multirow{3}{*}{\fL}      & best fit   & $-$0.15 & 0.0 \\
 &                            & 68\% \CL   & [$-$1.21, 0.16] & [$-$0.4,0.4] \\
                              & & 95\% \CL & [$-$19.5, 118.5]$\cup$[909.9, 964.1] & [$-$1.7,18.9] \\

\multirow{9}{*}{\HVV $~~~~~~~~~~~ \begin{cases} \\ \\ \\ \\ \\ \\ \\ \end{cases}$ } 
                                    &  \multirow{3}{*}{\fatwo} & best fit & $-$1.0  & 0.0  \\
\multirow{9}{*}{Approach 2}  &                          & 68\% \CL & [$-$104.1, 139.9] & [$-$31.1,39.8] \\
\multirow{9}{*}{(Float others)}  &                      & 95\% \CL & [$-$986.4, 981.2] & [$-$127.5,148.7] \\

 &  \multirow{3}{*}{\fathree} & best fit   & 0.34 & 0.0 \\
 &                            & 68\% \CL   & [$-$0.69, 3.4] & [$-$1.0,1.2] \\
                              & & 95\% \CL & [$-$201.3, 361.5] & [$-$4.3,5.3]  \\

 &  \multirow{3}{*}{\fL}      & best fit   & $-$0.1 & 0.0 \\
 &                            & 68\% \CL   & [$-$1.08, 3.78]$\cup$[7.2, 20.7]   & [$-$0.4,0.9] \\
                              & & 95\% \CL & [$-$994.8, 993.9] & [$-$1.9,21.4] \\

\multirow{3}{*}{\Hgg $~~~~~~~~~~~~~ \begin{cases} \\ \\ \end{cases}$ } 
                                    &  \multirow{3}{*}{\fagg} & best fit & $-$34 & 0  \\
                                    &                          & 68\% \CL & [$-$721, 383]  & [$-$1000, 1000] \\
                                    &                          & 95\% \CL & [$-$1000, 1000] & [$-$1000, 1000] \\

\end{tabular}

\label{tab:HVVsum}
\end{table*}

\begin{table}[htb]
\centering
\topcaption{
Summary of constraints on the SMEFT Higgs basis coupling parameters with the best fit values and 68\% \CL uncertainties. All four couplings are studied simultaneously.}
\renewcommand{\arraystretch}{1.3}
\begin{tabular}{ccc}
Coupling & Observed & Expected \\
\hline 
$\delta c_\text{z}$ & $-0.06^{+0.09}_{-0.16}$  & $0.00^{+0.08}_{-0.10}$ \\
$c_{\text{z}\Box}$   & $0.01^{+0.02}_{-0.06}$ & $0.00^{+0.02}_{-0.02}$ \\
$c_\text{zz}$     & $0.03^{+0.30}_{-0.52}$ & $0.00^{+0.23}_{-0.29}$ \\ 
$\tilde{c}_\text{zz}$ & $-0.17^{+0.42}_{-0.30}$ & $0.00^{+0.29}_{-0.32}$ \\ 
\end{tabular}
\label{tab:HVVsumHB}
\end{table}

\begin{table}[!htb]
\centering
\topcaption{
Summary of constraints on the SMEFT Warsaw basis coupling parameters with the best fit values and 68\% \CL uncertainties. 
Only one of $c_\text{HW}$ , $c_\text{HWB}$, and $c_\text{HB}$ is independent, the same is also true for $c_{\text{H}\tilde{\text{W}}}$, $c_{\text{H}\tilde{\text{W}}\text{B}}$, and $c_{\text{H}\tilde{\text{B}}}$.
Three independent fits to the data were performed with a different choice of four independent couplings in each.}
\renewcommand{\arraystretch}{1.3}
\begin{tabular}{ccc}
Coupling & Observed & Expected \\
\hline
 $c_{\text{H}\Box}$ & $-0.76^{+1.43}_{-3.43}$  & $0.00^{+1.37}_{-1.84}$ \\ 
 $c_\text{HD}$ & $-0.12^{+0.93}_{-0.32}$  & $0.00^{+0.43}_{-0.30}$  \\ 
 $c_\text{HW}$  & $0.08^{+0.43}_{-0.87}$ & $0.00^{+0.37}_{-0.48}$   \\
 $c_\text{HWB}$ & $0.17^{+0.88}_{-1.79}$ & $0.00^{+0.77}_{-0.96}$   \\ 
 $c_\text{HB}$ & $0.03^{+0.13}_{-0.26}$ & $0.00^{+0.11}_{-0.14}$  \\ 
 $c_{\text{H}\tilde{\text{W}}}$  & $-0.26^{+0.67}_{-0.50}$ & $0.00^{+0.48}_{-0.52}$    \\ 
 $c_{\text{H}\tilde{\text{W}}\text{B}}$ & $-0.54^{+1.37}_{-1.03}$ & $0.00^{+0.99}_{-1.07}$   \\ 
 $c_{\text{H}\tilde{\text{B}}}$ & $-0.08^{+0.20}_{-0.15}$ & $0.00^{+0.15}_{-0.16}$   \\ 
\end{tabular}
\label{tab:HVVsumWB}
\end{table}

\ifthenelse{\boolean{cms@external}}{\clearpage}{}

\section{Summary}
\label{sec:summary}

This paper presents a study of the anomalous couplings of the Higgs boson (\PH) with vector bosons, including $CP$ violating effects,
using its associated production with hadronic jets in gluon fusion, electroweak vector boson fusion, and associated production with a \PW or \PZ boson,
and its subsequent decay to a pair of \PW bosons. The results are based on the proton-proton collision data set collected by the CMS detector at the LHC during 2016--2018, corresponding to an integrated luminosity of 138\fbinv at a center-of-mass energy of 13\TeV.
The analysis targets the different-flavor dilepton ($\Pe\PGm$) final state, with kinematic information from associated jets combined using matrix element techniques to increase sensitivity to anomalous effects at the production vertex. Dedicated Monte Carlo simulation and matrix element reweighting provide modeling of all kinematic features in the production and decay of the Higgs boson with full simulation of detector effects. A simultaneous measurement of four Higgs boson couplings to electroweak vector bosons has been performed in the framework of a standard model effective field theory. All measurements are consistent with the expectations for the standard model Higgs boson and constraints are set on the fractional contribution of the anomalous couplings to the Higgs boson cross section. The most stringent constraints on the \HVV anomalous coupling cross section fractions are at the per mille level.
These results are in agreement with those obtained in the \HtoZZ and \HtoTT channels, and also significantly surpass those of the previous \HtoWW anomalous coupling analysis from the CMS experiment in both scope and precision.

\begin{acknowledgments}
We congratulate our colleagues in the CERN accelerator departments for the excellent performance of the LHC and thank the technical and administrative staffs at CERN and at other CMS institutes for their contributions to the success of the CMS effort. In addition, we gratefully acknowledge the computing centers and personnel of the Worldwide LHC Computing Grid and other centers for delivering so effectively the computing infrastructure essential to our analyses. Finally, we acknowledge the enduring support for the construction and operation of the LHC, the CMS detector, and the supporting computing infrastructure provided by the following funding agencies: SC (Armenia), BMBWF and FWF (Austria); FNRS and FWO (Belgium); CNPq, CAPES, FAPERJ, FAPERGS, and FAPESP (Brazil); MES and BNSF (Bulgaria); CERN; CAS, MoST, and NSFC (China); MINCIENCIAS (Colombia); MSES and CSF (Croatia); RIF (Cyprus); SENESCYT (Ecuador); ERC PRG, RVTT3 and MoER TK202 (Estonia); Academy of Finland, MEC, and HIP (Finland); CEA and CNRS/IN2P3 (France); SRNSF (Georgia); BMBF, DFG, and HGF (Germany); GSRI (Greece); NKFIH (Hungary); DAE and DST (India); IPM (Iran); SFI (Ireland); INFN (Italy); MSIP and NRF (Republic of Korea); MES (Latvia); LMTLT (Lithuania); MOE and UM (Malaysia); BUAP, CINVESTAV, CONACYT, LNS, SEP, and UASLP-FAI (Mexico); MOS (Montenegro); MBIE (New Zealand); PAEC (Pakistan); MES and NSC (Poland); FCT (Portugal); MESTD (Serbia); MCIN/AEI and PCTI (Spain); MOSTR (Sri Lanka); Swiss Funding Agencies (Switzerland); MST (Taipei); MHESI and NSTDA (Thailand); TUBITAK and TENMAK (Turkey); NASU (Ukraine); STFC (United Kingdom); DOE and NSF (USA).
    
\hyphenation{Rachada-pisek} Individuals have received support from the Marie-Curie program and the European Research Council and Horizon 2020 Grant, contract Nos.\ 675440, 724704, 752730, 758316, 765710, 824093, 101115353, and COST Action CA16108 (European Union); the Leventis Foundation; the Alfred P.\ Sloan Foundation; the Alexander von Humboldt Foundation; the Science Committee, project no. 22rl-037 (Armenia); the Belgian Federal Science Policy Office; the Fonds pour la Formation \`a la Recherche dans l'Industrie et dans l'Agriculture (FRIA-Belgium); the Agentschap voor Innovatie door Wetenschap en Technologie (IWT-Belgium); the F.R.S.-FNRS and FWO (Belgium) under the ``Excellence of Science -- EOS" -- be.h project n.\ 30820817; the Beijing Municipal Science \& Technology Commission, No. Z191100007219010 and Fundamental Research Funds for the Central Universities (China); the Ministry of Education, Youth and Sports (MEYS) of the Czech Republic; the Shota Rustaveli National Science Foundation, grant FR-22-985 (Georgia); the Deutsche Forschungsgemeinschaft (DFG), under Germany's Excellence Strategy -- EXC 2121 ``Quantum Universe" -- 390833306, and under project number 400140256 - GRK2497; the Hellenic Foundation for Research and Innovation (HFRI), Project Number 2288 (Greece); the Hungarian Academy of Sciences, the New National Excellence Program - \'UNKP, the NKFIH research grants K 124845, K 124850, K 128713, K 128786, K 129058, K 131991, K 133046, K 138136, K 143460, K 143477, 2020-2.2.1-ED-2021-00181, and TKP2021-NKTA-64 (Hungary); the Council of Science and Industrial Research, India; ICSC -- National Research Center for High Performance Computing, Big Data and Quantum Computing, funded by the NextGenerationEU program (Italy); the Latvian Council of Science; the Ministry of Education and Science, project no. 2022/WK/14, and the National Science Center, contracts Opus 2021/41/B/ST2/01369 and 2021/43/B/ST2/01552 (Poland); the Funda\c{c}\~ao para a Ci\^encia e a Tecnologia, grant CEECIND/01334/2018 (Portugal); the National Priorities Research Program by Qatar National Research Fund; MCIN/AEI/10.13039/501100011033, ERDF ``a way of making Europe", and the Programa Estatal de Fomento de la Investigaci{\'o}n Cient{\'i}fica y T{\'e}cnica de Excelencia Mar\'{\i}a de Maeztu, grant MDM-2017-0765 and Programa Severo Ochoa del Principado de Asturias (Spain); the Chulalongkorn Academic into Its 2nd Century Project Advancement Project, and the National Science, Research and Innovation Fund via the Program Management Unit for Human Resources \& Institutional Development, Research and Innovation, grant B37G660013 (Thailand); the Kavli Foundation; the Nvidia Corporation; the SuperMicro Corporation; the Welch Foundation, contract C-1845; and the Weston Havens Foundation (USA).    
\end{acknowledgments}

\bibliography{auto_generated}
\cleardoublepage \appendix\section{The CMS Collaboration \label{app:collab}}\begin{sloppypar}\hyphenpenalty=5000\widowpenalty=500\clubpenalty=5000
\cmsinstitute{Yerevan Physics Institute, Yerevan, Armenia}
{\tolerance=6000
A.~Hayrapetyan, A.~Tumasyan\cmsAuthorMark{1}\cmsorcid{0009-0000-0684-6742}
\par}
\cmsinstitute{Institut f\"{u}r Hochenergiephysik, Vienna, Austria}
{\tolerance=6000
W.~Adam\cmsorcid{0000-0001-9099-4341}, J.W.~Andrejkovic, T.~Bergauer\cmsorcid{0000-0002-5786-0293}, S.~Chatterjee\cmsorcid{0000-0003-2660-0349}, K.~Damanakis\cmsorcid{0000-0001-5389-2872}, M.~Dragicevic\cmsorcid{0000-0003-1967-6783}, P.S.~Hussain\cmsorcid{0000-0002-4825-5278}, M.~Jeitler\cmsAuthorMark{2}\cmsorcid{0000-0002-5141-9560}, N.~Krammer\cmsorcid{0000-0002-0548-0985}, A.~Li\cmsorcid{0000-0002-4547-116X}, D.~Liko\cmsorcid{0000-0002-3380-473X}, I.~Mikulec\cmsorcid{0000-0003-0385-2746}, J.~Schieck\cmsAuthorMark{2}\cmsorcid{0000-0002-1058-8093}, R.~Sch\"{o}fbeck\cmsorcid{0000-0002-2332-8784}, D.~Schwarz\cmsorcid{0000-0002-3821-7331}, M.~Sonawane\cmsorcid{0000-0003-0510-7010}, S.~Templ\cmsorcid{0000-0003-3137-5692}, W.~Waltenberger\cmsorcid{0000-0002-6215-7228}, C.-E.~Wulz\cmsAuthorMark{2}\cmsorcid{0000-0001-9226-5812}
% ...
\par}
\vskip\cmsinstskip
\dag:~Deceased\\
$^{1}$Also at Yerevan State University, Yerevan, Armenia\\
$^{2}$Also at TU Wien, Vienna, Austria\\
% ...
\end{sloppypar}
\end{document}